\documentclass[12pt,a4paper]{article}            
 \usepackage[skins,theorems]{tcolorbox}
\tcbset{highlight math style={enhanced,
  colframe=red,colback=white,arc=0pt,boxrule=1pt}}
  \usepackage[bookmarksopen, bookmarksnumbered, bookmarksopenlevel=2]{hyperref}
  \usepackage{tikz}
  \usepackage[all]{nowidow}
  \usepackage{tikz-3dplot}
 \usetikzlibrary{calc}
 \usetikzlibrary{decorations} %
 \usepackage[UKenglish]{babel}
 \usepackage[toc,page]{appendix}
 \usepackage{amsmath}
 \usepackage{amssymb}
 \usepackage{graphicx}
 \usepackage{hhline}
 \usepackage[bf]{caption}
\usepackage{cite}
\usepackage[vcentermath]{youngtab}
\usepackage{geometry}
\usepackage{slashed}
\usepackage{color}
\usepackage{stackrel}
\usepackage{tikz-cd} 
\usepackage{mathtools}
\usepackage{cancel} 
\usepackage{multirow}
\allowdisplaybreaks

\usepackage{empheq}
\usepackage{arydshln}

\newcommand{\bqa}{\begin{eqnarray}}
\newcommand{\eqa}{\end{eqnarray}}

\newenvironment{eqn*}{\begin{equation*}\begin{aligned}}{\end{aligned}\end{equation*}\noindent}
\hypersetup{
    pdftitle={},
    pdfauthor={},
    pdfsubject={}
}
\numberwithin{equation}{section}
\numberwithin{table}{section}

\makeatletter

\newcommand{\be}{\begin{equation}}
\newcommand{\ee}{\end{equation}}
\newcommand{\beq}{\begin{equation}}
\newcommand{\eeq}{\end{equation}}
\newcommand{\ba}{\begin{aligned}}
\newcommand{\ea}{\end{aligned}}

\newcommand{\bea}{\begin{eqnarray}}
\newcommand{\eea}{\end{eqnarray}}

\newcommand{\cO}{\mathcal{O}}
\newcommand{\cT}{\mathcal{T}}
\newcommand{\cE}{\mathcal{E}}
\newcommand{\cP}{\mathcal{P}} 

\newcommand{\cC}{\mathcal{C}}

\newcommand{\cL}{\mathcal{L}}

\newcommand{\cN}{\mathcal{N}}

\newcommand{\cF}{\mathcal{F}}

\newcommand{\cV}{\mathcal{V}}

\newcommand{\cM}{\mathcal M}

\newcommand\bi{\begin{itemize}}
\newcommand\ei{\end{itemize}}

\def\unit{{1\kern-.65ex {\rm l}}}
\def\1{{1\kern-.65ex {\rm l}}}

\def\CT{{\cal T}}

\def\CV{{\cal V}}

\def\CZ{{\cal Z}}

\newcount\hour \newcount\minute
\hour=\time \divide \hour by 60
\minute=\time
\def\now{%
\ifnum \hour<13
  \ifnum \hour=0 \advance \hour by 12 \number\hour:\else \number\hour:\fi%
     \ifnum \minute<10 0\fi%
     \number\minute%
\ A.M.%
\else \advance \hour by -12 \number\hour:%
  \ifnum \minute<10 0\fi%
  \number\minute%
  \ P.M.%
\fi%
}

\def\fnote#1#2{\begingroup\def\thefootnote{#1}\footnote{#2}
     \addtocounter{footnote}{-1}\endgroup}

\makeatother

\begin{document}

\vskip 40 pt
\begin{center}
{\large \bf
Light Strings and Strong Coupling in F-theory
} 

\vskip 11 mm

Max Wiesner

\vskip 11 mm

{\it Center of Mathematical Sciences and Applications, Harvard University, 20 Garden Street,\\ Cambridge, MA 02138, USA}

{\it Jefferson Physical Laboratory, Harvard University, Cambridge, MA 02138, USA}  \\[3 mm]

\fnote{}{mwiesner at cmsa.fas.harvard.edu}

\end{center}

\vskip 7mm

\begin{abstract}
We consider 4d $\cN=1$ theories arising from F-theory compactifications on elliptically-fibered Calabi--Yau four-folds and investigate the non-perturbative structure of their scalar field space beyond the large volume/large complex structure regime. We focus on regimes where the F-theory field space effectively reduces to the deformation space of the worldsheet theory of a critical string obtained from a wrapped D3-brane. In case that this critical string is a heterotic string with a simple GLSM description, we identify new strong coupling singularities in the interior of the F-theory field space. Whereas from the perturbative perspective these singularities manifest themselves through a breakdown of the perturbative $\alpha'$-expansion, the dual GLSM perspective reveals that at the non-perturbative level these singularities correspond to loci in field space along which the worldsheet theory of the critical D3-brane string breaks down and a 7-brane gauge theory becomes strongly coupled due to quantum effects. Therefore these singularities signal a transition to a strong coupling phase in the F-theory field space which can be shown to arise due to the failure of the F-theory field space to factorize between complex structure and K\"ahler sector at the quantum level. Such singularities are hence a feature of a genuine $\cN=1$ theory without a direct counterpart in $\cN=2$ theories in 4d. By relating our setup to recent studies of global string solutions associated to axionic strings we argue that the D3-brane string dual to the perturbative heterotic string leaves the spectrum of BPS strings when traversing into the strong coupling phase. The absence of the perturbative, critical heterotic string then provides a physical explanation for the breakdown of the perturbative expansion and the obstruction of certain classical infinite distance limits in accordance with the Emergent String Conjecture. 
\end{abstract}

\vfill
\setcounter{page}{0}
\thispagestyle{empty}
\newpage

\tableofcontents

\setcounter{page}{1}
\section{Introduction}
In the quest to uncover the fundamental nature of quantum gravity, string theory provides an ideal testing ground to identify general principles that are believed to be valid in any theory of quantum gravity. For that reason, string theory plays a key role in the so-called Swampland program, initiated in \cite{Vafa:2005ui}, that aims to find criteria that any effective theory needs to satisfy in order to arise as a low-energy approximation to quantum gravity. In that context, perturbative string theories provide the most striking evidence e.g. for the Distance Conjecture \cite{Ooguri:2006in} or the Weak Gravity Conjecture \cite{ArkaniHamed:2006dz} (cf. \cite{Brennan:2017rbf,Palti:2019pca,vanBeest:2021lhn,Harlow:2022gzl} for reviews). However, in order to have computational control over the string theory one typically requires the string to be weakly coupled, the effective theory to preserve a large amount of supersymmetry and the vevs of the scalar fields to be tuned to asymptotic regions in the scalar field space where the effective theory can be described by a string compactification on some geometric background. Unfortunately when aiming to understand the full nature of quantum gravity these requirements pose severe limitations. To obtain a more complete picture, one should hence also turn for instance to effective theories with at most $\cN=1$ supersymmetry in four dimensions which are not realized in strict asymptotic regions of the scalar field space in order to avoid the existence of an infinite tower of light states in these regimes as required by the Distance Conjecture. 

Due to the lack of computational control over the corrections to $\cN=1$ theories in four dimensions, the interior of the scalar field space of such theories is relatively unexplored.  Still, one might hope to encounter interesting structures once moving away from the strict asymptotic/weak coupling limits. In this paper, we want to partially address this question for the special case of F-theory compactifications on elliptically-fibered Calabi--Yau four-folds. Such setups give rise to four-dimensional effective theories with $\cN=1$ supersymmetry and thus provide an interesting setting to uncover the structure of the $\cN=1$ scalar field spaces. The asymptotic regions of the scalar field space for these theories have recently been investigated in the context of the flux compactifications \cite{Honma:2017uzn,Grimm:2019ixq, Marchesano:2021gyv, Grimm:2021ckh, Grana:2022dfw} as well as the Weak Gravity and Swampland Distance/Emergent String Conjecture \cite{Lee:2019jan,Lanza:2020qmt,Klaewer:2020lfg,Lanza:2021udy,Cota:2022yjw}. The latter conjecture \cite{Lee:2019oct} states that any infinite distance limit in a consistent theory of quantum gravity is either a limit in which a critical string becomes tensionless and weakly coupled, or a limit in which the theory effectively decompactifies. Among others this conjecture has been shown to hold in the K\"ahler sector of the F-theory scalar field space \cite{Lee:2019jan, Klaewer:2020lfg} and the strict differentiation between emergent string and decompactification limits has recently been stressed again in \cite{Cota:2022yjw}. More precisely, in asymptotic limits in the K\"ahler field space of $\cN=1$ four-dimensional F-theory compactifications  \cite{Cota:2022yjw} showed that there cannot be any tower with mass below the quantum gravity cut-off, i.e. the Planck scale or the species scale \cite{Dvali:2007hz}, that does \emph{not} arise from KK modes of a higher-dimensional theory or the excitations of a critical string. In particular any particle-like string excitations necessarily arise from weakly coupled, genuinely four-dimensional strings obtained by wrapping D3-branes on certain curves in the base of the elliptic CY four-fold which indeed can be shown \cite{Lee:2019jan,Klaewer:2020lfg,Cota:2022yjw} to be always dual to critical type II or heterotic strings. 

In this work we aim to investigate the interior of the F-theory scalar field space, $\cM^F$, away from strict weak coupling points. More precisely, our goal is to uncover the physics in corners of the scalar field space of genuine $\cN=1$ theories in 4d where the asymptotic, weakly-coupled description breaks down. We refer to the loci in field space where the asymptotic description breaks down as the border of the asymptotic region. In this context, it has already previously been noticed \cite{Klaewer:2020lfg} that, for instance, certain regimes in field space that classically look like an asymptotic emergent string region are obstructed due to a breakdown of the perturbative $\alpha'$-expansion. One of our goals in this work is to revisit these obstructions and give a physical explanation for the absence of emergent strings in these regions. 

As mentioned previously, asymptotic regions in the scalar field space of $\cN=1$ theories in 4d have the property that any tower of massive excitations with mass below the quantum gravity cutoff is either made up by KK modes or the excitations of a critical string \cite{Cota:2022yjw}. In this work, we want to exploit this property to find the borders of these asymptotic regimes in $\cM^F$.  For definiteness, we exclusively focus on the case where the light, massive states arise from a critical string. In the regimes of $\cM^F$ where this is the case, the full F-theory effectively reduces to a critical string theory. Such regimes are obtained in the case that a D3-brane wrapped on a curve becomes \emph{classically} lighter than any other stringy scale as the physics associated to these other scales effectively decouples. We are then left with a theory of a single string and its excitations. Though this is similar in spirit to the emergent string limits, unlike for emergent string limits we only require that the D3-brane string becomes light at the classical level and at this point are agnostic about whether it remains light and weakly coupled also at the quantum level. Still, the benefit of such regimes is that we are left with a residual scalar field space which can be identified with the deformation space of the string worldsheet theory. 

In the cases of interest for us, the light string is a critical string and we can thus investigate the properties of the residual scalar field space by studying the deformation space of a critical string in 4d. By the emergent string conjecture the existence of the asymptotic region and the presence of the perturbative excitations of this string are tightly related. In order to identify the borders of the asymptotic region in field space, the relevant question pertinent to the analysis in this paper is whether the light, critical string remains weakly coupled in the interior of the residual field space also at the non-perturabtive quantum level. To answer this question, in practice we restrict to the case that the critical string is a heterotic string whose worldsheet theory allows for a description as the low-energy limit of a $(2,2)$ (or $(0,2)$ deformation thereof) supersymmetric Gauged Linear Sigma Model (GLSM). This case corresponds to heterotic standard embedding and thus to a situation with space-time gauge theory $E_6\times E_8$. Though this is certainly a strong restriction, it enables us to explicitly study the FI-parameter space of the GLSM as a proxy for the quantum K\"ahler deformation space of the heterotic string (cf.  \cite{Morrison:1994fr} and \cite{McOrist:2008ji, Kreuzer:2010ph} for the $(0,2)$ case). Via heterotic/F-theory duality this then translates into a description of the residual F-theory K\"ahler field space in the light string limit.

In fact, the GLSM description allows us to identify a region in the residual scalar field space where the light string ceases to be weakly coupled. We arrive at this conclusion by considering the singular loci in the GLSM FI-parameter space and, more precisely, identify the principal component of the singular locus as being responsible for the light string failing to be weakly coupled. This is due to the fact that in the heterotic theory with standard embedding, the unbroken $E_8$ becomes strongly coupled along this locus in field space. Furthermore, as the correlators of the heterotic worldsheet theory become singular along this locus, also the worldsheet theory of the perturbative heterotic string breaks down entirely. Using heterotic/F-theory duality and employing the perturbative corrections to $\cM^F$ derived in \cite{Grimm:2013gma,Grimm:2013bha,Weissenbacher:2019mef,Klaewer:2020lfg}, we translate the structure of the FI-parameter space into the structure of $\cM^F$. Thereby we are able to identify a strong coupling phase also in the latter. By studying the perturbative corrections to $\cM^F$ it has already been noticed in \cite{Klaewer:2020lfg} that certain limits, in which a critical string becomes classically weakly-coupled, are obstructed since the perturbative $\alpha'$-expansion breaks down in F-theory. Our analysis shows that, at the non-perturbative level, this obstruction is due to the strong coupling phase in $\cM^F$ whose presence we infer via heterotic/F-theory duality and which is closely linked to the failure of $\cM^F$ to factorize in K\"ahler and complex structure sectors at the quantum level. 

Going further, our analysis provides a physical explanation for why this obstruction/break-down of the $\alpha'$-expansion occurs. Though this effect is a genuine property of a $\cN=1$ theory without a direct counterpart in $\cN=2$ theories, we can still draw an analogy to a similar situation in the vector multiplet moduli space of CY threefold compactifcations of type II string theory. More precisely our approach to study the interior of the F-theory scalar field space in regions where the full F-theory reduces to a string theory can be viewed as the $\cN=1$ analogue of studying point particle limits of $\cN=2$ compactifications of type II string theory \cite{Kachru:1995fv}. In these setups the $\alpha'\rightarrow 0$ limit of the type II moduli space can be identified with the Coulomb branch of $\cN=2$ $SU(2)$ SYM theory with D-brane states playing the role of the $W^\pm$-bosons. The conifold singularity of the full type II vector multiplet moduli space can then be identified with the singularities on the Coulomb branch at which the gauge theory becomes strongly coupled. In our $\cN=1$ version of this, the relevant BPS objects are not particles but $\frac12$-BPS strings. Still, in a similar spirit, the theory associated to the $\frac12$-BPS string becomes strongly coupled at the singularity in the field space and we expect a strong coupling phase beyond that point. In fact, following the approach of \cite{Marchesano:2022avb}, we can also identify the singularity with a non-critical, non-geometric string that becomes light along that locus replacing the D3-brane string as the fundamental BPS object. Since the critical string fails to be weakly-coupled and to be part of the BPS spectrum in the strong coupling phase, it also does not provide us with a tower of perturbative string excitations. The absence of such a tower then implies that we also reached the border of the asymptotic region in field space and any attempt to extend the asymptotic region into the strong coupling region should be obstructed.  Hence, the failure of the critical string to be part of the BPS-string spectrum provides a physical explanation why certain classically allowed emergent string limits in $\cM^F$ are obstructed \cite{Klaewer:2020lfg}.  \\

This paper is structured as follows: In section \ref{sec:setup} we introduce the general setup and review basic properties of the scalar field space of $\cN=1$ F-theory compactifications to 4d. We further introduce the relevant string theory limits of F-theory and draw the analogy to the field theory limits of $\cN=2$ string theory compactifications. In section \ref{sec:dual} we then take a closer look at the deformation space of the worldsheet theory of the light string to which the F-theory scalar field space reduces in the string theory limit. In particular, we identify candidates for strong coupling singularities in the associated GLSM FI-parameter space that can be responsible for obstructions to the classical light string limits.  Based on heterotic/F-theory duality, we show in section \ref{sec:strongcouplingF} that these strong coupling singularities also arise in the F-theory scalar field space. We further discuss the physical interpretation of the F-theory strong coupling phases in terms of the $\frac12$-BPS string spectrum. We present our conclusions in section \ref{sec:discussion}. The appendices \ref{sec:hetFduality} and \ref{app:GLSMs} provide some background information about heterotic/F-theory duality and GLSMs. 
\section{Light string limits in F-theory}\label{sec:setup}
In this section, we set the stage for our analysis in the rest of the paper. In particular, we review certain light string limits in F-theory that play a central role throughout our analysis. Such light string limits have been investigated previously in \cite{Lee:2019jan, Klaewer:2020lfg}.  Here, we would like to revisit such limits in order to get insights into the non-perturbative structure of the F-theory scalar field space, $\mathcal{M}^F$. In section \ref{sec:lightstringlimits} we start by giving some background on the F-theory scalar field space and introduce the classical light string regimes pertinent to the analysis in this paper. In section \ref{sec:N2comparison} we then discuss the analogy between these \textit{string theory limits} in four-dimensional $\cN=1$ compactifications of F-theory and \textit{field theory limits} in four-dimensional $\cN=2$ compactifications of type II string theory. This analogy then serves as a guide to the general strategy to analyze the properties of the F-theory scalar field space in such string theory limits which we summarize in Section~\ref{sec:generalprinciple}. 

\subsection{General setup}\label{sec:lightstringlimits}
In this work, we are primarily interested in F-theory compactifications on elliptically-fibered Calabi--Yau four-folds. The resulting effective four-dimensional theory has $\cN=1$ supersymmetry and its effective action has been derived in detail in \cite{Grimm:2010ks}. Let us review the central aspects: The scalar fields of this effective theory are part of chiral multiplets  and can be associated to the complex structure deformations of the CY four-fold $X_4$ and the (complexified) K\"ahler deformations of the base, $B_3$, of the elliptic fibration $\pi: X_4\rightarrow B_3$. In the limit of large volume and large complex structure, the scalar field space effectively factorizes as 
\begin{equation}\label{eq:factM}
 \mathcal{M}_{\rm chiral}^F \longrightarrow \mathcal{M}_{\rm c.s.}^F \times \mathcal{M}_{\rm cK}^F\,. 
\end{equation}
The first factor corresponds to the complex structure deformations of $X_4$ whereas the second factor is spanned by the complexified K\"ahler deformations parametrized by 
\begin{equation}\label{eq:defSi}
 S_i = \frac{1}{2} \int_{D_i} J\wedge J + \int_{D_i} C_4\,. 
\end{equation}
Here $D_i$, $i=1,\dots, h^{1,1}(B_3)$, are generators of the cone of \textit{effective} divisors on $B_3$, $J$ is the K\"ahler form on $B_3$ and $C_4$ the type IIB RR four-form. The classical factorization \eqref{eq:factM} translates into a factorized form for the K\"ahler potential 
\begin{equation}\label{eq:Kahlerpot}
 K= - \log \left(\int_{X_4} \Omega_4\wedge \bar{\Omega}_4 \right) -2 \log \left(\mathcal{V}_{B_3}\right)\,,
\end{equation}
where $\Omega_4$ is the holomorphic $(4,0)$-form on $X_4$ whose variation is associated to complex structure deformations and $\mathcal{V}_{B_3}$ is the volume of the base $B_3$ as a function of the $S_i$. Unlike in e.g. Calabi--Yau three-fold compactifications of type II string theory, this factorization of the scalar field space does not necessarily survive in the interior of the scalar field space since quantum effects can mix the K\"ahler and complex structure sectors. However, in the large volume/large complex structure limit this factorization is approximately realized. Furthermore, in the large volume limit the K\"ahler potential enjoys a couple of shift symmetries. Among others 
\begin{equation}
 S_i \rightarrow S_i +ic_i \,,\qquad c_i\in \mathbb{R}\,,
\end{equation}
implying that the classical K\"ahler potential only depends on $\text{Re}\,S_i$. This shift symmetry allows to dualize the chiral multiplets into linear multiplets with scalar component given by the real fields
\begin{equation}\label{eq:defLi}
 L^i = -\frac12 \frac{\partial K}{\partial \text{Re}\,S_i}\equiv \frac12 \frac{l^i}{\mathcal{V}_{B_3}}\,,  
\end{equation} 
where $l^i$ are the volumes of the curves $C^i\in H_2(B_3)$ dual to the generators, $D_i$, of the cone of effective divisors. Furthermore, the axion $\text{Im}\,S_i$ gets dualized into a two-form obtained by reducing $C_4$ along $C^i$. Instead of working with the linear multiplets, for most of the paper we directly refer to the volumes $l^i$ which can be thought of as rescaled versions of the linear multiplets $l^i=2\mathcal{V}_{B_3} L^i$. 

Notice that unlike for $\cN=2$ theories, the scalar field space of $\cN=1$ theories is not necessarily an actual moduli space since the scalar fields can become massive due to the presence of a non-trivial scalar potential
\begin{equation}
 V=e^K \left(g^{i\bar j}D_i W \bar{D}_{\bar j}\bar W -3|W|^2\right)\,,
\end{equation}
where $W$ is a superpotential and $D$ is the K\"ahler covariant derivative. In general, $W$ receives non-perturbative contributions due to D3-brane instantons. In fact, it is expected \cite{Palti:2020qlc} that $W$ only vanishes identically if it is protected by supersymmetry which is the case if the $\cN=1$ theory is related to an $\cN=2$ theory. Hence in general at least some directions of $\cM^F$ correspond to massive scalar fields. In the following, however, we treat $\cM^F$ as a quasi-moduli space and for most of the discussion are agnostic about the presence of non-perturbative contributions to the scalar potential.  

The main focus of this work is on the K\"ahler sector of $\cM^F$ although the mixing with the complex structure sector in the interior of $\cM^F$ is going to play a crucial role later on. There are two different parameterizations for $\cM_{\rm cK}^F$. On the one hand one can consider the scalar fields $S_i$ as defined in \eqref{eq:defSi} as the complexified volume of \textit{effective} divisors. On the other hand, in analogy to the $S_i$ we can define a different set of complex scalar fields given by  
\begin{equation}\label{eq:defTi}
 T_i = \frac{1}{2} \int_{J_i} J\wedge J + \int_{J_i} C_4\,. 
\end{equation}
These can be thought of as the complexified volumes of the generators $J_i$ of the K\"ahler cone. Accordingly, we can also expand the K\"ahler form $J$ of $B_3$ in two ways 
\begin{equation}\begin{aligned}
 J& = l^i D_i\,,\\
 J&= v^i J_i \,.
\end{aligned}\end{equation}
Here $l^i$ are the curve volumes appearing in \eqref{eq:defLi} whereas $v^i$ are the volumes of the curve classes generating the Mori cone. We are now interested in limits in the classical K\"ahler cone where the effective F-theory reduces to a string theory. This is achieved if $l^j \rightarrow 0$ for some $j\in \{1,\dots, h^{1,1}(B_3)\}$ while keeping $l^i>1$  for all $i\neq j$.  To see this, notice that a D3-brane wrapped on a holomorphic curve $C=a_iC^i$  in $B_3$, where $\{C^i\}$ is a basis of movable curves, gives rise to a string in 4d. The tension of this string is classically determined by the volume of the holomorphic curve, i.e.
\begin{equation}
\frac{T}{M_{\rm IIB}^2}= 2\pi \mathcal{V}_C = 2\pi \,a_i l^i \,. 
\end{equation} 
More precisely, we are interested in regimes where a \textit{unique} D3-brane string becomes classically lighter than any other string. We call such a regime the \textit{string theory limit} of F-theory because in this case we classically have a string that is lighter than any other mass scale in the theory.\footnote{Notice that this is related, but not equivalent to the emergent string limits defined in \cite{Lee:2019oct}. The difference between our string theory limit and the emergent string limit is made more precise below.} In such a regime, $\cM^F$  effectively reduces to the deformation space of the light string which, in favorable cases, can be studied from the worldsheet perspective of the string. A particularly interesting case for us arises if the light string obtained in the limit $l^j\rightarrow 0$ is a critical, perturbative string since in this case we \textit{do} have a worldsheet description.\footnote{In fact it can be shown that this is the only possibility for $l^j\rightarrow 0$ while keeping $l^i>1$ for all $i \neq j$ \cite{Cota:2022yjw}.} This possibility has been investigated in detail in \cite{Lee:2019jan, Klaewer:2020lfg} where it was shown that the light string is indeed either a critical heterotic or a type II string.  

To obtain such a situation, one needs $B_3$ to be either a rational or a genus-one fibration \cite{Lee:2019jan}. For this to be the case there has to exist (at least) one K\"ahler cone generator $J_1$ satisfying
\begin{equation}
 J_1^3= 0\,, \qquad J_1^2\neq 0\,.   
\end{equation}
In this case,\footnote{There is also the possibility that $J_1^2=0$, dubbed $J$-class B in \cite{Lee:2019jan}, but we focus on the $J_1^2\neq 0$ situation here.} $B_3$ can be viewed as a fibration $\rho:C^0 \rightarrow B_2$ with $C^0=J_1.J_1$ the fibral class with genus 
\begin{equation}\begin{aligned}
 g(C^0)&=0 \,,\qquad \text{if} \qquad \bar{K}_{B_3}.C^0=2\,,\\
 g(C^0)&=1\,,\qquad \text{if} \qquad \bar{K}_{B_3}.C^0=0\,,
\end{aligned}\end{equation}
 where $\bar{K}_{B_3}$ is the anti-canonical divisor of $B_3$. For this fibration, the set of K\"ahler cone generators can be split into two different sets $\mathcal{I}_1$ and $\mathcal{I}_3$ according to their intersections with $J_1$ \cite{Lee:2019jan}: 
\begin{equation}\begin{aligned}\label{eq:intersections}
  J_\mu . J_1^2 \neq 0\,,\qquad & \forall \mu \in \mathcal{I}_1\,,\\
  J_a . J_1^2 \neq 0\,,\qquad & J_a.J_b.J_1=0 \,,\qquad \forall a,b\in \mathcal{I}_3\,. 
\end{aligned}\end{equation}
Notice that in particular $1\in \mathcal{I}_3$.\footnote{In general, we can interpret the K\"ahler cone generators $J_a$ with $a\in \mathcal{I}_3$ as vertical divisors with respect to $\rho$ obtained by pulling back K\"ahler cone generators of $B_2$ to the full fibration.} From the above we infer that the volume of $C^0$ is given by 
\begin{equation}
\mathcal{V}_{C^0} = \sum_{\mu \in \mathcal{I}_1} \kappa_{11\mu} v^\mu\,. 
\end{equation}
Since $C^0$ is the class of the generic fiber of $B_3$ it is in fact a generator of the cone of movable curves. The dual generator of the cone of effective divisors is given by the zero section $D_0$ of the fibration $\rho$. We can therefore identify 
\begin{equation}
\mathcal{V}_{C^0} = l^0 \,. 
\end{equation}
The string theory limit in which we are interested corresponds to the limit $l^0\rightarrow 0$. In order not to spoil the F-theory description from the get-go, we need to ensure that $\mathcal{V}_{B_3}$ remains finite in this limit. Given the intersection numbers \eqref{eq:intersections} of $B_3$, one finds that to leading order in the limit $l^0\rightarrow 0$
\begin{equation}\label{VB3approx}
  \mathcal{V}_{B_3} = \frac12 l^0 (v^1)^2 + \dots\,,
\end{equation} 
where the dots stand for terms including $v^r$ for $1\neq r\in \mathcal{I}_3$. In order to keep $\mathcal{V}_{B_3}$ finite we thus need to ensure \cite{Lee:2019jan} 
\begin{equation}\label{eq:relativescaling}
 l^0\succsim \frac{1}{(v^1)^2}\,,
\end{equation}
which implies $v^1\rightarrow \infty$ as $l^0\rightarrow 0$.\footnote{We may need to blow-up additional $v^a$ with $a\in \mathcal{I}_3$. To keep the discussion simple here we focus only on the single modulus $v^1$.} This, in turn, implies that the volume of the effective divisor $D_0$ dual to $C^0$, given by \eqref{eq:defSi}, also blows up as $\text{Re}\,S_0\sim \frac12 (v^1)^2$. This ensures that, at the classical level, the light string is also weakly coupled. Since the string is weakly-coupled it can be treated perturbatively and $\cM^F$ indeed reduces to the deformation space of the worldsheet theory on the string.

Notice that in this limit, $\mathcal{V}_{B_3}$ factorizes and by \eqref{eq:Kahlerpot} the scalar field space also factorizes further. This factorization is a crucial property of a weakly-coupled string theory limit of F-theory since one factor of the field space can be interpreted as the string coupling of the light string whereas the remainder can be treated as the deformation space of the worldsheet theory of the light string. Such a factorization is characteristic of a perturbative string theory. Our analysis in the following is based on the fact that at the classical level this factorization holds such that in the limit $(l^0,v^1)\rightarrow (0, \infty)$ the full F-theory reduces to a critical string theory and we can use the duality to a critical string theory to analyze the non-perturbative aspects of $\cM^F$.

It is interesting to notice that the kind of light string regimes in $\cN=1$ theories considered here are analogous to certain field theory limits in the moduli space of four-dimensional $\cN=2$  theories obtained from Calabi--Yau compactifications of type IIA string theory. Since the comparison to the $\cN=2$ field theory limits is quite illuminating we are going to review these limits next.

\subsection{\texorpdfstring{Comparison to $\cN=2$ field theory limits}{Comparison to N=2 field theory limits}}\label{sec:N2comparison}
Let us briefly digress from our discussion of string theory limits in 4d $\cN=1$ theories and turn to similar limits in $\cN=2$ type II string compactifications. In particular we are interested in limits in which the full string theory effectively reduces to a field theory and the stringy moduli space reduces to the Coulomb branch of an $\cN=2$ SYM theory. Such limits can be viewed as the analogues of the limits in $\cM^F$ where the theory reduces to a string theory associated to a single, critical, $\frac12$-BPS string. The field theory limits of the string theory in question have first been studied in \cite{Kachru:1995fv} for the case of type IIA compactifications on K3-fibered Calabi--Yau three-folds. 

We are now going to review the key features of the analysis of \cite{Kachru:1995fv}. Therefore, consider type IIA string theory compactified on a (suitably blown-up) degree-12 hypersurface in the weighted projective space $\mathbb{P}_{1,1,2,2,6}$. The resulting smooth Calabi--Yau three-fold has $h^{1,1}=2$ and can be viewed as a one-parameter K3-fibration over $\mathbb{P}^1$. Let us denote the complexified K\"ahler parameters by $(S_{\rm IIA},T_{\rm IIA})$, where $S_{\rm IIA}$ is the volume of the base $\mathbb{P}^1$ and $T_{\rm IIA}$ the volume of a holomorphic curve in the K3-fiber. In a certain corner of the moduli space, this theory effectively reduces to a field theory. More precisely, this happens in the vicinity of the large volume divisor $S_{\rm IIA}\rightarrow i\infty$ which effectively corresponds to the limit $\alpha'\rightarrow 0$, i.e. the point-particle limit of string theory. Of particular interest is the intersection of this large volume divisor and the discriminant divisor given by the vanishing of 
\begin{equation}\label{DeltaIIA}
\Delta_{\rm IIA} =  \left(1-1728 q_T\right)^2 - 4\cdot 1728^2 q_S q_T^2 = 0\,.
\end{equation}
Here, we introduced the exponentiated K\"ahler parameters $q_T=e^{2\pi iT_{\rm IIA}}$ and $q_S=e^{2\pi i S_{\rm IIA}}$. In these coordinates, there is point of tangency between the loci $\{\Delta_{\rm IIA}=0\}$ and $\{S_{\rm IIA}=i \infty\}$ at
\begin{equation}\label{qTqS}
 (q_T,q_S) = \left( \frac{1}{1728},0\right)\,.  
\end{equation}
In the vicinity of this point one can identify $S_{\rm IIA}$ with the gauge coupling of an $\cN=2$ $SU(2)$ SYM theory whereas $T_{\rm IIA}$ corresponds to the vev of the $\mathfrak{su}(2)$-adjoint scalar $\Phi=a\sigma_3$, with $\sigma_3$ the third Pauli matrix, via 
\begin{equation}\label{heteroticnaivecoordinates}
 a =\frac{T_{\rm IIA}-i}{T_{\rm IIA}+i}\,. 
\end{equation}
From the field theory perspective, one would classically expect the $W^\pm$ bosons of $SU(2)$ to become massless at the point $a=0$, i.e. $T_{\rm IIA}=i$. In the  string theory realization, these states correspond to D4-branes wrapped on the K3-fiber with two units of D0-brane charge (cf. \cite{Lee:2019oct} for a detailed discussion). However, from Seiberg--Witten theory \cite{Seiberg:1994rs} we know that non-perturbative effects become important in the vicinity of $a=0$ and that the point $a=0$ can in fact never be reached at finite coupling. Instead, once $a^2\sim \Lambda^2$, where $\Lambda$ is the dynamically generated scale of the gauge theory, the theory reaches a strong coupling phase and the magnetic monopole becomes light and part of the BPS spectrum. In string theory, this monopole is given by the D6-brane wrapped on the entire Calabi--Yau three-fold and the singularity of the Coulomb branch can be identified with a point on the singular divisor $\{\Delta_{\rm IIA}=0\}$ given by \eqref{DeltaIIA} for some fixed $S_{\rm IIA}$. Accordingly, the Coulomb branch is a good approximation to the local moduli space in the vicinity of $T_{\rm IIA}=i$ with the presence of the singularity $\{\Delta_{\rm IIA}=0\}$ spoiling a weak-coupling description in accordance with the expectation from field theory. From SW theory we know that away from the classical, perturbative limit ($a\rightarrow \infty$), the coordinate $a$ in \eqref{heteroticnaivecoordinates} should be replaced by its dual, $a_D$. The local moduli space around $T_{\rm IIA}=i$ is thus best described by this coordinate. And indeed, as shown in \cite{Kachru:1995fv} one can identify the classical moduli $S_{\rm IIA}$ and $a$ as the A-periods of the Calabi--Yau in a suitable basis 
\begin{equation}\label{Aperiods}
 \text{A-periods}: \qquad (1,S_{\rm IIA}, a)\,,
\end{equation} 
while the dual periods are given by 
\begin{equation}
\text{B-periods}:\qquad (2\cF - S_{\rm IIA}\partial_{S_{\rm IIA} } \cF- a\partial_a F, \partial_{S_{\rm IIA}} \cF,  \partial_a \cF)\,,
\end{equation}
where the prepotential $\cF$ reads
\begin{equation}\label{prepHEt1}
 \cF = S_{\rm IIA} a^2 + \sum_n c_n a^{2-4n}\Lambda^{4n} \exp(-n S_{\rm IIA})\,.
\end{equation}  
Up to linear combinations the B-periods can be identified with the coordinates on the $SU(2)$ Coulomb branch 
\begin{align*}
(a_D, u , uS_{\rm IIA})\,,
\end{align*}
where the central charge of the monopole is given by $a_D=\partial_a \mathcal{F}$ and $u$ is the gauge invariant Casimir $u(a)=\text{Tr} \,\Phi^2=2a^2+\dots$. In fact to get the correct normalization of the Coulomb branch one has to perform the rescaling 
\begin{equation}\label{rescalingSW}
 \tilde u = \frac{u}{\Lambda^2 \exp(-\hat S_{\rm IIA})}\,,
\end{equation} 
in which case the singular locus is located at $\tilde u=1$. Here $\hat{S}_{\rm IIA}$ is related to the actual gauge coupling $S_{\rm IIA}$ as 
\begin{equation}\label{hatSIIA}
 S_{\rm IIA} =  \hat S_{\rm IIA} - \log(\Lambda^4 \alpha'^2)\,. 
\end{equation}
The coordinate $\tilde u$ can then be identified with the coordinate on the blow-up $\mathbb{P}^1$ in the mirror complex structure moduli space used to resolve the point of tangency \eqref{qTqS}. The singularity described by \eqref{rescalingSW} corresponds to the intersection between the blow-up $\mathbb{P}^1$ and the discriminant locus $\{\Delta_{\rm IIA}=0\}$. Notice that in the strict $S_{\rm IIA}\rightarrow i\infty$ limit, the entire strong coupling phase of the Coulomb branch gets mapped to a single point as expected for a point of tangency. 

The setup just reviewed illustrates that, indeed, in the vicinity of the point \eqref{qTqS} the full string theory reduces to a field theory. In particular, the singular locus $\{\Delta_{\rm IIA}=0\}$ reproduces the obstruction to reaching the gauge enhancement point in the $SU(2)$ SYM theory. We now want to compare this situation to the setup we are interested in. Instead of the point-particle limit of string theory, we are now interested in the 'string theory limit' of F-theory. As described in the previous section, by this limit we mean the limit where the quasi-moduli space can effectively be described by the moduli space of the theory living on the worldsheet of a weakly-coupled, critical string. More precisely, we are interested in the non-perturbative structure of the F-theory moduli space in the vicinity of the point where classically the string decouples from all other scales of the theory, i.e. the limit $l^0\rightarrow 0$ in the convention of section \ref{sec:lightstringlimits}. This is the analogue of the limit $a\rightarrow 0$ in the field theory case. In particular, we may ask whether we also observe a non-perturbative obstruction against reaching this limit at finite coupling similar to the obstruction of reaching the gauge enhancement point in $SU(2)$ SYM theories at finite coupling. 

To make this analogy more concrete, we should think of the $\frac12$-BPS string with tension $l^0$ in the $\cN=1$ theory as the analogue of the BPS particle (the W-boson) with mass $2a$ in the $\cN=2$ theory. We should further think of the modulus $v^1$ appearing in \eqref{VB3approx} as the analogue of the tree-level gauge coupling of the $SU(2)$ gauge theory, $S_{\rm IIA}$. To summarize the analogy identifies\footnote{Notice that this identification should not be taken literally but merely to provide some intuition for the $\cN=1$ case. In particular, the field space on the $\cN=2$ side is complex whereas the linear multiplet space on the $\cN=1$ side is real.}
\begin{equation}\label{eq:identifyA}
 \left(\text{Im}\,S_{\rm IIA},  \text{Im}\,a=\text{Im}\left(\frac{T_{\rm IIA}-i}{T_{\rm IIA}+i}\right)\right) \quad \longleftrightarrow \quad (v^1, l^0)\,. 
\end{equation} 
On the other hand, in the vicinity of the strong coupling singularity of the $\cN=2$ gauge theory, we are advised to use the dual coordinates $a_D$ and replace the Coulomb-branch parameter $a$ by $u\simeq2 a^2+\dots$. In the $\cN=1$ case, instead of the (rescaled) linear multiplets we should then consider the dual coordinates $T_1$ and $S_0$ defined in \eqref{eq:defTi} and \eqref{eq:defSi}, respectively. Comparing with the $\cN=2$ case, we can draw the analogy 
\begin{equation}\label{eq:identifyB}
 ( a_D, u , u S_{\rm IIA}) \quad {\longleftrightarrow} \quad ( S_0, T_1, \mathcal{V}_{B_3})\,,
\end{equation} 
such that the good coordinate on the residual moduli space is $T_1$. Given the analogy to the $\cN=2$ case we expect $S_0$ to vanish on a possible strong-coupling singularity. 

\subsection{General strategy}\label{sec:generalprinciple}
The goal of this work is to identify a possible strong coupling singularity in $\cM^F$ in the limit $M_{\rm string}/M_{\rm IIB} \rightarrow 0$, where $M_{\rm string}$ is set by the tension of the light, critical string. This strong coupling singularity should be thought of as the analogue of $\alpha'\rightarrow 0$ in the $\cN=2$ string theory discussed above. To that end, we first want to identify a strong coupling singularity in the residual field space orthogonal to the weak coupling limit. As alluded to before, this field space can be identified with the deformation space of the theory realized on the worldsheet of the light string. In general the worldsheet theory of a D3-brane wrapped on a curve inside the base of an elliptically-fibered CY four-fold corresponds to a Non-Linear Sigma Model (NLSM) with $(0,2)$ supersymmetry in two dimensions, cf. \cite{Lawrie:2016axq} for a detailed discussion. We are particularly interested in the case that the D3-brane string can be identified with a heterotic string in which case the NLSM is specified by a target space given by an elliptically-fibered CY three-fold together with a choice of gauge bundle \cite{Lee:2019jan}. Of particular interest to us is further the case that this NLSM can be viewed as the low-energy limit of a Gauged Linear Sigma Model (GLSM) with some Abelian gauge group $U(1)^n$ (cf. appendix \ref{app:GLSMs} for a review of essentials of GLSMs). The benefit of this restriction is that in this case the deformation space of the worldsheet theory can be described quite explicitly even at the quantum level. More precisely, the quantum K\"ahler field space $\cM_{\rm qK}^H$ of the heterotic string can be described in terms of the FI-parameter space associated to the $U(1)^n$ gauge group once the K\"ahler deformations are identified with the FI-parameters via (a $(0,2)$ version of) the mirror map.

As reviewed in Appendix~\ref{sec:hetFduality}, F-theory/heterotic duality relates the volumes of $\rho$-vertical divisors of the F-theory base $B_3$ to the K\"ahler moduli, i.e. curve volumes in the base of the elliptically-fibered CY target space of the heterotic NLSM. These can then be identified with the GLSM FI-parameters. We thus have the chain of identifications
\begin{center}
\begin{tabular}{ccccc}
 \begin{tabular}{@{}c@{}}F-theory divisor \\  volumes \end{tabular}& $\stackrel{\text{F/het-duality}}{\longleftrightarrow}$ &  \begin{tabular}{@{}c@{}}Heterotic curve\\ volumes \end{tabular}&  $\stackrel{\text{$(0,2)$ Mirror map}}{\longleftrightarrow}$ &\begin{tabular}{@{}c@{}}GLSM \\ FI-parameters \end{tabular}
\end{tabular} 
\end{center}
where the F-theory divisors are $\rho$-vertical and the heterotic curves correspond to curves the base $B_2$ of the elliptically-fibered CY three-fold. From \eqref{eq:identifyB} we then conclude that the FI parameters of the GLSM take over the role of the Coulomb branch parameter $u$ in the field theory limit of $\cN=2$ string theory discussed in the previous section. In order to infer a possible strong coupling singularity in $\cM^F$ we are thus advised to first study the singularity structure of the FI-parameter space: In general this space has a number of singular loci corresponding to the zero set of the discriminant
\begin{equation}
 \Delta^H = \Delta^H_1(q_H) \cdots \Delta^H_n(q_H)=0\,.
\end{equation} 
Here $q_H$ stands for the exponentiated FI parameters of the heterotic GLSM. Via the $(0,2)$ mirror map these translate into singularities on the heterotic quantum K\"ahler field space. Just as the Coulomb branch singularity at $u=\Lambda^2$ is related to the type IIA discriminant locus $\{\Delta_{\rm IIA}=0\}$, via F-theory/heterotic duality we expect the discriminant on the heterotic quantum K\"ahler field space to translate into a singularity on $\cM^F$. More precisely, for a component $\Delta_i^H$ of $\Delta^H$ we expect a component of the F-theory singular locus given by 
\begin{equation}
 \Delta_i^F(S_i)=0\,.
\end{equation}
Here, $S_i$ are the F-theory scalar fields defined in \eqref{eq:defSi}. In particular, some components of $\Delta_i^F$ can depend on $S_0$, i.e. on the coupling of the light string. Such a coupling-dependence can, however, not be inferred directly from the GLSM because the GLSM description is, strictly speaking, only valid on the $S_0=\infty$ locus. Still, for fixed $S_0$ the singularity $\Delta^F_i$ should reduce to a copy of the GLSM discriminant $\Delta^H_i$. Since we cannot infer the $S_0$ dependence from the worldsheet theory we need to calculate this directly within $\cM^F$. To that end, one can use the perturbative $\alpha'$-corrections to the F-theory scalar field space calculated in\cite{Grimm:2013gma,Grimm:2013bha,Weissenbacher:2019mef} (cf. \cite{Klaewer:2020lfg} for a discussion in the present context) and match them with perturbative corrections to the heterotic moduli space geometry. To summarize, our strategy is to employ the GLSM description to identify the singularity of $\cM^H_{qk}$ and then use the perturbative corrections to the F-theory field space to first match them onto $\cM^F$ and in particular to infer the dependence of the singularities on $S_0$. 
 
From the analogy to the $\cN=2$ case, we are particularly interested in components of $\Delta^H$ that can be interpreted as strong coupling singularities for the heterotic string, i.e. loci in field space at which $S_0$ vanishes. Suppose we find such a component, $\Delta_P^H$, of $\Delta^H$. Then at the level of the relevant singularities we can summarize the analogy between the string theory limit of $\cN=1$ F-theory and the field theory limit of $\cN=2$ string theory as
\begin{center}
\begin{tabular}{c|c}
$\cN=2$ field theory limit & $\cN=1$ string theory limit \\ \hline \hline 
Strong coupling singularity $u=\Lambda^2$ & GLSM singularity $\Delta^H_P=0$ \\ \hline
Type IIA discriminant $\Delta_{\rm IIA}=0$ & F-theory singularity $\Delta^F_P=0$\\ \hline
\end{tabular}
\end{center}

One may further ask for a physical interpretation of the singularities. Recall that in the point-particle limit of type IIA, the singularity signals the presence of additional light states and that the $W^\pm$-bosons, i.e. the purely electrically charged particles, leave the spectrum of BPS states. Given the analogy between the string and field theory limits, this suggests that a strong coupling singularity in the string theory limit of $\cM^F$ also signals the presence of additional light states, which are now expected to be additional light strings spoiling the worldsheet description. Similar to the electrically charged states leaving the BPS spectrum in the vicinity of the strong coupling singularity one might expect that also the D3-brane wrapped on the curve $C^0$ itself leaves the BPS string spectrum. We come back to this question in section \ref{sec:4dEFTstring}. 
\section{Dual Description}\label{sec:dual}
In the previous section, we argued that, in the vicinity of the locus $\{l^0=0\}$, $\mathcal{M}^F$ effectively reduces to the deformation space of the worldsheet theory of the light, critical heterotic string. In particular, we argued that the singularity and phase structure of $\cM^F$ in the vicinity of this point is inherited from the structure of the dual heterotic field space. In this section, we exclusively focus on this dual perspective before coming back to the F-theory description in the next section. For simplicity, here we restrict to models that have a description in terms of a GLSM.  We refer to appendix \ref{app:GLSMs} for some background on GLSMs useful for the discussion presented in this section. More precisely, we mainly consider heterotic models with standard embedding on the $(2,2)$ locus. However, in order to exclude that all our statements about the structure of the field space are a consequence of the enhanced $(2,2)$ worldsheet supersymmetry, we further consider deformations away from the $(2,2)$ locus. The benefit of specifying to models on (deformations of) the $(2,2)$ locus is that the singularity structure of the GLSM parameter space can be obtained explicitly; cf. \cite{Morrison:1994fr} or \cite{McOrist:2008ji, Kreuzer:2010ph} for the $(0,2)$ case. For these models the deformation space of the worldsheet theory is in fact an exact moduli space as the contributions of worldsheet instantons to the non-perturbative superpotential vanish, even though the full non-perturbative superpotential does not vanish identically due to the contribution from gauge instantons/NS5-brane instantons. 

In section \ref{sec:standardembedding} we start by considering heterotic models with gauge bundle given by the tangent bundle of the compactification space. Apart from discussing the structure for general compactification manifolds that are elliptically-fibered CY three-folds, we focus on two example with $B_2=\mathbb{P}^2, \mathbb{F}_1$ the first of which being the main example throughout the rest of this paper. In section \ref{sec:tangentdeformation} we then consider simple deformations of the tangent bundle in order to test whether our statements are merely a consequence of enhanced worldsheet supersymmetry. 

\subsection{Heterotic string with standard embedding}\label{sec:standardembedding}
Let us start with the simplest case corresponding to the heterotic $E_8\times E_8$ string compactified on a Calabi--Yau three-fold $Z_3$ with  gauge bundles given by 
\begin{equation}\label{V1V2standemb}
 V_1 =\mathcal{O}_{Z_3} \,,\qquad V_2 = TZ_3 \,,
\end{equation}
i.e. the trivial bundle and the tangent bundle of $Z_3$. In this case the quantum K\"ahler moduli space $\cM^H_{\rm qK}$ of the heterotic string compactified on $Z_3$ is well-known and is identical to the vector multiplet moduli space of type IIA compactified on the same manifold $Z_3$. Let us denote the complexified K\"ahler moduli of $Z_3$ by
\begin{equation}
 t^a = b^a + i s^a = \int_{C_a}\left(B+iJ_H\right)\,,
\end{equation}
where $\{C_a\}$ are the generators of the Mori cone, $J_H$ the K\"ahler form on $Z_3$ and $B$ the heterotic 2-form field.  
 As for the singularity structure of this moduli space, there exist standard techniques to identify the singular locus ~\cite{Morrison:1994fr}. The singular locus of this moduli space is given by a complex co-dimension one hypersurface $\{\Delta=0\}\subset \cM_{qK}^H$. In general this locus splits into multiple components 
\begin{equation}\label{deltageneral}
\Delta = \Delta_1 \cdot \Delta_2 \cdot \dots {\cdot}  \Delta_n\,,
\end{equation}
giving rise to an intricate network of singularities. For reasons becoming clear shortly,  the factor of $\Delta$ that corresponds to the so-called principal component of the discriminant locus is of particular interest to us. We denote the principal component of $\Delta$ by $\Delta_P$ such that 
\begin{equation}\label{deltaPR}
\Delta = \Delta_P \cdot \Delta_R\,,
\end{equation}
where $\Delta_R$ is the remainder of the discriminant locus that can in principle factorize further. One can give an explicit expression for $\Delta_P$ using, e.g., the methods of \cite{Morrison:1994fr}. To describe the locus $\{\Delta_P=0\}$, one can employ the language of GLSMs for which we review some basics in appendix \ref{app:GLSMs}. This description is particularly well-suited for our case since the relation to the quantum K\"ahler moduli space of the heterotic string is most direct in this case. 

For simplicity, we assume that the heterotic compactification manifold $Z_3$ is a smooth Weierstrass model over some base $B_2$ such that there exists a holomorphic zero section
\begin{equation}
\begin{aligned}
e: \quad B_2 & \rightarrow \mathbb{P}_{2,3,1}\\
b&\mapsto [1:1:0]\,,
\end{aligned}
\end{equation}
where $\mathbb{P}_{2,3,1}$ is the ambient space of the elliptic fiber. Let us denote the divisor associated to this section by $E_-$. The twist of the elliptic fiber over the base $B_2$ is governed by a holomorphic line bundle $\cL$. By supersymmetry, the first Chern class of this line bundle is given by the anti-canonical class of $B_2$, i.e., it satisfies $c_1(\cL) = c_1(B_2)\,.$ The first Chern class of $B_2$ can be expanded as 
\begin{equation}
c_1(B_2) = c_1^a j_a\,,
\end{equation}
where $j_a$ are the K\"ahler cone generators of $B_2$. If $Z_3$ is the anti-canonical hypersurface in a $d$-dimensional projective toric variety $\mathbb{P}^d_{*,\dots,*}$ and since $Z_3$ is a smooth Weierstrass model, the generator of the Mori cone corresponding to the elliptic fiber is associated to one of the $h^{1,1}$ $U(1)$ gauge factors of the GLSM with charges 
\begin{equation}
\mathfrak{l}^{(0)} = (-6;2,3,1,\underbrace{0,...,0}_{d-2})\,,
\end{equation}
for the matter fields $\Phi_i$, $i=0,\dots, d$\,, of the GLSM (cf. Appendix \ref{app:GLSMs}). The other generators of the Mori cone are then associated to the remaining $U(1)$ gauge factors with charges given by 
\begin{equation}
\mathfrak{l}^{(a)} = (0;0,0,-c_1^a,Q^a_1,...,Q_a^{d-2})\,,
\end{equation}
where the $Q^a_i$ are the GLSM charges of the fields matter fields $\Phi_i$ under the $a$th $U(1)$ gauge factor subject to the constraint
\begin{equation}
c^a_1 =\sum_{i=1}^{d-2} Q^a_i \,.
\end{equation} 
The precise values for the $Q_i^a$ depend on the details of the choice for $B_2$. 

From this data, one is able to find the principal component of the discriminant divisor by searching for $\sigma$-vacua, i.e. solutions to the equations \eqref{app:singularcondition}, which more generally read \cite{Morrison:1994fr}:
\begin{equation}
 \prod_{i|Q_i^a>0} \left(Q_i^b \sigma_b\right)^{Q_i^a} = q_a \prod_{i | Q_i^a <0} \left(Q_i^b \sigma_b\right)^{-Q_i^a}\,. 
\end{equation}
Here the $\sigma_a$ are the leading components of the neutral 2d chiral  superfields defined in appendix~\ref{app:GLSMs}, and \begin{equation}
q_{a} = e^{2\pi i(\theta^{a} + i r^{a})}\,,
\end{equation}
the exponentiated FI-parameters of the GLSM. For an arbitrary number of moduli this condition gets quite involved to solve, but in order to understand its general form let us reduce it to a two-parameter system by picking one generator $\mathfrak{l}^{(a_0)}$ and setting $q_a = 0\,,\; \forall a\neq a_0$ . In this way, we can easily find the dependence of $\Delta_P$ on $q_0$ and $q_{a_0}$. The condition $q_a = 0$ translates to $\sigma_a = 0\,,\;\forall a\neq a_0$, such that we end up with the reduced system of equations 
\begin{equation}
(\sigma_0-c_1^{a_0}\sigma_{a_0} )=432q_0 \sigma_0\,,\qquad  (\sigma_{a_0} )^{c_1^{a_0}} =(\sigma_0 -c_1^{a_0}\sigma_{a_0} )^{c_1^{a_0} }q_{a_0} ,
\end{equation} 
which for $c_1^{a_0}\neq 0$ can be solved to give
\begin{equation}\label{Delta1general}
\left. \Delta_P\right|_{(q_a=0 \,|\, a\neq a_0)} \simeq \left[\frac{1}{c_1^{a_0}} \left(1 - 432q_0\right)\right]^{c_1^{a_0}} - (432q_0)^{c_1^{a_0}}  q_{a_0} = 0\,.  
\end{equation}
Here, ``$\simeq$'' indicates that the actual principal component $\{\Delta_P=0\}$ can contain additional factors that are independent of $q_{a_0}$. Similarly, the actual discriminant might be an integer power of $\left. \Delta_P\right|_{(q_a=0 \,|\, a\neq a_0)} $. The expression above shows quite clearly that for $q_a\rightarrow 0$ for all $a$ the singular locus asymptotes to $q_0=1/432$ which corresponds to the boundary of the classical K\"ahler cone at which the volume of the elliptic fiber of $Z_3$ vanishes.\footnote{This can be seen by noticing that for $q_a=0$, $\forall a$, the mirror map is encoded in the modular $j$-function as \cite{Aspinwall:1999ii}
\begin{equation*}
 j(t^0) = \frac{1}{1728 q_0 (1-432q_0)}\,.
\end{equation*}
Therefore the moduli space at $q_a=0$ is a double cover of the $SL(2,\mathbb{Z})$ fundamental domain with $q_0=0$ corresponding to $t^0\rightarrow i \infty$ and $q_0=1/432$ the S-dual $t^0=0$.} Away from $\{q_a=0\}$ the topology of the singular locus is affected by the heterotic worldsheet instantons. For $c_1^{a_0}>1$ the fully factorized equation at $q_{a_0}=0$ turns into a general polynomial of degree $c_1^{a_0}$. Therefore, for $q_{a_0}>0$ the singular locus generally splits into multiple components. This can be interpreted as a non-zero quantum volume for the elliptic fiber away from the divisor $q_{a_0} = 0$ (cf. \cite{Mayr:1996sh,Greene:1996tx}).  For fixed $q_{a_0}$, and $c_1^{a_0}>1$, the induced quantum volume is parametrically given by
\begin{equation}\label{t0QM}
 t^0_{\rm QM} \sim q_{a_0} = e^{2\pi i(\theta^{a_0} + i r^{a_0})}\,. 
\end{equation}
Here, $\tau^{a_0}\equiv \theta^{a_0}+i r^{a_0}$ is the complexified GLSM FI-parameter corresponding to the $a_0$-th $U(1)$ factor which is related to the heterotic K\"ahler modulus $t^{a_0}$ via the mirror map. 

Let us now explain why, from the space-time perspective, the principal component of $\Delta$ corresponds to a strong coupling singularity for the heterotic string and is therefore particularly relevant for us. To that end, let us first consider type IIA string theory compactified on the same elliptically-fibered Calabi--Yau three-fold $Z_3$. In this case  the central charge of the  BPS state obtained by wrapping a D6-brane on $Z_3$ vanishes along $\{\Delta_P = 0\}$. At the perturbative level, the central charge of the D6-brane is given by
\begin{equation}\label{centralcharge}
 Z(\cO_{Z_3})= \int_{Z_3} e^{B+iJ_H} \sqrt{{\rm Td}(Z_3)} \Gamma_\mathbb{C}(Z_3) {\rm ch}\left(\cO_{Z_3}\right)^\vee \,,
\end{equation}
where $\text{Td}(Z_3)$ and $\Gamma_\mathbb{C}(Z_3)$ are the Todd- and complex Gamma-class of $Z_3$ which can be expanded in terms of the Chern classes of $Z_3$ as 
\begin{equation}
\text{Td}(Z_3) = 1 + \frac{c_2(Z_3)}{12}\,,\qquad \Gamma_\mathbb{C}(Z_3)=1 -\frac{\zeta(3)}{(2\pi i)^3} c_3(Z_3)\,. 
\end{equation}
 For the structure sheaf, $\cO_{Z_3}$, the Chern character is simply given by
\begin{equation}
{\rm ch}(\cO_{Z_3} ) = 1 \,.
\end{equation}
We can identify the Mukai vector, $\mu(\mathcal{E})$, of a generic sheaf, $\mathcal{E}$, as
\begin{equation}
\mu(\cE) = {\rm ch}(\cE)^\vee\sqrt{{\rm Td}(Z_3)}\,,
\end{equation}
such that 
\begin{equation}
\mu(\cO_{Z_3}) = \left(1,0, -\frac{1}{24} c_2^a(Z_3),0\right)\,. 
\end{equation}
Since in the heterotic theory the space-time supersymmetry is reduced by half as compared to the type IIA theory, there are no BPS particle states and in particular we do not have a 6-brane state whose central charge vanishes along $\{\Delta_P=0\}$. For the heterotic string we thus require a different interpretation of this singularity. To that end, recall that in heterotic M-theory the two non-Abelian gauge factors arise from end-of-the-world branes located at the fixed point of the compactification on $S^1/\mathbb{Z}_2$. If we further compactify on $Z_3$, the complexified gauge coupling of the end-of-the-world branes is classically given by the (complexified) volume of the end-of-the-world branes wrapped on $Z_3$. At the quantum level, we should thus identify the complexified gauge coupling with the central charge of a D6-brane in type IIA. Under this identification, the gauge bundle $V_i$ in the respective $E_8$ factor specifies the analogue of the induced D4-D2-D0-brane charges which can be summarized in the Mukai-vector
\begin{align*}
\mu(V_i) = &\left[{\rm rk}\, V_i, -c_1(V_i), -\frac{1}{2} \left(c_1(V_i)^2-2 c_2(V_i)\right)-\frac{1}{24} c_2(Z_3), \right. \\
&\left. -\frac{1}{6}\left(c_1(V_i)^3-3c_1(V_i) c_2(V_i)+ 3c_3(V_i)\right)\right]\,. 
\end{align*}
On the (2,2) locus the coupling at the perturbative level is then given by \eqref{centralcharge}. Using \eqref{V1V2standemb} we find the Mukai vectors
\begin{equation}
\mu(V_1) = (1, 0, -\frac{1}{24} c_2^a(Z_3),0)\,, \qquad \mu(V_2)=\left(3,0,(1-\frac{1}{24})c_2^a(Z_3), \frac{1}{2}\chi_3(V_2)\right)\,. 
\end{equation}
Accordingly, on the $(2,2)$ locus, the central charge of the physical type IIA D6-brane is identified with the (complexified) gauge coupling of the unbroken $E_8$ group in the heterotic theory. As a consequence, the principal component $\Delta_P$ of the singular locus in the heterotic theory corresponds to a strong coupling singularity for the gauge coupling of the unbroken $E_8$ factor. On the other hand, the gauge coupling of the $E_6$ factor of the gauge group remains finite as does the volume of $Z_3$. Still, the locus $\{\Delta_P=0\}$ qualifies as a candidate for the strong coupling singularity in the heterotic moduli space that we are after. 

In principle there could be other components of $\{\Delta=0\}$ that could account for a strong coupling singularity of the heterotic string. Such additional components of $\Delta$  are part of $\Delta_R$ in \eqref{deltaPR} and correspond to mixed Higgs-$\sigma$ branches in the language of GLSMs. In general there is one component of $\Delta$ for each boundary component of the K\"ahler cone. To get the full picture of the singularity structure of the quantum K\"ahler moduli space, we need to take into account the singular loci associated to $\Delta_R$ as well. The precise form of these remaining singular loci, however, depends on the details of the base $B_2$. In order to judge whether these singular loci can play the role of strong coupling singularities for the heterotic string, we need to understand whether the physics in the vicinity of the loci $\{\Delta_R=0\}$ differs significantly from the physics at $\{\Delta_P=0\}$. In the following, we are going to discuss two simple examples of bases $B_2$ for which we can analyze the full structure of the moduli space explicitly. 

\subsubsection{Example I: $B_2=\mathbb{P}^2$} 
As our prime example in this and the following section, let us consider the (blow-up of) the degree-18 hypersurface in $\mathbb{P}_{1,1,1,6,9}$ \cite{Candelas:1994hw}. The resulting Calabi--Yau three-fold can be viewed as a smooth Weierstrass model over $B_2=\mathbb{P}^2$. The GLSM description of this example requires the introduction of seven chiral field $\Phi^i$, $i=0,\dots, 6$, with charges specified by the Mori-cone generators of the Calabi--Yau hypersurface
\begin{equation}\begin{aligned}\label{MoriconeP2}
\mathfrak{l}^{(0)} &= (-6;2,3,1,0,0,0)\,,\\
\mathfrak{l}^{(1)} &= (0;0,0,-3,1,1,1)\,.
\end{aligned}\end{equation}
The phase structure of the GLSM parameter space on the (2,2) locus has been discussed in detail in \cite{Aspinwall:2001zq}. The conditions to find a $\sigma$-vacuum of the associated GLSM are
\begin{equation}
(\sigma_0 -3\sigma_1) = 432q_0\sigma_0\,,\qquad  \sigma_1^3 = q_1(-3\sigma_1 + \sigma_0)^3\,,
\end{equation}
which have a solution provided
\begin{equation}
\Delta_P \equiv(1-432q_0)^3 -432^3 \cdot 27q_0^3q_1 =0\,,
\end{equation}
which we identify as the principal component of the singular locus. In addition, there is a mixed $\sigma$-Higgs branch  given by
\begin{equation}
\Delta_R =1+27q_1 =0\,.
\end{equation}
The singularity $\Delta_R=0$ corresponds to the corrected K\"ahler cone boundary at which, for type IIA compactified on $Z_3$, the central charge of a D4-brane wrapped on the zero section $E_-$ vanishes. Classically, this boundary of the K\"ahler cone corresponds to $t^1=0$. However, at the quantum level, this is not the case anymore. This can be seen by considering the one-parameter system described by $q_1$ only: Since $\Delta_R$ does not depend on $q_0$, we can take the limit $q_0\rightarrow 0$ and study the resulting one-modulus system given by the single Mori cone generator $\mathfrak{l}^{(1)}$ corresponding to the hyperplane class in  $\mathbb{P}^2$.  This system is governed by the Picard--Fuchs operator 
\begin{equation}
\mathfrak{L}^{(1)} = \left( z \frac{d}{dz}\right)^3 - z \left( z \frac{d}{dz}\right)\left(z \frac{d}{dz}+\frac{1}{3} \right) \left(z \frac{d}{dz}+\frac{2}{3}\right)\,,
\end{equation} 
where we rescaled $z=-27 q_1$. The non-trivial solution to $\mathfrak{L}^{(1)}f=0$ then specifies the mirror map relating $\theta^1 +ir^1 = \frac{1}{2\pi i} \log q_1$ to the complexified volume $t^1=b^1 + i s^1$ of the curve in the hyperplane class $H$ of $\mathbb{P}^2$. We can identify the one-modulus system with $\mathcal{O}(-3)\rightarrow \mathbb{P}^2$, i.e. the resolved orbifold $\mathbb{C}^3/\mathbb{Z}_3$ for which the solutions to $\mathfrak{L}^{(1)}f=0$ were obtained in \cite{Aspinwall:1993xz}. Apart from the constant solution, a second solution for $|z|<1$ is given by
\begin{equation}
 f(z) = \log(z) + C +\frac{2}{9}z+\frac{5}{81} z^2 + \dots\,,
\end{equation}
with $C$ some constant and for $|z|>1$ by
\begin{equation*}
 f(z) = -3\frac{\Gamma\left(\frac{1}{3}\right)}{\Gamma^2\left(\frac{2}{3}\right)} e^{-\frac{\pi i }{3}} z^{-1/3} {}_3F_2\left(\frac{1}{3},\frac{1}{3},\frac{1}{3};\frac{2}{3},\frac{4}{3};z^{-1}\right)+ \frac{9}{2}\frac{\Gamma\left(\frac{2}{3}\right)}{\Gamma^2\left(\frac{1}{3}\right)}  e^{-\frac{2\pi i }{3}} z^{-1/3} {}_3F_2\left(\frac{2}{3},\frac{2}{3},\frac{2}{3};\frac{4}{3},\frac{5}{3};z^{-1}\right) \,. 
\end{equation*}
We are interested in the value of $s^1$ for $\Delta_R =0$, i.e. at  $z=1$. With the above information, \cite{Aspinwall:1993xz} calculates 
\begin{equation}\label{BiJP2}
s^1 |_{z=1} = 0.465  \neq 0\,.
\end{equation}
Therefore, along $\Delta_R=0$ the volume of a curve in the hyperplane class $H$ is finite while (in type IIA language) the central charge of a D4-brane on the zero section $E_-$ vanishes.  

We may now ask whether the locus $\{\Delta_R=0\}$ may as well correspond to a strong coupling singularity.  Therefore let us compare it to $\{\Delta_P=0\}$: First, we notice that $\{\Delta_P=0\}$ lies at the boundary between a geometric and a Landau--Ginzburg phase. By contrast $\{\Delta_R=0\}$ lies between two geometric phases: the Calabi--Yau phase and the $\mathbb{C}^3/\mathbb{Z}_3$ orbifold phase. This already indicates that the consequences of $\{\Delta_R=0\}$ are less severe than those associated to $\{\Delta_P=0\}$. In particular, notice that the presence of $\{\Delta_R=0\}$ itself does not constitute an obstruction against reaching the point $s^1=0$, which indeed is reached for $(q_0,q_1)\rightarrow (0,\infty)$, i.e. at the orbifold point \cite{Aspinwall:1993xz}.

To see that $\{\Delta_R=0\}$ does indeed not correspond to a strong coupling singularity, consider the behavior of the worldsheet theory in the vicinity of the two singularities. To that end, let us calculate the correlators of the worldsheet theory using the GLSM techniques developed in \cite{Morrison:1994fr} reviewed in appendix \ref{app:GLSMs}. Using \eqref{quantumrestriction}, we find\footnote{We thank I. Melnikov for pointing out a mistake in the expressions for the correlators in a previous version.} 
\begin{equation}\begin{aligned}\label{correlators}
 \langle \sigma_0^3 \rangle_{Z_3} &= \frac{9}{\Delta_P}\,,\\
 \langle \sigma_0^2\sigma_1 \rangle_{Z_3}  &= \frac{3 - 1296 q_0}{\Delta_P}\,,\\
 \langle \sigma_0\sigma_1^2 \rangle_{Z_3}  &= \frac{(1 - 432 q_0)^2}{\Delta_P}\,,\\
  \langle \sigma_1^3 \rangle_{Z_3}  &= \frac{9 (1 + 1296 q_0 (-1 + 432 q_0)) q_1}{\Delta_P\Delta_R} \,.
\end{aligned}\end{equation}
We see that \emph{all} correlators are singular along the locus $\{\Delta_P=0\}$ but only the last correlator is singular along $\{\Delta_R=0\}$. Therefore only a sub-sector of the theory becomes singular along $\{\Delta_R=0\}$ with the singular correlator corresponding to the Yukawa coupling that vanishes classicallly, i.e. in the $q_i\rightarrow 0$ limit. As a consequence, the singularity $\{\Delta_R=0\}$ is less severe than $\{\Delta_P=0\}$ and merely indicates that we are moving from one geometric phase, the CY phase, to another such phase, the orbifold phase. Notice also that the heterotic gauge theory remains weakly coupled along $\{\Delta_R=0\}$ since \eqref{centralcharge} does not vanish on this component of the singularity. We can thus conclude that in the simple example with $B_2=\mathbb{P}^2$ only $\{\Delta_P=0\}$ signals the transition into a strongly-coupled phase whereas the theory is better behaved along $\{\Delta_R=0\}$. Thus, $\{\Delta_P=0\}$ is indeed the relevant strong coupling singularity we are looking for.  

As already mentioned before the boundary of the K\"ahler cone corresponding to small base volume is given by $\{\Delta_R=0\}$. Since we identified $\{\Delta_P=0\}$ as the strong coupling singularity, we may now ask whether we reach this singularity within the K\"ahler cone. To answer this question, we need to know the relative position of the loci $\{\Delta_R=0\}$ and $\{\Delta_P=0\}$. Therefore notice that the two branches of the discriminant get exchanged under the involution 
\begin{equation}
 q_0 \rightarrow \frac{1}{432} - q_0' \,,\qquad q_1 \rightarrow -\left(\frac{432 q_0'}{1-432 q_0'}\right)^3 q_1'\,.
\end{equation}
This involution corresponds to the geometric part of the Fourier--Mukai transform discussed in appendix \ref{sec:hetFduality} and thus amounts essentially to two T-duality transformations along the fiber of the CY three-fold. Accordingly at the point in moduli space at which the two branches of the discriminant intersect
\begin{equation}
q_0 = \frac{1}{864}\, ,\qquad q_1 =- \frac{1}{27} \,,
\end{equation}
the volume of the fiber is given by the self-dual value, i.e. $s^0=1$. For $|q_0|>\frac{1}{864}$ we hence first enter the strong coupling region signaled by the presence of $\Delta_P=0$ before reaching the corrected boundary of the K\"ahler cone $\{\Delta_R=0\}$. Strictly speaking, we do not necessarily hit the singular locus $\{\Delta_P=0\}$ within the K\"ahler cone due to the dependence of the singular locus on the phases of $q_0$, and $q_1$. However, since for $|q_0|>1/864$ the amoeba of $\{\Delta_P=0\}$ lies within the K\"ahler cone (cf. Figure 3b in \cite{Marchesano:2022avb}) we know that the strong coupling phase associated to $\{\Delta_P=0\}$ is necessarily reached within the K\"ahler cone.\footnote{This can be compared with $\cN=2$ SYM theory with gauge group $SU(2)$ where for $u\notin \mathbb{R}$ we also do not hit the strong coupling singularity but still reach the strong coupling phase.} For simplicity (also when performing the duality to F-theory in the next section) in the following we may assume $q_0\in \mathbb{R}_{>0}$ for which we indeed always hit $\{\Delta_P=0\}$ within the K\"ahler cone. As a consequence the minimal quantum volume of the curve in $\mathbb{P}^2$ gets an extra contribution for $1/432>q_0>1/864$ due to the presence of the singularity $\Delta_P=0$ of the order\footnote{To see this, notice that the mirror map identifies $q_0=1/432$ with $t^0=0$. Turning on $0<|q_1|\ll 1$, we can solve $\Delta_P=0$ for
\begin{equation*}
e^{-2\pi s^0}= 1 - 2\pi s^0 + \mathcal{O}\left[(s^0)^2\right]= 1- 3q_1^{1/3} + O(q_1^{2/3}) \,,
\end{equation*} 
from which we can deduce \eqref{t1min} at leading order.}  
\begin{equation}\label{t1min}
s^1_{\rm min.} \sim \log s^0\,.
\end{equation}  
To summarize, for $1/432>|q_0|>1/864$ we encounter a singularity inside the corrected K\"ahler cone along which the worldsheet theory of the heterotic string ceases to be perturbative. On the other hand, for $|q_0|<1/864$ the worldsheet theory remains weakly-coupled inside the entire corrected K\"ahler cone and importantly does not entirely break down at the boundary corresponding to $\Delta_R=0$.

\subsubsection{Example II: $B_2 =\mathbb{F}_1$} 
\begin{figure}[!t]
\centering 
\includegraphics[width=.7\textwidth]{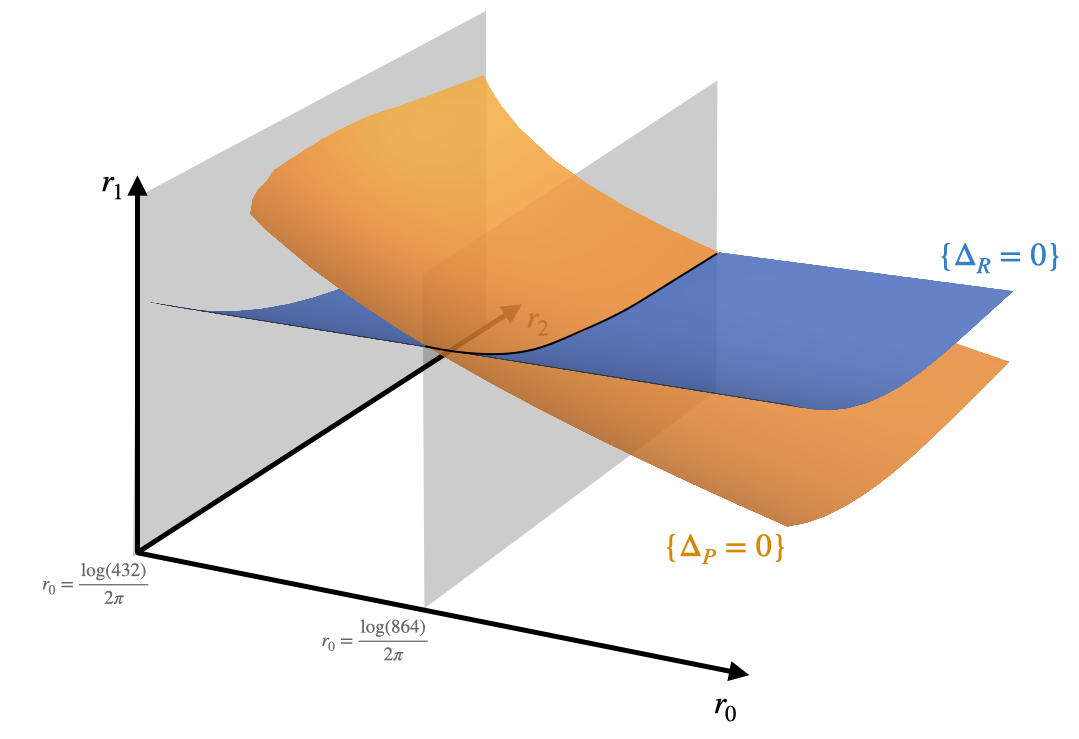}
\caption{This figure shows the boundary of the amoeba associated to the singular locus $\{\Delta=0\}$ for the smooth Weierstrass model over $B_2=\mathbb{F}_1$ projected to the real part, $r_i$, of the GLSM FI-parameters. Shown is only the quadrant describing the CY phase of the GLSM. The orange plane shows the amoeba for the principal singularity $\{\Delta_P=0\}$ whereas the blue plane corresponds to the amoeba of $\{\Delta_R=0\}$. The K\"ahler cone corresponds to the space bounded by the  $\{\Delta_R=0\}$-plane, the plane spanned by the $(r_0,r_1)$ axes, and the grey plane corresponding to constant $r_0=\frac{\log 432}{2\pi}$. We further indicated the plane $r_0=\frac{\log 864}{2\pi}$ where the boundaries of the amoebas of $\Delta_P$ and $\Delta_R$ intersect.}\label{fig:amoeba}
\end{figure}

As a second example, take $Z_3$ to be a smooth Weierstrass model over $B_2=\mathbb{F}_1$. Since here the FI parameter space of the associated GLSM has higher dimension, also its singularity structure is richer. The GLSM associated to $Z_3$ in this case is defined by the charge vectors 
\begin{equation}\begin{aligned}
\mathfrak{l}^{(0)} &= (-6;2,3,1,0,0,0,0)\,,\\
\mathfrak{l}^{(1)} &=(0;0,0,-2,1,0,1,0) \,,\\
\mathfrak{l}^{(2)} &= (0;0,0,-1,0,1,-1,1)\,.
\end{aligned}\end{equation}
The condition for the $\sigma$-vacua is given by
\begin{equation}\begin{aligned}
(\sigma_0 - \sigma_2 - 2\sigma_1) &= 432q_0\sigma_0\,,\\
\sigma_1(\sigma_1 -\sigma_2)&=q_1(-2\sigma_1 -\sigma_2 +\sigma_0)^2\,, \\
\sigma_2^2 &= q_2 (-\sigma_2 + \sigma_1) (-2\sigma_1 -\sigma_2 + \sigma_0) \, .
\end{aligned}\end{equation}
From here one can compute the principal component of the singular locus which now takes the form
\begin{equation}\begin{aligned}\label{Delta1F1}
 \Delta_P = \frac{1}{16} + 27 q_0 &\bigg\{-4 + q_2 + 
    432 q_0 \Big[6 - 3 q_2 - 8 q_1 - 
       432 q_0 \big[4 - 3 q_2 + 4 (-4 + 9 q_2) q_1 \\&
          +432 q_0 (-1 + q_2 + (8 + 9 q_2 (-4 + 3 q_2)) q_1 - 16 q_1^2)\big]\Big]\bigg\}\,,
\end{aligned}\end{equation}
which is complemented by a second component of the discriminant locus given by
\begin{equation}
 \Delta_R = (1-4q_1)^2 +(-1 + 36q_1)q_2 +(-27 q_1)q_2^2 =0\,.
\end{equation} 
This second component is nothing but the principal component of the discriminant locus of the base $\mathbb{F}_1$ itself and accordingly describes the locus in moduli space along which the zero section $E_-$ vanishes. As before, this locus in moduli space can be interpreted as the corrected boundary of the K\"ahler cone. Since, again, the heterotic gauge coupling does not vanish along $\{\Delta_R=0\}$ it does not correspond to a strong coupling singularity. The relevant singularity is thus again $\{\Delta_P=0\}$. In figure~\ref{fig:amoeba} we show the boundary of the amoeba of the singular loci in the CY phase of the FI-parameter space. 

We may now ask whether the strong coupling phase associated to $\{\Delta_P=0\}$ lies again inside the corrected K\"ahler cone associated to the base $B_2$ and thus induces a quantum volume for the curves in the base $B_2$. To get some intuition for the structure of $\cM_{\rm qK}^H$, consider the two extreme cases $(q_2 = 0, q_1\neq 0)$ and $(q_2\neq  0, q_1 = 0)$. For the first case, we find
\begin{equation}
\left.\Delta_P\right|_{q_2=0} =  \left(\frac{1}{4}(1 - 432q_0)^2 - 432^2q_0^2q_1\right)^2 \,, \qquad \left. \Delta_R\right|_{q_2=0} = 1-4q_1\,. 
\end{equation}
 This agrees with our general expectation \eqref{Delta1general} up to the overall square. Accordingly, on the locus $q_2=0$ the classical singularity at $q_0 = 1/432$ splits into two singularities for $q_1 > 0$. Both components correspond to double zeros of $\Delta_P$. For $|q_0|>1/864$ the amoeba of the principal component $\{\Delta_P=0\}$ lies again inside the corrected K\"ahler cone with boundary determined by $\{\Delta_R=0\}$ such that we have a strong coupling singularity within the corrected K\"ahler cone as illustrated in figure~\ref{fig:amoeba}. As in the previous case, this implies that there is a non-zero quantum volume induced for the Mori cone generator associated to $\mathfrak{l}^1$ proportional to $\log s^0$, cf. \eqref{t1min}.
 
Notice that, compared to the case $B_2=\mathbb{P}^2$, the locus $\{\Delta_R=1-4q_1=0\}$ has different properties. In fact the one-parameter system corresponding to $q_1$ is a $\mathbb{Z}_2$ orbifold of $\mathbb{C}^2$. As calculated in \cite{Aspinwall:1993xz} the complexified K\"ahler modulus associated to the exceptional divisor for this case is given by 
\begin{equation}\label{eq:t1}
t^1\equiv b^1+is^1 = \frac{1}{2\pi i} \log \left[\frac{1-2q_1 -2\sqrt{1-4q_1}}{2q_1}\right]\,,
\end{equation} 
which vanishes for $q_1=1/4$. Thus for $q_2=0$ a non-zero quantum volume is only introduced through the presence of $\{\Delta_P=0\}$. Notice that this is only true on the locus $q_2=0$ since $q_2\neq 0$ introduces a non-zero quantum value for $t^1$ on the singular locus $\Delta_R = 0$ \cite{Mayr:1996sh}. 

In the  $(q_2\neq  0, q_1 = 0)$ case, we find 
\begin{equation}
\left.\Delta_P\right|_{q_2=0} = \left(1-432q_0\right)^3\left[(1 - 432q_2) + 432q_0q_2\right] = 0\,, \qquad \left.  \Delta_R\right|_{q_1=0} = 1-q_2\,. 
\end{equation}
Thus, the singular locus splits into multiple components away from the $q_2=0$ locus. However, this splitting is different from the previous limit. Now, a triple zero of $\Delta_P$ remains at $q_0=1/432$ for any value of $q_2$ whereas one root of $\Delta_P$ moves to 
\begin{equation}
 q_0 =\frac{1}{432(1-q_2)} > \frac{1}{432} \,,\qquad \text{for}\qquad |q_2|<1\,.
\end{equation}
Since it is located at $q_0>\frac{1}{432}$ this singularity does not lie within the classical K\"ahler cone. On the other hand, the singularity at $q_0=1/432$ indicates that in this case $\Delta_P=0$ does not induce a quantum volume for the elliptic fiber. This implies that for all values of $q_0$ we \emph{can} reach the K\"ahler cone boundary $t^2=0$, i.e. the point at which the base $\mathbb{P}^1$ of $\mathbb{F}_1$ shrinks to zero size, without crossing into a strong coupling phase. Phrased differently, for $q_1=0$, the singularity $\Delta_P=0$ does not induce a quantum volume for the Mori cone generator associated to $\mathfrak{l}^{(2)}$. 

Coming back to the general case, i.e. $q_1,q_2\neq0$, we notice that in this case $\{\Delta_P=0\}$ splits into four components giving a non-zero contribution to the minimal quantum volume to both Mori cone generators associated to $\mathfrak{l}^{(1,2)}$. For $|q_0|>\frac{1}{864}$, the strong coupling phase associated to the singular locus $\{\Delta_P=0\}$ lies within the K\"ahler cone, cf. figure~\ref{fig:amoeba}.  Due to the presence of this strong coupling phase it is thus impossible to reach the small base volume limit, which would require $q_1,q_2 > 0$, within the perturbative weak-coupling regime for $|q_0|> \frac{1}{864}$. Similar as in the previous example one can show that it is the strong coupling singularity that is accompanied by a singularity in all correlators \cite{Morrison:1994fr}.
\newline 

Before we move on and consider deformations of the tangent bundle, let us summarize the insights we got from studying the $(2,2)$ locus. Though the two examples discussed here are far from being exhaustive, they confirm our general expectation that the principal component of the singular locus $\{\Delta_P=0\}$ is associated to a strong coupling singularity. From the space-time perspective the gauge theory of the heterotic string becomes strongly coupled at this point whereas at the worldsheet level, the perturbative worldsheet description breaks down. On the other hand, on the locus $\{\Delta_R=0\}$ we do not expect a strong coupling singularity for any of the gauge groups and also the worldsheet theory is not necessarily singular. Accordingly, $\{\Delta_R=0\}$ does not spoil the description of the weakly-coupled heterotic string. 

Importantly, for small torus volume, i.e. $q_0>1/864$, the amoeba of the singular locus $\{\Delta_P=0\}$ necessarily lies within the K\"ahler cone bounded by $\{\Delta_R=0\}$. Therefore, in this regime the strong coupling singularity prevents us from reaching the K\"ahler cone boundary and thereby also from reaching the small base limit at weak coupling. This difference between $\{\Delta_P=0\}$ and $\{\Delta_R=0\}$ is crucial and should be reflected in the structure of the F-theory moduli space that we are going to discuss in section \ref{sec:strongcouplingF}.

\subsection{Deformations of the tangent bundle}\label{sec:tangentdeformation}
So far we exclusively discussed heterotic models for which the gauge bundle is given by the tangent bundle. However, this is a very restrictive choice and one might wonder whether all our findings are a consequence of the enhanced $(2,2)$ worldsheet supersymmetry. To address this concern let us consider genuine $(0,2)$-models. For simplicity, we restrict to the case that the heterotic gauge bundle is a deformation of the tangent bundle since in this case similar techniques as on the $(2,2)$ locus can be used. In the following, we want to investigate whether the singularity structure of the GLSM FI-parameter space of the deformed theory still resembles the structure on the $(2,2)$ locus. More precisely, we want to see whether there are additional singularities appearing that could account for a strong coupling singularity or whether the (deformed version of) the principal component $\Delta_P$ of the discriminant is still the relevant strong coupling singularity for the $E_8$ gauge group and induces a logarithmic correction to the minimal volume of a base curve as in \eqref{t1min}. If this is the case, we can still interpret the presence of $\{\Delta_P=0\}$ as an obstruction to reaching the small base volume limit. 

Before investigating the singularity structure of the moduli space of the deformed theory, let us give some details of the kind of bundle deformations we are considering here: For definiteness, we focus on the case $B_2=\mathbb{P}^2$ and describe the deformations of the tangent bundle in GLSM language. We are interested in the deformation space of the bundle $\mathcal{E}$ on the Calabi--Yau $Z_3$. Our strategy here follows \cite{McOrist:2008ji,Kreuzer:2010ph} and we first want to study deformations of the tangent bundle of the toric ambient space $V$ defined by the charges 
\begin{equation}\begin{aligned}\label{chargesV}
\tilde{\mathfrak{l}}^{(0)} &= (2,3,1,0,0,0)\,,\\
\tilde{\mathfrak{l}}^{(1)} &= (0,0,-3,1,1,1)\,,
\end{aligned}\end{equation}
and then subsequently restrict to $Z_3$. The tangent bundle on $V$ can be viewed as the quotient 
\begin{equation}
  0 \;\longrightarrow \; \mathcal{O}^r \;\stackrel{Q_i^a z^i}\longrightarrow \;\oplus_i \mathcal{O}(D_i) \;\longrightarrow \;T_V \;\longrightarrow \;0\,,
\end{equation}
which can be deformed to a more general $(0,2)$ bundle by replacing the function $Q_i^a z^i$ by some map $E$ and then consider the quotient 
\begin{equation}\label{bundleEV}
0 \;\longrightarrow \; \mathcal{O}^r \;\stackrel{E}\longrightarrow \;\oplus_i \mathcal{O}(D_i) \;\longrightarrow \;\mathcal{E} \;\longrightarrow \;0\,.
\end{equation}
By $(0,2)$ supersymmetry we need to ensure $\sum E^i \mathcal{J}_i=0$ for the $\mathcal{J}-$parameters of the GLSM (see appendix \ref{app:GLSMs} for a review). In the GLSM associated to $V$ we can choose $\mathcal{J}_i=0$ and consider the deformations encoded in $E$ only. The deformation of $Q_i^a z^i $ into a more general $E$ can only mix chiral field $\Phi_i$ that have the same charge under all $U(1)$. In the light of \eqref{chargesV} the deformation can thus can only mix the last three fields. Grouping the chiral fields into sets of same charge, we can have the deformations 
\begin{equation}\begin{aligned}
E^1 &= i2\sqrt{2} \Sigma_0 \Phi^1\,,\\
E^2&= i 3\sqrt{2}\Sigma_0 \Phi^2\,,\\
E^3& = i\sqrt{2}\left[\Sigma_0\Phi^3-3\Sigma_1\Phi^3\right]\,,\\
 E^4  &= i \sqrt{2} \left[\Sigma_1 \Phi^4 + \Sigma_0(\epsilon_1 \Phi^4 +\epsilon_2 \Phi^5  + \epsilon_3 \Phi^6)\right]\,,\\
 E^5  &= i \sqrt{2} \left[\Sigma_1 \Phi^5 + \Sigma_0(\gamma_1 \Phi^4 +\gamma_2 \Phi^5  + \gamma_3 \Phi^6)\right]\,,\\
E^6 &= i \sqrt{2} \left[\Sigma_1\Phi^3 + \Sigma_0(\kappa_1 \Phi^4 +\kappa_2 \Phi^5 + \kappa_3 \Phi^6)\right]\,,
\end{aligned}\end{equation}
which we can conveniently summarize with the matrices (cf. \eqref{app:malpha})
\begin{equation}\begin{aligned}\label{matricesM}
M^{(1)} &= 2\sigma_0 \,,\quad M^{(2)} = 3\sigma_0 \,,\quad M^{(3)} = \sigma_0 - 3\sigma_1\,. \\
M^{(4)} &=\left(\begin{matrix}\sigma_1+ \epsilon_1 \sigma_10& \epsilon_2 \sigma_0 & \epsilon_3 \sigma_0 \\ \gamma_1 \sigma_0& \sigma_1 +\gamma_0 \sigma_1 & \gamma_3 \sigma_0 \\ \kappa_1 \sigma_0 & \kappa_2 \sigma_0 & \sigma_1 +\kappa_3\sigma_0 \end{matrix}\right)\,.
\end{aligned}\end{equation}
We then obtain the quantum cohomology relations 
\begin{equation}\label{quantumcohomology}
 (\sigma_0 - 3\sigma_1)=432q_0\sigma_0 \,,\qquad \text{det}M^{(4)} = q_1 \left(-3\sigma_1 + \sigma_0\right)^3\,.  
\end{equation}
Using these relations one can calculate the correlators of the $V$-model 
\begin{equation}
 \langle \sigma_0^{a} \sigma_1^{4-a}\rangle = \sum_{\sigma|\eqref{quantumcohomology} }\sigma_0^a \sigma_1^{4-a} \left[\text{det} \tilde J_{ab} \prod_{\alpha =1}^4 \text{det} M^{(\alpha)} \right]^{-1}\,,
\end{equation}
where the sum is taken over all values for $\sigma_a$ that satisfy \eqref{quantumcohomology}. Given the deformation of the bundle on the toric ambient space $V$, we can now restrict to the Calabi--Yau $Z_3$ and consider the deformations of the tangent bundle of $Z_3\subset V$. Therefore, we need to introduce an additional field $\Phi_0$ such that $\sum_i Q_i^a \Phi^i =0$ for all $a$ which, in the current case, amounts to introducing a field $\Phi_0$ with charge $(Q_0^0, Q_0^1)=(-6,0)$. The deformed bundle can now be viewed as the cohomology of the short-exact sequence 
\begin{equation}
0 \;\longrightarrow \; \mathcal{O}^r|_X \;\stackrel{E}\longrightarrow \;\oplus_i \mathcal{O}(D_i) |_X\;\stackrel{\mathcal{J}}\longrightarrow \;\mathcal{O}(\sum_i D_i)|_X \;\longrightarrow \;0\,.
\end{equation} 
Here $E$ are the same deformation functions as for the bundle in \eqref{bundleEV} but we now have the additional deformation parameters $\mathcal{J}$ which we cannot take to zero in the $Z_3$-model. In fact, prior to deformations, the $\mathcal{J}_i$ derive from a superpotential $\mathcal{J}_i = \partial W/\partial\Phi^i$ with $W$ on the $(2,2)$ locus given by the polynomial\footnote{There are more general choices for the polynomial $P$ corresponding to complex structure deformations of the CY three-fold $Z_3$. Since the final result does not depend on the complex structure of $Z_3$ we restrict here to this simple choice.}
\begin{equation}
W =\Phi^0 P \,,\qquad \text{with} \qquad P = \left[\left(\Phi^4\right)^{18}+\left(\Phi^5\right)^{18}+\left(\Phi^6\right)^{18}\right]\left(\Phi^3\right)^6 + \left(\Phi^1\right)^3+\left(\Phi^2\right)^2\,,
\end{equation}
thus ensuring that $\sum_i E^i \mathcal{J}_i =0$. When turning on the deformations, we can ensure $\sum_i E^i \mathcal{J}_i=0$ by simply deforming $\mathcal{J}_0 = P +\Delta \mathcal{J}_0$ with 
\begin{equation}
 \Delta \mathcal{J}_0 = 3\left[\left(\Phi^4\right)^{17}\sum_{i=1}^3 \epsilon_i \Phi^{i+3} + \left(\Phi^5\right)^{17}\sum_{i=1}^3 \gamma_i \Phi^{i+3}+\left(\Phi^5\right)^{17}\sum_{i=1}^3 \kappa_i \Phi^{i+3}  \right]\,. 
\end{equation}
To keep the expressions manageable let us consider a simplification and set $\epsilon_3=\gamma_2=\gamma_3=\kappa_1=\kappa_2=\kappa_3=0$. From \eqref{quantumcohomology} we can then compute the deformed principal component of the discriminant to be given by
\begin{equation}\label{Deltacorr}
 \Delta_{\epsilon_1, \epsilon_2,\gamma_1} = (1-432q_0)^3+3\epsilon_1 \left(1-432q_0\right)^2 -9\epsilon_2 \gamma_1 \left(1-432q_0\right)  -27q_1\,432^3 q_0^3 =0\,.  
\end{equation}

We now want to check whether this component still gives the relevant strong coupling singularity for the heterotic string. In the deformed case, we do not have the type IIA analogy to identify the gauge couplings with central charges of type IIA D-branes. Instead we identify a strong coupling singularity through singularities of the correlators. Using the quantum restriction formula for the $(0,2)$ case \cite{McOrist:2008ji, Kreuzer:2010ph} we find
\begin{equation}
 \langle \sigma_0^{3-a} \sigma_1^{a} \rangle|_{Z_3} =  \langle \frac{6\sigma_0^{4-a}\sigma_1^a}{1-6^6\sigma_0}\rangle_{V} =  \sum_{z|P(z)=0} \underbrace{\frac{6z^a}{(1-3z-432q_1)H(1,z)}}_{=:G(z)}\,,
\end{equation}
with 
\begin{equation}\begin{aligned}
 P(z)&=z^3 - \epsilon_1 z^2 -\epsilon_2\gamma_1 z + q_2(1-3z)^3\,,\\
 H(1,z) &= -36 (\epsilon_2 \gamma_1 (1 + 6 z) - z (2 \epsilon_1 + 3 (1 + \epsilon_1) z))\,,
\end{aligned}\end{equation}
and where we defined $z=\sigma_1/\sigma_0$ and used \eqref{quantumcohomology}. We can evaluate the sum over the zeros of $P(z)$ as in \cite{Kreuzer:2010ph} as a sum over residues 
\begin{equation}
\langle \sigma_0^{3-a} \sigma_1^{a} \rangle|_{Z_3} = -\left[\text{Res}_{z= \frac{1-432q_0}{3}} + \text{Res}_{z=z_\pm}+\text{Res}_{z=\infty} + \text{Res}_{z=0}\right] G(z)\frac{P'(z)}{P(z)}\,,
\end{equation}
where we defined 
\begin{equation}
z_\pm= \frac{-\epsilon_1 + 3 \epsilon_e \gamma_1 \pm \sqrt{
 \epsilon_1^2 + 3 \epsilon_2 \gamma_1 - 3 \epsilon_1 \epsilon_2 \gamma_1 + 9 \epsilon_2^2 \gamma_1^2}}{3 (1 + \epsilon_1)}\,.\end{equation}
As a result we find the following correlators 
\begin{equation}\begin{aligned}\label{correlatorsdeformed}
 \langle \sigma_0^3 \rangle_{Z_3} &= \frac{9}{\Delta_{\epsilon_1,\epsilon_2,\gamma_1}}\,,\\
 \langle \sigma_0^2\sigma_1 \rangle_{Z_3}  &= \frac{3 - 1296 q_0}{\Delta_{\epsilon_1,\epsilon_2,\gamma_1}}\,,\\
 \langle \sigma_0\sigma_1^2 \rangle_{Z_3}  &= \frac{(1 - 432 q_0)^2}{\Delta_{\epsilon_1,\epsilon_2,\gamma_1}}\,,\\
  \langle \sigma_1^3 \rangle_{Z_3}  &= \frac{-\epsilon_1 (1 - 432 q_0)^2 + 9 q_1- 3 (-1 + 432 q_0) (\epsilon_2 \gamma_1 + 3888 q_0 q_1)}{\Delta_{\epsilon_1,\epsilon_2,\gamma_1}\Delta_R} \,.
\end{aligned}\end{equation}
The structure of the correlators is very similar to the undeformed case. In particular, the deformed version of the principal component of the singular locus $\{\Delta_{\epsilon_1, \epsilon_2, \gamma_1}=0\}$ signals a complete break-down of the worldsheet theory and hence a strong coupling singularity.  On the other hand, as before the locus associated to $\Delta_R=0$ only yields a singularity for the last correlator indicating that, again, only a subsector of the heterotic theory becomes singular.\footnote{However, unlike in the undeformed theory the last correlator is non-zero also in the classical $q_i\rightarrow 0$ limit.} The locus $\{\Delta_R=0\}$, and hence the boundary of the corrected K\"ahler cone, itself is not affected by the deformation. This can be already seen on the level of the $\sigma$-vacuum equation \eqref{quantumcohomology}. If we decouple $q_0$ and look at the one-parameter model corresponding to the base $\mathbb{P}^2$ then the second equation just reduces to 
\begin{equation}
1+27q_1=0\,,
\end{equation} 
which we recognize as the equation $\Delta_R=0$. Accordingly in this example the boundary of the K\"ahler cone is not sensitive to the bundle moduli considered here.\footnote{For more general deformations or bundles that are not deformations of the tangent bundle this does not necessarily need to be the case. } 

To summarize, we identify the principal component of the discriminant as the relevant strong coupling singularity. Furthermore, for $q_0\gtrsim 1/864$ the strong coupling phase associated to the singularity $\{\Delta_{\epsilon_1,\epsilon_2,\gamma_1}=0\}$ still lies within the K\"ahler cone. Thus, as before, this component of the singular locus obstructs the small base volume limit by inducing a strong coupling singularity for the $E_8$ group. Moreover, the worldsheet theory still breaks down at this locus as is clear from \eqref{correlatorsdeformed}. As in the undeformed case, the presence of this singularity effectively induces a non-zero minimal quantum volume for the saxion $s^1$ that, as in \eqref{t1min}, depends logarithmically on $s^0$.  This crucial property is important in section \ref{sec:strongcouplingF} when matching the perturbative corrections to the F-theory scalar field space with the heterotic K\"ahler moduli space.

\section{Strong Coupling Singularities in F-theory}\label{sec:strongcouplingF}
In this section we aim to translate the structure of $\cM_{qK}^H$ into the quantum geometry of $\cM^F$ in the string theory limits discussed in section \ref{sec:lightstringlimits}. Due to its interpretation as a strong coupling singularity for the heterotic string, the embedding of the  principal component of the singular locus $\{\Delta_P=0\}\subset \cM_{qK}^H$ into $\cM^F$ is of particular relevance for us. 

On the F-theory side not much is known about the properties of the scalar field space beyond the large volume/large complex structure regime, though perturbative corrections to the classical K\"ahler potential have been calculated \cite{Grimm:2013gma,Grimm:2013bha,Weissenbacher:2019mef}. The effect of these perturbative corrections on classical emergent string limits has already been discussed in \cite{Klaewer:2020lfg}. Since we rely on these corrections to match the structure of $\cM^F$ and $\cM_{qK}^H$, we first give a brief review of their effect in section \ref{Ftheorymodulispacesingularities}. Based on the perturbative corrections and employing heterotic/F-theory duality, we then explain in section \ref{ssec:strongcoup} how a strong coupling singularity associated to $\{\Delta_P=0\}\subset \cM_{\rm qK}^H$ arises in $\cM^F$. As a consistency check, we reproduce the logarithmic quantum volume of the curves in the base $B_2$ observed on the heterotic side. Matching the perturbative F-theory corrections with the structure of the heterotic GLSM parameter space, we can further infer the dependence of the singular divisor on the string coupling. This allows us to identify a codimension one locus in $\mathcal{M}^F$ corresponding to this strong coupling singularity and signaling the transition into a strong coupling phase for the $\cN=1$ theory. In section \ref{sec:4dEFTstring} we then use these results to show that in the vicinity of the strong coupling singularity, the string obtained by wrapping a D3-brane on the fibral curve $C^0$ of $B_3$ leaves the spectrum of light strings which we further interpret in the context of the emergent string conjecture and the results of \cite{Marchesano:2022avb} concerning global string solutions associated to axionic strings. 

\subsection{Perturbative corrections to F-theory field space}\label{Ftheorymodulispacesingularities}
We are interested in the structure of the F-theory scalar field space in the limit in which the tension of a D3-brane wrapped on the generic fiber of a rationally fibered base $\rho: B_3\rightarrow B_2$ decouples from any other quantum gravity scale. From section \ref{sec:lightstringlimits} we recall that this corresponds to the limit in which the volume of the fiber of $B_3$ vanishes. As before, we denote the fibral curve by $C^0$ and its volume by $l^0$ and refer to the limit as $l^0\rightarrow 0$. In order not to leave the supergravity regime, we need to further ensure that \eqref{eq:relativescaling} is satisfied. The crucial insight of \cite{Klaewer:2020lfg} is that, even if \eqref{eq:relativescaling} is fulfilled, one cannot shrink the volume $l^0$ of the curve $C^0$ arbitrarily fast compared to the volume of the base of $B_3$. More precisely, we need to ensure that $l^0\succsim 1/v^a$, where $v^a$ is the volume of a curve in the base of $B_3$ (cf. section \ref{sec:lightstringlimits}). To see this \cite{Klaewer:2020lfg} analyzed the corrections to the K\"ahler potential obtained from higher derivative terms in the 11d M-theory action upon compactification on $Y_4$ and subsequent uplift to F-theory. Since this is a crucial ingredient for our analysis, let us briefly review how perturbative effects correct the F-theory scalar field space, referring to \cite{Grimm:2013gma,Grimm:2013bha,Weissenbacher:2019mef} for the original derivations. First, at the perturbative $(\alpha')^2$-level, the K\"ahler potential $K$ remains to be given by 
\begin{equation}
 K = -2\log \mathcal{V}_{B_3}\,.
\end{equation}
However, $\mathcal{V}_{B_3}$ receives $(\alpha')^2$-corrections:
\begin{equation}
 \mathcal{V}_{B_3} = \mathcal{V}_{B_3}^0+ \alpha^2 \left[(\tilde \kappa_1 +\tilde \kappa_2) \mathcal{Z}  + \tilde \kappa_2 \mathcal{T} \right]\,.
\end{equation}
Here $\tilde \kappa_1$ and $\tilde \kappa_2$ are constants (cf. the discussion in appendix C of \cite{Klaewer:2020lfg}) and $\mathcal{Z}$ and $\mathcal{T}$ are given by 
\begin{equation}\begin{aligned}\label{defZalpha}
 \mathcal{Z}_i &= \int_{Y_4} c_3(Y_4) \wedge \pi^*(J_i)\,, \hspace{6.4cm} \mathcal{Z} = \mathcal{Z}_i v^i\,, \\
 \mathcal{T}_i &=  -18(1+\alpha_2)\frac{1}{{\rm Re}T_i^0}\left( \int_{J_i} c_1(B_3) \wedge J \right)\left(\int_{J_i} J \wedge J_i\right)\,, \qquad \mathcal{T}_i = v^i\mathcal{T}_i\,,
\end{aligned}\end{equation} 
where the constant $\alpha_2$ remains undetermined after dimensional reduction \cite{Weissenbacher:2019mef}. Perturbative control over the $\alpha'$-expansion now requires that 
\begin{equation}\label{eq:perturbativecontrol}
 \frac{\mathcal{Z}}{\mathcal{V}_{B_3}^0} \ll 1 \,,\qquad \text{and} \qquad  \frac{\mathcal{T}}{\mathcal{V}_{B_3}^0} \ll 1\,. 
\end{equation}
This condition might be spoiled in case $\mathcal{Z}_0\neq 0$ or $\mathcal{T}_0\neq 0$ which, as analyzed in \cite{Klaewer:2020lfg}, is generically the case if $C^0$ is a rational curve. An actual breakdown of the F-theory perturbation theory in the small fiber limit occurs whenever the relative scaling of $l^0$ and $v^1$ satisfies
\begin{equation}\label{obstructedlimits}
 l^0 \prec \frac{1}{v^1}\,,
\end{equation}
whereas a relative scaling of $l^0 \sim 1/v^1$ is marginally allowed meaning that \eqref{eq:perturbativecontrol} is satisfied but not parametrically \cite{Klaewer:2020lfg}. The loss of perturbative control was interpreted in \cite{Klaewer:2020lfg} as an obstruction towards taking the limits satisfying \eqref{obstructedlimits}. Still, at this point it is not clear what exactly happens once the perturbative $\alpha'$-expansion breaks down and whether losing perturbative control over the F-theory effective action indeed corresponds to an obstruction to taking the limit $l^0\rightarrow 0$. Instead it could be possible that at the non-perturbative level such a limit in fact \emph{does} exist. The goal of this and the following section is to answer this question by exploiting our findings about the heterotic dual setup of the previous section.  

Recall from appendix \ref{sec:hetFduality} that the F-theory chiral multiplets defined in \eqref{eq:defTi} and the K\"ahler moduli $b^a + is^a$ of the heterotic base $B_2$ are related via 
\begin{equation}
 T_a = -i \eta_{ab} (b^b + i s^b)\,, \quad\text{for}\quad a\in \mathcal{I}_3\,. 
\end{equation}
We observe that limits $l^0\rightarrow 0$ satisfying \eqref{obstructedlimits} lead to
\begin{equation}
 \text{Re}\,T_a \rightarrow 0 \,, \quad \text{for} \quad a\in\mathcal{I}_3\,.
\end{equation}
Accordingly, light string limits in the F-theory scalar field space satisfying \eqref{obstructedlimits} correspond, on the heterotic side, to limits in which the K\"ahler moduli of the base $B_2$ become small. Given our analysis of the heterotic GLSM parameter space, this is precisely the regime of the field space where we encounter an interesting network of singular divisors. 

To relate the analysis of the structure of the heterotic GLSM parameter space to the F-theory scalar field space, we need to ensure that we consider a region in F-theory field space where we can trust the duality reviewed in appendix \ref{sec:hetFduality}. To that end, we first need that both, the heterotic CY $Z_3$ and the F-theory base $B_3$, are adiabatic fibrations over a common base $B_2$ and second, we need to take the stable degeneration limit on the F-theory side. On the heterotic side this latter condition ensures that we can encode the bundle in terms of its spectral data, i.e. in terms of a spectral cover together with a spectral line bundle in the limit of large torus volume (cf. appendix \ref{sec:hetFduality}).  This spectral data needs then to be translated into a Monad bundle in order to define the GLSM associated to the heterotic compactification. The spectral cover of Monad bundles is, however, in general degenerate \cite{Aspinwall:1998he,Bershadsky:1997zv}, i.e. it is given by an equation of the form
\begin{equation}
 \mathcal{C}: \quad z^3 f(u,v)=0\,,
\end{equation}
where $z$ is a projective coordinate on the fibral $\mathbb{P}_{2,3,1}$ and $u,v$ are coordinates on the base $B_2$ of the heterotic CY $Z_3$. In particular this is the case for standard embedding \cite{Aspinwall:1998he} and the spectral cover for the $E_8$ gauge group that is broken to $E_6$ has the form
\begin{equation}
 \mathcal{C}_+: \quad z^3 \Delta(u,v)=0\,,
\end{equation}
where $\Delta$ is the discriminant divisor of the heterotic Weierstrass model. Accordingly, the class of the two spectral covers associated to the gauge bundles inside the two $E_8$ factors are given by 
\begin{equation}\label{spectralcover}
 [\mathcal{C}_-] = [E_-] \,,\quad \text{and} \quad [\cC_+] = 3[E_-] + 12 c_1(B_2)\,,
\end{equation}
where we used $[\Delta]=12c_1(B_2)$.  Here $E_-$ is the zero section of the elliptic fibration. This degenerate spectral cover corresponds to an F-theory Weierstrass model described by 
\begin{equation}\label{undeformed}
 y^2 = x^3 + f_4 \tilde u^4 \tilde v^4 xz^4 + (\Delta(u,v) \tilde u^5 \tilde v^7 + g_6  \tilde u^6 \tilde v^6 + g_7  \tilde u^7 \tilde v^5)z^6\,,
\end{equation} 
where $\tilde u, \tilde v$ are coordinates on the fibral $\mathbb{P}^1$ of $\rho: B_3 \rightarrow B_2$, and $(f_4, g_{6,7})$ are functions of $(u,v)$. Naively one would hence expect two $E_8$ singularities at $\tilde u=0$ and $\tilde v=0$. However, for standard embedding one of the $E_8$'s has to be necessarily broken to $E_6$. On the heterotic side this breaking is achieved by a non-trivial line bundle which on the F-theory side translates into the data of a T-brane \cite{Anderson:2013rka}. 

To relate the structure of the heterotic K\"ahler moduli space to the F-theory field space, we need at least to have some information about the perturbative corrections in F-theory. To calculate these, let us perform a smoothing deformation of the Weierstrass model \eqref{undeformed} parametrized by two parameters $\epsilon, \delta$. The deformed Weierstrass model is now given by
\begin{equation}\label{deformedWeierstrass}
  y^3 = x^3 + f_4 \tilde u^4 \tilde v^4 x z^4 + \epsilon f_3 \tilde u^3 \tilde v^5 x z^4 + (\delta g_4(u,v) \tilde u^4 \tilde v^8  +\Delta(u,v) \tilde u^5 \tilde v^7 + g_6  \tilde u^6 \tilde v^6 + g_7  \tilde u^7 \tilde v^5)z^6\,.
\end{equation}
This deformation breaks $E_8\rightarrow E_6$ at $\tilde u=0$ and in the limit $\epsilon, \delta\rightarrow 0$ correctly reduces to the degenerate case  \eqref{undeformed}. Since we do not expect any additional states to become massless at the point $\epsilon=\delta=0$ in complex structure moduli space, the topology of the Calabi--Yau does not change under the deformation. 

Fortunately, at  the \textit{perturbative} level, the corrections in F-theory just depend on the topology of the four-fold such that it is save to work with the deformed Weierstrass model. That this is indeed the case can be also seen via heterotic/F-theory duality. Therefore recall from our discussion of the heterotic side that, at the perturbative level, the gauge coupling and the corrections to it also just depend on the topology of the bundle. In particular, when considering the deformations of the tangent bundle as in section \ref{sec:tangentdeformation} we do not change the Mukai vectors since we keep 
\begin{equation}
c_2(V_1)= 0\,,\qquad c_2(V_2) = c_2(T_V)\,. 
\end{equation} 
On the other hand, the change in the discriminant locus due to the deformations of the tangent bundle, are a consequence of the non-trivial mixing between the K\"ahler modulus of the heterotic fiber and the bundle moduli. As we have seen in section \ref{sec:tangentdeformation} these corrections do not change the singularity structure considerably such that the physics is qualitatively the same irrespective of whether the bundle deformations are turned on or not. Therefore also the F-theory behaviour should not significantly depend on the actual deformation in \eqref{deformedWeierstrass}.  Accordingly, in order to calculate the topological quantities that govern the $\alpha'$-corrections we can safely work with the deformed Weierstrass model. \\

Let us study the perturbative corrections for the effective theory obtained by compactifying F-theory on the four-fold with an $E_6$ singularity on $\mathcal{S}_+$ and an $E_8$ singularity at $\mathcal{S}_-$. For definiteness, in the following we focus on the concrete example with $B_2 = \mathbb{P}^2$ whose heterotic dual has already featured prominently in section \ref{sec:dual}.\footnote{Since other bases can be obtained from this $\mathbb{P}^2$ through a series of blow-ups, most of the relevant features of a general model are already present in this example which allows us to keep the discussion relatively simple.}  F-theory/heterotic duality relates the class of the spectral covers $\mathcal{C}_\pm$ to the twist of $\mathbb{P}^1$ over $\mathbb{P}^2$. In the present case,  \eqref{spectralcover} corresponds to a twist bundle $\mathcal{T}$ with 
\begin{equation}
c_1(\mathcal{T}) =6c_1(B_2) = 18 H\,.
\end{equation}
Here, $H$ is the hyperplane class of $\mathbb{P}^2$. We denote its pull-back to a vertical divisor of $B_3$ by $J_1=\rho^*(H)$. The model-dependent perturbative $\alpha'$-corrections are governed by the characteristic integral\footnote{As shown in \cite{Klaewer:2020lfg} the leading correction $\mathcal{T}_i$ is model independent and in general non-vanishing.}
\begin{equation}\label{eq:cZ1}
\CZ_1 =  \int_{Y_4} c_3(\tilde Y_4) \wedge J_1\,.
\end{equation}
For a Weierstrass model with gauge groups on non-intersecting divisors, this integral can be evaluated using the results of \cite{Esole:2018tuz}. In our case, the relevant gauge groups are realized on the exceptional section $\mathcal{S}_-$ and the section at infinity $\mathcal{S}_+=\mathcal{S}_-+c_1(\mathcal{T})$ of the $\mathbb{P}^1$-fibration. These satisfy 
\begin{equation}
\mathcal{S}_-\cdot \mathcal{S}_+ =0 \,,
\end{equation}
such that  we can directly apply the results of \cite{Esole:2018tuz} to evaluate \eqref{eq:cZ1}: We first get a generic contribution depending only on the geometry of $B_3$
\begin{equation}
 \CZ_1^{\rm gen} = -60 \int_{B_3} c_1(B_3)^2 \wedge J_1= -60 \int_{B_3} (2\mathcal{S}_- + 21J_1)^2 \wedge J_1 = -720\,.
\end{equation}
From the $E_8$ gauge group on $\mathcal{S}_-$ we get a contribution given by 
\begin{equation}
\CZ_1^- = \int_{B_3}\left[ 120(2\mathcal{S}_-+21J_1)\wedge \mathcal{S}_- -60 \mathcal{S}_-^2\right] \wedge J_1= -720\,,
\end{equation}
and the contribution from the $E_6$ gauge group on $\mathcal{S}_+$  
\begin{equation}
\CZ_1^+ = \int_{B_3}\left[ 90(2\mathcal{S}_+ +21J_1)\wedge \mathcal{S}_+ - 36\mathcal{S}_+^2\right] \wedge J_1= 1242\,,
\end{equation}
such that in total we find 
\begin{equation}
 \CZ_1= \CZ_1^{\rm gen.} + \CZ_1^+ + \CZ^-_1 = -198 \,.
\end{equation}
We notice that this correction is negative.\footnote{Note that in case we had worked with the undeformed Weierstrass model \eqref{undeformed} we would have had two $E_8$ groups on $\mathcal{S}_\pm$. In this case the topological term $\CZ_1$ would vanish identically. However, in this case we would have to deal with the T-brane data which is not captured by the known perturbative $\alpha'$-corrections. We avoided dealing with the T-brane data by considering the deformed Weierstrass model. If we completely ignored the T-brane and had not deformed the Weierstrass model, our F-theory model would be dual to the heterotic string with a bundle corresponding to point-like instantons on the heterotic side. Since these point-like instantons can be thought of as stacks of NS5-branes wrapping the class $18 H \subset \mathbb{P}^2$, we would have additional degrees of freedom corresponding on the  F-theory side to blowing up curves in the base $B_2$ which changes the topology of the four-fold $Y_4$ and hence $\mathcal{Z}_1$.} As a consequence for the current model, close to the point $l^0=0$ and, depending on the scaling of the other linear multiplets, we indeed expect to reach a region in moduli space where we lose perturbative control once we try to reach the locus $\text{Re}\,T_a=0$.

\subsection{F-theory singularity structure}\label{ssec:strongcoup}
Let us now explain the obstruction to reach the locus $\text{Re}\,T_a=\frac12\int_{J_a} J\wedge J=0\subset \cM^F$ from a non-perturbative perspective. To that end, we exploit the full non-perturbative structure of the GLSM parameter space, $\cM_{\rm qK}^H$, associated to the dual heterotic compactification analyzed in section \ref{sec:dual}. More precisely, we use the perturbative F-theory corrections discussed in the previous section as a guideline to match the structure of the heterotic GLSM parameter space onto $\cM^F$. We will show how this matching works in five steps which we first summarize briefly:
\begin{enumerate}
\item At first sight, one may identify the locus in $\cM^F$ at which the perturbative expansion breaks down with the corrected boundary of the K\"ahler cone at finite $\text{Re}\,T_a$ replacing the locus $\text{Re}\,T_a=\frac{1}{2} \int_{J_a} J\wedge J =0\subset \cM^F$. In analogy to the locus $\{\Delta_R=0\}\subset \cM_{\rm qK}^H$ describing the corrected K\"ahler cone boundary in the heterotic GLSM parameter space, we denote the corrected K\"ahler cone boundary in $\cM^F$ by $\{\Delta_R^F=0\}$. 

\item Recalling from section~\ref{sec:dual} that the heterotic gauge coupling remains unaffected by the presence of $\{\Delta_R=0\}$ one realizes that this identification cannot be quite correct: The perturbative corrections to $\cM^F$ also affect the gauge coupling on the 7-branes in the vicinity of $\{\Delta_R^F=0\}$ indicating that the breakdown of the perturbative expansion cannot just be a consequence of reaching the corrected boundary of the K\"ahler cone in $\cM^F$.

\item Using heterotic/F-theory duality we show that the regime of $\cM^F$ where we have a light, perturabtive heterotic string gets mapped to a region in the GLSM parameter space where the locus $\{\Delta_P=0\}$ lies within the corrected K\"ahler cone bounded by $\{\Delta_R=0\}$. On the F-theory side this implies the existence of a singular locus $\{\Delta_P^F=0\}$ inside the corrected K\"ahler cone along which the 7-brane gauge coupling diverges. Before reaching $\{\Delta_R^F=0\}$ one hence enters a strong coupling phase of F-theory. Therefore the breakdown of the perturbative expansion is in fact triggered by the singularity $\{\Delta_P^F=0\}$ and the presence of the strong coupling phase. 

\item As a consistency check we use the perturbative corrections to $\cM^F$ to calculate the corrections to the minimal value of $\text{Re}\,T_a$ induced by the presence of the strong coupling singularity in $\{\Delta_P^F=0\}\subset\cM^F$. This in fact reproduces the logarithmic correction to the minimal quantum volume of heterotic base curves observed in section~\ref{sec:dual}, cf. \eqref{t1min}. 

\item Finally, we use the perturbative corrections to $\cM^F$ to infer the dependence of the location of the singular locus $\{\Delta_P^F=0\}$ on the perturbative coupling, $S_0$, of the 7-brane gauge theory. This dependence leads to a singularity structure of $\cM^F$ in the string theory limit that is very reminiscent of the structure of the $\cN=2$ moduli space in the field theory limits of type II Calabi--Yau compactifications as anticipated in section \ref{sec:N2comparison}. 
\end{enumerate}
With this summary, let us now explain the above steps in some more detail: 

\subsubsection{Corrected K\"ahler cone boundary} 
To start, let us examine how the boundary of the classical K\"ahler cone $\text{Re}\,T_a=0$ gets affected by the perturbative quantum corrections. From the discussion of the heterotic dual in section~\ref{sec:dual} we recall that at the actual boundary of the K\"ahler cone, corresponding to $\{\Delta_R=0\}$, the central charge of the zero section $E_-$ of the heterotic CY $Z_3$ vanishes.  As a consequence of the non-perturbative corrections, this locus in moduli space does not correspond to the point $s^a=0$ but is shifted to finite $s^a$ as is clear from \eqref{BiJP2}. Via duality we hence expect also the point $\text{Re}\,T_a=0$ not to lie within the corrected K\"ahler cone on the F-theory side. 

To see this notice that the corrections proportional to $\CZ_i$ and $\CT_i$ only become relevant in case the classical volumes of \textit{all} vertical divisors $\rho^*(C_a)$ are taken to small values. The corrected volume of a $\rho$-vertical divisor is given by~\cite{Grimm:2013gma,Grimm:2013bha,Weissenbacher:2019mef} 
\begin{equation}\label{ReT1corr}
 \text{Re} \,T_a = \text{Re}\,T_a^{(0)}\left[1 +  \alpha^2\left((\kappa_3+\kappa_5)\frac{\mathcal{Z}}{{\mathcal V}_{B_3}^{(0)}} + \kappa_5 \frac{\mathcal{T}}{{\mathcal V}_{B_3}^{(0)}}\right)\right]+\alpha^2\left( \tilde{\mathcal{Z}_a}  \log \mathcal{V}_{B_3}^{(0)}+\kappa_6 \mathcal{T}_a + \kappa_7 {\cal Z}_a \right)\,. 
\end{equation}
Perturbative control requires
\begin{equation}\label{singularconditionF}
 \left| \alpha^2\left((\kappa_3+\kappa_5)\frac{\mathcal{Z}}{{\mathcal V}_{B_3}^{(0)}} + \kappa_5 \frac{\mathcal{T}}{{\mathcal V}_{B_3}^{(0)}}\right)\right| \stackrel{!}{<} 1 \,. 
\end{equation}
Since $\CZ_a<0$ and $\kappa_3+\kappa_5>0$, perturbative control is lost once the term proportional to $\text{Re}\,T_a^{(0)}$ in \eqref{ReT1corr} vanishes. As a consequence, the actual volume of $J_a$ at the corrected boundary of the K\"ahler cone is given by 
\begin{equation}\label{ReT1star}
\text{Re}\, T_a^* = \alpha^2\left( \tilde{\mathcal{Z}_a}  \log \mathcal{V}_{B_3}^{(0)}+\kappa_6 \mathcal{T}_a + \kappa_7 {\cal Z}_a\right)\,,
\end{equation}
up to higher order and non-perturbative corrections. By matching the F-theory $\alpha'$-corrections to heterotic loop corrections, \cite{Klaewer:2020lfg} showed that heterotic/F-theory duality constrains $\tilde{\CZ}_a=0$. Accordingly, the value of ${\rm Re}\,T_a^*$ is determined by the constants $\cT_a$ and $\CZ_a$. From the perturbative perspective we expect ${\rm Re}\,T_a^*>0$, since $\kappa_7 = -(\kappa_3+\kappa_5)$ and $\CZ<0$. Thus, indeed, the corrections shift the K\"ahler cone boundary and we can identify $\text{Re}\,T_a^*>0$ as the volume of $J_a$ along this boundary which we denote by $\{\Delta_R^F=0\}$. This matches with the heterotic result where the corrected K\"ahler cone boundary, $\{\Delta_R=0\}$, also corresponds to $s^a>0$, though the precise value depends on non-perturbative effects and possibly higher order corrections in $\alpha'$ that we did not take into account in our F-theory analysis. Notice that both the value of  $\text{Re}\,T_a^*$ and the value of $s^a$ along $\{\Delta_R=0\}$ are independent of the value of the other moduli in the theory including the deformations discussed in section \ref{sec:tangentdeformation}. Thus, at this level, the perturbative analysis of the F-theory moduli space is consistent with the analysis of the heterotic quantum K\"ahler moduli space to the extend that in both cases the boundary of the K\"ahler cone is shifted at the quantum level.

\subsubsection{Perturbative corrections to string coupling} 
From the dual heterotic perspective we expect that the corrections shifting the boundary of the K\"ahler cone only affect the moduli $T_a$ since the component $\Delta_R$ of the discriminant is independent of e.g. $q_0$. In particular in the heterotic theory we can still freely tune the heterotic gauge couplings along $\{\Delta_R=0\}$. By contrast, on the F-theory side this is not the case as can be seen by considering the corrections to the volume of the exceptional divisor $D_0\equiv \mathcal{S}_-$:
\begin{equation}\label{ReS-}
 \text{Re}\, S_0 = \text{Re} S_0^{(0)}\left[1 +  \alpha^2\left((\kappa_3+\kappa_5)\frac{\mathcal{Z}}{{\mathcal V}_{B_3}^{(0)}} + \kappa_5 \frac{\mathcal{T}}{{\mathcal V}_{B_3}^{(0)}}\right)\right]+\alpha^2\left( \tilde{\mathcal{Z}_0}  \log \mathcal{V}_{B_3}^{(0)}+\kappa_6 \mathcal{T}_0 + \kappa_7 {\cal Z}_0 \right)\,,
\end{equation}
where $S_0$ is defined in \eqref{eq:defSi}. As reviewed in appendix~\ref{sec:hetFduality}, we can interpret $\text{Re}\,S_0$ as the gauge coupling of the unbroken $E_8$ gauge group. Classically, one can choose $\text{Re}\,T_a^{(0)}$ and  $\text{Re}\, S_0^{(0)}$ independently such that one can tune the gauge coupling for the $E_8$ gauge group on $\mathcal{S}_-$ to arbitrarily small values while keeping $\text{Re}\,T_a^{(0)}$ finite. In the vicinity of the locus $\text{Re}\,T_a=\text{Re}\,T_a^*$, i.e. $\{\Delta_R^F=0\}$, this is obviously not the case anymore as we have 
\begin{equation}
 \text{Re} \,S_0^* = \alpha^2\left( \tilde{\mathcal{Z}_0}  \log \mathcal{V}_{B_3}^{(0)}+\kappa_6 \mathcal{T}_0 + \kappa_7 {\cal Z}_0 \right) \,,
\end{equation}
which is not proportional to the the tree-level value $\text{Re}\, S_0^{(0)}$. We are thus unable to tune the gauge coupling to arbitrarily small values at that point. Let us stress again that this should be contrasted to the situation along the locus $\{\Delta_R=0\}$ in the heterotic GLSM parameter space at which the perturbative heterotic gauge theories remain weakly coupled. We thus conclude that the break-down of the perturbative expansion in F-theory cannot merely be a consequence of the shift of the K\"ahler cone encoded in $\{\Delta_R=0\}$.  Instead, as we show in the following, this perturbative obstruction is in fact associated to the singularity $\{\Delta_P=0\}$ in the heterotic GLSM parameter space. 

\subsubsection{Strong coupling via heterotic/F-theory duality}
To see the relevance of $\{\Delta_P=0\}$, let us first notice that in order for the field space around $l^0=0$ to be identified with the stringy moduli space of a weakly-coupled heterotic string, we need to ensure that we can express the heterotic bundle data in terms of the spectral cover plus spectral line bundle as required for standard F-theory/heterotic duality. As reviewed in appendix \ref{sec:hetFduality} this requires us to take the stable degeneration limit for the four-fold $Y_4$ on the F-theory side. To take this limit, we should replace the Weierstrass model in \eqref{undeformed} by\footnote{Here we work with the undeformed Weierstrass model in order to keep the expressions simple.} 
\begin{equation}\label{stabledegeneration_s}
 y^2 = x^3 + f_4 \tilde u^4 \tilde v^4 xz^4 +g_6  \tilde u^6 \tilde v^6 z^6+ \xi \, (\Delta(u,v) \tilde u^5 \tilde v^7 +  g_7  \tilde u^7 \tilde v^5)z^6\,,
\end{equation} 
and consider the regime $\xi \ll 1$. Notice that we are not taking the strict $\xi \rightarrow 0$ limit which would be another infinite distance limit but just consider fixed $\xi \ll 1$ in order to trust the duality to the heterotic string.

On the heterotic side of the duality the stable degeneration parameter $\xi$ corresponds to the volume modulus, $s^0$, of the elliptic fiber of the heterotic CY three-fold $Z_3$. The regime $\xi\ll1$ on the F-theory side now translates into the limit $s^0\rightarrow \infty$ for the heterotic string on $Z_3$ equipped with spectral cover and spectral line bundle. To trust the F-theory/heterotic duality we further have to impose
\begin{equation}\label{adiabatichierarchy}
s^0 \ll \text{vol}B_2^H\,,
\end{equation}
which ensures an adiabatic limit. At the classical level, this hierarchy ensures that the heterotic K\"ahler potential factorizes as 
\begin{equation}\label{classicalKahlerpot}
 K= -\log(s^0) - \log \left(\frac{1}{2} \cV_{B_2}\right)+\dots\,.
\end{equation}
However, we cannot directly apply the GLSM analysis of section \ref{sec:dual} to the heterotic string with spectral cover and line bundle. Instead, we first need to translate the spectral data obtained from F-theory in the stable degeneration limit into the bundle data that enters the GLSM description. As we review in appendix \ref{sec:hetFduality} the bundle data can be obtained from the spectral data by means of a Fourier--Mukai transform. This transform acts as a double T-duality on the elliptic fiber of $Z_3$. Hence, when applying the GLSM description, the stable degeneration limit $\xi\ll 1$ on the F-theory complex structure moduli space should in fact be viewed as the regime $s^0\ll \mathcal{O}(1)$, or $q_0\ll 1/864$. From our discussion of the GLSM parameter space in section \ref{sec:dual}, we recall that in this regime of the heterotic GLSM parameter space, the amoeba of the singularity $\{\Delta_P=0\}$ necessarily lies within the boundary of the corrected K\"ahler cone set by $\Delta_R=0$. Therefore, in the light string limit of F-theory we also expect to first reach a singularity associated to $\Delta_P=0$ before we reach the K\"ahler cone boundary $\{\Delta_R^F=0\}$. Let us denote the singular locus in $\cM^F$ by $\{\Delta_P^F=0\}$

Recall from section \ref{sec:dual} that along the singularity $\Delta_P=0$ the gauge coupling of the unbroken $E_8$ diverges. This implies that on the F-theory side the gauge theory on $\mathcal{S}_-$ becomes strongly coupled along $\{\Delta_P^F=0\}$. The fact that the perturbative corrections also affect the gauge coupling can be seen as a remnant of this strong coupling singularity  at the perturbative level. The upshot is thus that, from heterotic/F-theory duality, the obstruction to reach ${\rm Re}\,T_a=0$ is due to a strong coupling singularity for the gauge theory on $\mathcal{S}_-$. 

\subsubsection{Logarithmic quantum volume}
In fact, we can give further support for this observation by considering the contribution to the quantum volume for $\rho$-vertical divisors $J_a$ induced by the presence of $\{\Delta_P^F=0\}$. Therefore recall that on the heterotic side the contribution to the minimal quantum volume of a curve in the base $B_2$ due to $\{\Delta_P=0\}$ is roughly given by 
\begin{equation}
s^a \sim - \log s^0+\dots \,, 
\end{equation}
up to details encoded in the mirror map and the choice of bundle deformations. Since on the F-theory side $s^0$ has to be identified with $\xi$ introduced in \eqref{stabledegeneration_s}, we also expect that the minimal quantum volume for $\text{Re}\,T_a$ depends logarithmically on the complex structure parameter $\xi$. To see this, we notice that the $s^0$-dependence of the minimal volume of the curves in $B_2$ given by $s^a_\text{min} \propto \log(s^0)$ can be interpreted as arising from a 1-loop correction to the worldsheet instanton action. Thus this should be an effect already visible at the perturbative level in F-theory. By duality, ${\rm Re}\,T_a=\eta_{ab} s^b$ is the (real part of the) action of a D3-brane instanton in F-theory and we therefore expect to also observe a logarithmic dependence on the stable degeneration parameter from studying the perturbative $\alpha'$-corrections to ${\rm Re} \,T_a$. 

Given the expression \eqref{ReT1corr}, it is not immediately clear where such a dependence on the complex structure modulus $\xi$ dual to $s^0$ comes from. To identify such a dependence, let us exploit the vanishing of the gauge coupling of the unbroken $E_8$ on the singular locus given by $\Delta_P^F=0$, as argued for above. Imposing the strong coupling condition, ${\rm Re} \,{S_0} =0$, yields a condition similar to \eqref{singularconditionF} up to loop corrections that are encoded in 
\begin{equation}
 \Xi_{\rm 1-loop} = \alpha^2\left( \tilde{\mathcal{Z}_0}  \log \mathcal{V}_{B_3}^{(0)}+\kappa_6 \mathcal{T}_0 + \kappa_7 {\cal Z}_0 \right)\,. 
\end{equation} 
Since we argued that the logarithmic correction to ${\rm Re}\, T_a^*$ should be a one-loop effect, we need to understand these loop terms. To that end, we can use the result of \cite{Klaewer:2020lfg} interpreting the logarithmic term in $\Xi_{\rm 1-loop}$ as a one-loop term in the heterotic theory which fixes 
\begin{equation}
 \tilde{\mathcal{Z}}_0 = \frac{b}{8\pi}\,,
\end{equation}
with $b$ the $\beta$-function coefficient of the gauge group realized on $\mathcal{S}_-$. One obtains this result by noticing that the non-holomorphic threshold corrections to the heterotic gauge coupling are given by 
\begin{equation}\label{Delta1}
 \tilde\Delta_P = \frac{c}{8\pi} K_H + \ldots \,,
\end{equation}
where $K_H$ is the K\"ahler potential for the heterotic K\"ahler moduli and $c$ another one-loop coefficient. In the present case the relevant contribution to $K_H$ is classically given by (cf. \eqref{classicalKahlerpot})
\begin{equation}\label{KMMhet}
 K_H \supset  - \log(\mathcal{V}_{Z_3}) = -\log \left[\frac{1}{2} \eta_{ab} s^a s^b\right] - \log s^0\,. 
\end{equation}
On the other hand one finds
\begin{equation}\label{VB3}
\log \mathcal{V}_{B_3}^{(0)}  =   \log \left(\frac{M^2_{\rm het}}{M_S^2} \mathcal{V}_{B_2}^F \right) = \log \left(\frac{M^2_{S}}{M_{\rm het}^2}\right) + \log\left[\frac{1}{2} \eta_{ab} s^a s^b\right]\,,
\end{equation} 
where in the last step we re-expressed all curve volumes in terms of the heterotic string scale. Comparing \eqref{VB3} with \eqref{Delta1} and using \eqref{KMMhet} we find that the non-holomorphic threshold corrections evaluated at $M_{\rm het}^2$ agree with the F-theory correction proportional to $\log \CV_{B_3}$ up to a term proportional to $\log s^0$ which can be identified with $\log \xi$ upon F-theory/heterotic duality. Let us denote the renormalized gauge coupling of the gauge theory on $\mathcal{S}_-$ by ${{\rm Re} \,S_0^{(1)}}$. Using that the singularity corresponds to strong coupling, i.e. $ {\rm Re}\,{S_0} =0$, we find from \eqref{ReS-}
\begin{equation}
\left.\left(1 + \alpha^2 (\kappa_3 + \kappa_5) \frac{\CZ}{\CV_{B_3}^{(0)}}+\kappa_5 \frac{\mathcal{T}}{ {\mathcal V}_{B_3}^{(0)}} \right)\right|_{\rm sing.} = -\frac{\alpha^2}{ \text{Re}\,S_0^{(1)}}\left(\frac{b}{8\pi} \log \xi +\kappa_6 \mathcal{T}_0 + \kappa_7 {\cal Z}_0 \right) \,. 
\end{equation}
We can insert this into \eqref{ReT1corr} to obtain 
\begin{equation}\label{ReS0dependence}
  \frac{1}{\alpha^2}{\rm Re}\, T_a\bigg|_{\rm sing.}= -\frac{{\rm Re} T_a^{(0)}}{{\rm Re}\,S_0^{(1)}} \left[\frac{b}{8\pi} \log \xi+\kappa_6 \mathcal{T}_0 + \kappa_7 {\cal Z}_0 \right] + \kappa_6 \CT_a +\kappa_7 \CZ_a \,.
\end{equation}
This expression agrees with ${\rm Re}\, T_a^*$ given in \eqref{ReT1star} up to the first term which logarithmically depends on $\xi$. Hence, for ${\rm Re}\,S_0^{(1)}\rightarrow \infty$, the value of  ${\rm Re}\, T_a\big|_{\rm sing}$ asymptotes to a constant that at the perturbative level is determined by $\kappa_6  \CT_a +\kappa_7 \CZ_a$, but in general is sensitive to the logarithmic corrections. We can extract the information about this correction by considering  
\begin{equation}\label{defu1}
\mathfrak{u}_a\equiv  {\rm Re} T_a - {\rm Re}\, T_a^* \,.
\end{equation}
At the singularity we find 
\begin{equation}\label{u1sing}
 \mathfrak{u}_a\bigg|_{\rm sing.} =  -\frac{{\rm Re} T_a^{(0)}}{{\rm Re}S_0^{(1)}} \frac{b}{8\pi} \log \frac{\xi}{\exp\left(\Xi'\right)}\,,
\end{equation}
where we absorbed $\Xi'= \Xi-\alpha^2 \tilde{\mathcal{Z}_0}  \log \mathcal{V}_{B_3}^{(0)}$ into the logarithm. The coordinates $\mathfrak{u}_a$ thus parametrize the effect of the singularity $\Delta_P^F=0$ on the quantum volume of the vertical divisor $J_a$. The perturbative F-theory results therefore reproduce perturbative information encoded in the heterotic singular locus $\Delta_P=0$, such as the $\log\xi$ dependence reflected in \eqref{u1sing}. 

\subsubsection{Coupling dependence}
\begin{figure}[!t]
\centering 
\includegraphics[width=.6\textwidth]{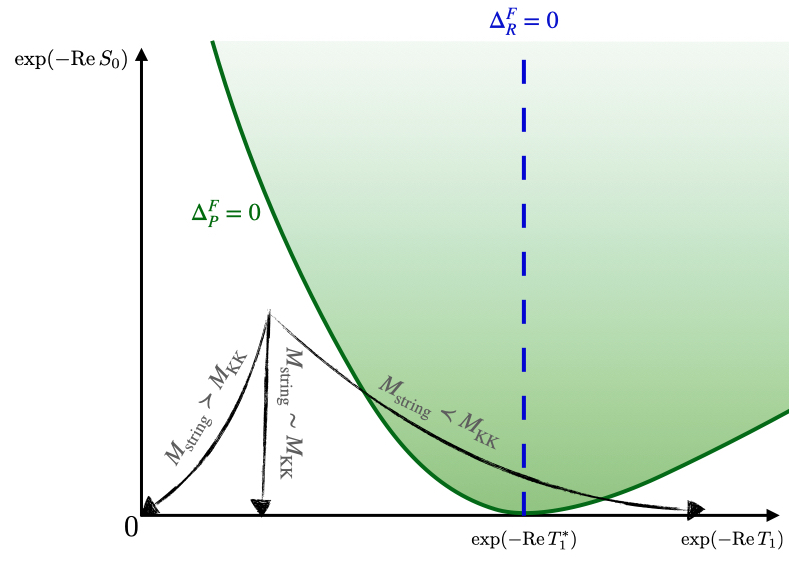}
\caption{This figure shows a sketch of the F-theory K\"ahler moduli space projected to the real part of the chiral fields for the example of $B_2=\mathbb{P}^2$. Shown are the two directions corresponding to the (exponentiated) volume $\text{Re}\,S_0$ of the exceptional divisor $\mathcal{S}_-$ and the volume $\text{Re}\,T_1$ of pull-back of the hyperplane $H$ in $\mathbb{P}^2$ to $B_3$. We also show the singularities whose existence we inferred from duality to the heterotic GLSM parameter space. The projection of $\Delta_P^F=0$  to this plane is indicated by the green solid line and corresponds to a strong coupling singularity for the $E_8$ gauge group realized on the exceptional divisor $\mathcal{S}_-$ whereas the singularity $\Delta_R^F=0$ (blue, dashed line) corresponds to the boundary of the K\"ahler cone. The region shaded in green corresponds to a strong coupling phase. In this figure, the Calabi--Yau phase of the perturbative heterotic string corresponds to the lower left corner. The intersection of $\Delta_R^F=0$ and  $S_0=\infty$ is a point of tangency between $\Delta_P^F=0$ and $S_0=\infty$. The arrows indicate the different types of classical infinite distance limits as in \eqref{possiblelimits} (see also \cite{Klaewer:2020lfg}) originating from a common point in the geometric phase.}\label{fig:modulispace}
\end{figure}

Using duality, so far we have established the existence of a strong coupling singularity in $\cM^F$ which we identify as the source for the breakdown of the perturbative expansion. We further explained how to match the perturbative corrections to the F-theory effective action with the heterotic expectation. We now use these results to further investigate the structure of $\cM^F$ in the light string limit. In general the strong coupling singularity is expected to correspond to a complex co-dimension one locus $\{\Delta_P^F=0\}$ in $\mathcal{M}^F$ whose position is encoded in $\mathfrak{u}_a\big|_{\rm sing}$ via \eqref{u1sing}. And indeed, from \eqref{u1sing} we can extract important information about $\Delta_P^F$: First, as alluded to before, it is sensitive to the complex structure sector of the F-theory compactification. This reflects that in an $\cN=1$ compactification of F-theory the K\"ahler and complex structure deformation spaces do not factorize at the quantum level. Notice that in the F-theory effective theory this complex structure dependence is crucial to differentiate between $\Delta_P=0$ and $\Delta_R=0$ and therefore for the interpretation of the obstruction to reach the classical light string limit as a consequence of a strong coupling singularity. Second, the location of the singularity in the $\mathfrak{u}_a$-hyperplane depends on the value of the perturbative string coupling $\text{Re}\,S_0^{(1)}$. Accordingly, the locus $\{\Delta_P^F=0\}\subset \cM^F$ also has such a coupling dependence. Notice that this coupling dependence cannot be inferred from the heterotic GLSM analysis directly, though the perturbative F-theory corrections allow us to uncover this dependence.\footnote{Let us stress that this is very different from the GLSM for Type IIA on the same CY. In this case we do not have a hidden dependence of $\{\Delta_P=0\}$ on the 4d string coupling of type IIA. This is due to the fact that for Type IIA on CY three-folds the 4d string coupling resides in a hypermultiplet which is independent from the vector multiplet moduli space described by the Type IIA K\"ahler moduli. On the other hand for Calabi--Yau compactifications of Type IIB, the K\"ahler moduli and the 4d string coupling are part of the same hypermultiplet moduli space which does not factorize at the quantum level. This non-factorization then leads to the obstruction of certain infinite distance limits as discussed in \cite{Marchesano:2019ifh, Baume:2019sry}. Similarly, the fact that for the heterotic string there is a coupling dependence of $\Delta_P$ reflects the fact that in $\cN=1$ theories there is no factorization between the different parts of the scalar field space. } However, at this point, we only have access to the dependence of the singular locus on $\text{Re}\,S_0$. In general there is an additional dependence on $\text{Im}\,S_0$ which we do not discuss here. Still, the projection to the real part of $\text{Re}\,S_0$ provides us with important information about the singularity structure of the scalar field space. This was already the case in the heterotic FI-parameter space where the projection of the singular loci to the real plane gave rise to the amoeba of the singularity from which the location of the strong coupling phases can be read off (cf. figure~\ref{fig:amoeba}). For real $S_0$ the location of $\{\Delta_P^F=0\}$ and the strong coupling phase in the string theory limit of $\cM^F$ are illustrated schematically in figure~\ref{fig:modulispace} for the simple case $B_2=\mathbb{P}^2$.

This structure is very reminiscent of the $\cN=2$ point-particle limit of type IIA string theory reviewed in section \ref{sec:N2comparison}. There the relevant singularity corresponds to the Coulomb branch singularities of the $\cN=2$ SYM theory at 
\begin{equation}
   u\,\big|_{\rm sing.} =\pm \Lambda^4 \exp(-\hat S_{\rm IIA}) \,. 
\end{equation} 
Here $u$ is the $SU(2)$ Coulomb branch parameter. Notice that $u\,|_{\rm sing.}$ depends on $\hat S_{\rm IIA}$, i.e. the tree-level gauge coupling of the $SU(2)$ gauge theory and the dynamically generated 1-loop scale $\Lambda$ of the field theory. Both are related to the actual gauge coupling as in \eqref{hatSIIA}. The analogy between the $\cN=2$ field theory limit and the $\cN=1$ string theory limit identifies 
\begin{equation} 
 u\, \big|_{\rm sing.}\quad \longleftrightarrow \quad \mathfrak{u}_a\big|_{\rm sing.}\,. 
\end{equation} 
In the same way as $\{\Delta_{\rm IIA}=0\}$ can be generated from $u\,|_{\rm sing.} =\pm \Lambda^4 \exp(-\hat S_{\rm IIA}) $ by varying the modulus $S_{\rm IIA}$, we obtain the F-theory singular locus $\{\Delta_P^F=0\}\subset \cM^F$ by varying $\text{Re}\, S^{(1)}_0$. In addition to the locus $\{\Delta_P^F=0\}\subset \cM^F$ we also have the locus $\{\Delta_R^F=0\}$. From our analysis around \eqref{ReT1star} we recall that $\{\Delta_R^F=0\}$ is located at $\text{Re}\,T_a=\text{Re}\,T_a^*$ and is therefore independent of $\text{Re}\, S_0^{(1)}$. Hence $\{\Delta_R^F=0\}$ is given by a locus orthogonal to the weak-coupling divisor $\text{Re}\,S^{(1)}_0= \infty$. Importantly, we observe a point of tangency between the large volume/weak coupling divisor $\text{Re}\,S_0\rightarrow \infty$ and the singular locus $\{\Delta_P^F=0\}$: for $\text{Re}\, S^{(1)}_0\rightarrow \infty$ we find $\mathfrak{u}_a|_{\rm sing.}\rightarrow 0$ such that the two loci $\{\Delta_P^F=0\}$ and $\{\Delta_R^F=0\}$ coincide asymptotically. Equivalently, the entire strong coupling phase, i.e. the shaded region in figure \ref{fig:modulispace}, reduces to a single point. This is precisely what happens in the $\cN=2$ setup as well, where in the vicinity of $S_{\rm IIA}\rightarrow i\infty$ the entire strong coupling phase of the $\cN=2$ SYM theory gets mapped to $T_{\rm IIA}=i$. Notice that for this to happen it is crucial that $\Delta_P^F$ depends on the coupling $S_0$. This effect is therefore not visible in the heterotic GLSM parameter space that does not take into account any coupling dependence. To retrieve the structure of the heterotic GLSM one needs to compensate for this coupling dependence by rescaling $\mathfrak{u}_a \rightarrow \tilde{\mathfrak{u}}_a$ as
\begin{equation}
 \tilde{\mathfrak{u}}_a=\left(\frac{8\pi}{b} \frac{{\rm Re} {S}_-^{(1)}}{{\rm Re} T_a^{(0)}}\right) \mathfrak{u}_a
 \end{equation} 
Using this rescaling we can then identify $\tilde{\mathfrak{u}}_a \leftrightarrow \eta_{ab}\left(s^b-s^b_{0}\right)$ where $s^b_{0}$ is the quantum volume of the curve $H\in H_2(B_2)$ on the corrected K\"ahler cone boundary induced by non-perturbative effects (cf. e.g. \eqref{BiJP2}). At the singularity we then find 
\begin{equation}
 \tilde{\mathfrak{u}}_a\big|_{\rm sing.} \sim - \log \frac{\xi}{\exp\left(\Xi'\right)} \,.
\end{equation} 
The rescaling $\mathfrak{u}_a\rightarrow \tilde{\mathfrak{u}}_a$ can be viewed as the analogue of the rescaling $u\rightarrow \tilde u$ in the $\cN=2$ field theory limit of type IIA string theory as in \eqref{rescalingSW} where $\tilde u$ is the coordinate on the $\mathbb{P}^1$ used to resolve the point of tangency in moduli space.

Thus, as anticipated in section \ref{sec:generalprinciple}, at the level of the field space structure, we see a clear analogy between the point-particle limit of type IIA string theory studied in \cite{Kachru:1995fv} and the string theory limit of F-theory. We can further exploit this analogy when searching for an interpretation of $\Delta^F_P$ in terms of the light string spectrum. Therefore, recall that in the $\cN=2$ case the $W^\pm$ bosons, which classically become massless at $T=i$, leave the BPS spectrum at the field theory singularity. On the other hand, the magnetic monopole becomes massless at $u=\Lambda^4\exp(-\hat S_{\rm IIA})$.  Similarly, in the F-theory setup we expect the D3-brane wrapped on $C^0$ to leave the spectrum of BPS strings at the singularity $\Delta^F_P$ and to be replaced by some other BPS string. In the next section we are going to discuss this perspective in more detail. \\

Before discussing the spectrum of BPS strings, let us put the structure of the field space into the context of the obstruction to certain classical emergent string limits observed in \cite{Klaewer:2020lfg}. In that work classical infinite distance limits were classified according to the relative asymptotic scaling between the tension $M_{\rm string}^2$ of the emergent string and the KK-scale leading to the three possibilities 
\begin{equation}\label{possiblelimits}
 M_{\rm string} ^2\succ M_{\rm KK}^2 \,,\qquad M_{\rm string}^2\sim M_{\rm KK} \,,\quad \text{or}\quad  M_{\rm string}^2\prec M_{\rm KK}^2\,. 
\end{equation}  
The first case was identified as a decompactification limit whereas both, the second and the third case, are classically emergent string limits. However, as shown in \cite{Klaewer:2020lfg}, the latter limits are obstructed at the perturbative limit such that only the second case survives as actual emergent string limits at the quantum level. In figure~\ref{fig:modulispace} limits of this second kind correspond to trajectories orthogonal to the horizontal axis. On the other hand, decompactification limits asymptote towards the origin in figure~\ref{fig:modulispace}. Finally, those classical emergent string limits, that are obstructed at the quantum level, intersect the horizontal axis at $\exp(-\text{Re}\,T_1)=1$, i.e. to the right of the $\{\Delta_R^F=0\}$. Therefore any classical limit that asymptotically leads to $M_{\rm string}\prec M_{\rm KK}$ necessarily intersects the singular divisor $\{\Delta_P^F=0\}$ such that this singularity can indeed be interpreted as obstructing the unphysical limits where the tension of a 4d critical string decouples parametrically from the geometric KK-scale. Notice that the locus along which $M_{\rm het} = M_{\rm KK}$ may in fact lie to the left of $\{\Delta_R^F=0\}$ such that $M_{\rm string}\lesssim M_{\rm KK}$ can in principle be achieved though not parametrically. This is, however, not a problem since we only need an obstruction to \emph{parametrically} separate the scales $M_{\rm string}$ and $M_{\rm KK}$ which is indeed the consequence of the presence of $\{\Delta_P^F=0\}$. 
\subsection{Singularities and the string spectrum}\label{sec:4dEFTstring}
We now want to interpret the structure of $\cM^F$ and in particular the strong coupling singularity in terms of the BPS string spectrum. We do this from two different persepctives: first from the perspective of the worldsheet of the string and second by relating the structure of the scalar field space directly to the spectrum of BPS strings, following the ideas of \cite{Marchesano:2022avb}. In addition, we compare our results to the perhaps more familiar setups of F-theory compactifications to 6d.

\subsubsection{Worldsheet perspective}
Starting with the worldsheet perspective, we would like to argue that in the vicinity of the strong coupling singularity $\Delta_P^F=0$ the string obtained from the D3-brane wrapped on $C^0$ leaves the spectrum of light BPS strings. To that end, we want to show that the supersymmetric worldvolume theory on the D3-brane string breaks down at $\Delta_P^F=0$. To that end, we first recall from \cite{Lee:2019jan} that on the worldsheet level the identification between the D3-string and the heterotic string can be achieved by first reducing the worldvolume theory on a D3-brane, i.e. $\cN=4$ SYM theory with varying gauge coupling, along the curve $C^0$. Using the results of \cite{Lawrie:2016axq} one finds the following spectrum of fields on the worldsheet \cite{Lee:2019jan}:\footnote{In \cite{Martucci:2022krl} the worldsheet theories of more general D3-brane strings in 4d $\cN=1$ F-theory have been discussed in order to derive quantum gravity constraints from anomaly inflow on these strings.}
\begin{itemize}
\item First there are two chiral multiplets $(\phi_i, \mu_+^i)$, $i=1,2$, from the reduction of the scalar and the gaugino in the $\cN=4$  gauge multiplet. The two scalars $\phi_i$ describe the motion of the D3-brane in the internal directions transverse to $C^0$, i.e. the complex coordinates on the base $B_2$. 
\item In addition one has a chiral multiplet $(g, \gamma_+)$ also arising from the reduction of the $\cN=4$ multiplet scalar. The scalar in this multiplet describes the motion of the string in the two extended directions transverse to the string. 

\item Moreover, one finds a chiral multiplet $(a,\psi_+)$ for which the scalar is identified with the Wilson line of the $\cN=4$ vector field over $C^0$. On the heterotic side this scalar can be identified with the coordinates of the torus fibered over $B_2$ in the CY $Z_3$ that is the compactification manifold of the heterotic string. 
\item Finally there are 16 Fermi multiplets $\lambda_-$ associated to the bundle degrees of freedom of the heterotic string. 
\end{itemize} 
Summarizing the spectrum as 
\begin{equation}
\underbrace{(g, \gamma_+)}_\text{2d extended} \,, \qquad  \underbrace{2 \times(\phi, \mu_+) }_{B_2}\,, \qquad  \underbrace{(a,\psi_+)}_{T^2}\,, \qquad  \underbrace{\lambda_-}_\text{bundle}\,,
\end{equation}
the worldsheet theory of the  D3-brane wrapped on $C^0$ can then be identified with a non-linear sigma model (NLSM) with target space $Z_3: T^2\rightarrow B_2$ describing a CY compactification of the critical heterotic string in the large volume phase. This NLSM preserves $(0,2)$ supersymmetry in 2d and thus (at least in the large volume limit) the D3-brane string with this NLSM as its worldsheet theory can be viewed as a genuine BPS object of the $\cN=1$ EFT. In order for this to be the case also away from the large volume limit, we need to ensure that the worldsheet theory on the D3-brane string remains well-defined. To that end we can consider correlators of the form
\begin{equation}\label{eq:correlators}
 \langle \mathcal{O}_{i_1} \dots \mathcal{O}_{i_n}\rangle\,,
\end{equation} 
of the worldsheet theory and check whether these remain finite. Here the $\mathcal{O}_i$ are some operators on the worldsheet and singular correlators would imply that the underlying worldsheet theory is ill-behaved. Since we identified the worldsheet theory with the heterotic NLSM we can use the heterotic theory to calculate these correlators explicitly. The relevant scalar field space is spanned by 
\begin{equation}
 \cM_{\rm string} = \langle S_0 , T_a\rangle\,,
\end{equation} 
which, via heterotic/F-theory duality, are identified as the gauge coupling of the unbroken $E_8$ heterotic gauge theory and the heterotic K\"ahler moduli $t^a$ of the base $B_2$. For the heterotic NLSM the $\mathcal{O}_i$ appearing in the correlators are associated to elements of $H^2(Z_3)$. In section \ref{sec:dual} we discussed the GLSM correlators, see \eqref{correlators} for the example $B_2=\mathbb{P}^2$ . These GLSM correlators can be expanded in terms of the exponentiated FI-parameters $q_i$ which should be interpreted as GLSM gauge instantons.  In the geometric phase of the GLSM, the correlators of the A(A/2)-twisted heterotic NLSM agree with these GLSM correlators and the gauge instanton expansion can be rephrased as a worldsheet instanton expansion upon applying the mirror map
\begin{equation}
2\pi i\, t^a= \log q_a + \mathcal{O}(q_a)\,,
\end{equation} 
relating the K\"ahler moduli of the NLSM target space to the GLSM FI-parameters. Importantly, irrespective of the details of this map the singularities of the GLSM correlators translate into singularities for those of the NLSM. In general one finds that the GLSM correlators have the form 
\begin{equation}
 \langle \sigma_{a_1} \sigma_{a_2} \sigma_{a_3} \rangle_{Z_3} \sim \frac{f(q_0,q_a)}{\Delta_P}\,,
\end{equation}
implying that, as was already anticipated in section \ref{sec:dual}, the heterotic NLSM becomes singular along $\{\Delta_P=0\}$ when identifying the GLSM operators $\sigma_a$ with the corresponding operators $\mathcal{O}_a$ of the NLSM. 

As argued in this section, the singularity $\{\Delta_P=0\}$ in the heterotic K\"ahler moduli space has a counterpart $\{\Delta_P^F=0\}$ in the F-theory scalar field space. Since we can identify the worldsheet theory on the D3-brane wrapped on $C^0$ with the heterotic NLSM we conclude from our discussion above that the worldsheet theory on this D3-brane string becomes singular along $\{\Delta_P^F=0\}$.  At and beyond the locus $\{\Delta_P^F=0\}$ reducing the D3-brane action on $C^0$ hence does not yield a consistent supersymmetric  worldsheet theory anymore signaling that the D3-brane wrapped on $C^0$ leaves the spectrum of BPS strings. This is to be contrasted to the geometric phase, where the NLSM living on the wrapped D3-brane is a well-defined $(0,2)$ supersymmetric theory confirming that the D3-brane on $C^0$ is a BPS object in this regime. Thus, from the worldsheet perspective we expect a string other than the D3-brane on $C^0$ to be the fundamental BPS state once we move into the phase beyond $\{\Delta_P^F=0\}$. To summarize, the worldsheet perspective thus provides us with a physical interpretation of the singularity $\Delta_P^F=0$ as the point in field space where the critical D3-brane string leaves the spectrum of BPS strings. In the following, we want to present further support for this picture from a space-time point of view.

\subsubsection{4d EFT string perspective}
Apart from the worldsheet perspective, we can directly relate the structure of $\cM^F$ to the BPS string spectrum following \cite{Marchesano:2022avb}. In there the fate of strings in 4d EFTs at strong coupling is analyzed. Most importantly, \cite{Marchesano:2022avb} relates the singularity structure of the moduli space to the existence of strings in the 4d EFT and identifies a way to estimate the tension of $\frac12$-BPS strings away from weak coupling limits. Before we apply the discussion of \cite{Marchesano:2022avb} to our F-theory setup, let us review their main setup and results. \newline

Therefore consider a four-dimensional $\cN=2$ or $\cN=1$ supersymmetric EFT. The scalar field space $\mathcal{M}$ of this theory is spanned in general by complex scalar fields $t^i$ for which the effective action is given by
\begin{equation}\label{eq:S4d}
S_{4d}= M_{\rm P}^2 \int \left(\frac12 R*1 + g_{a\bar b}dt^a \wedge *d\bar{t}^b\right)\,. 
\end{equation} 
Here $g_{a\bar b}$ is the field space metric and for the moment we assume that any contribution to the scalar potential is negligible such that we can treat the scalar field space spanned by the $t^a$ as an actual moduli space. In $\cN=2$ theories this is ensured by supersymmetry whereas in a genuine $\cN=1$ theory this is generically not the case as e.g. in string theory constructions fluxes or non-perturbative effects can generate a scalar potential. 

The scalar fields $t^a$ are periodic such that 
\begin{equation}\label{eq:axionshift}
 t^a \sim  t^a +i \,,
\end{equation}
In certain regions of the scalar field space a continuous version of this shift symmetry can be approximately realized. In this case the imaginary part of $t^a$ can be treated as an axion. The objects magnetically charged under these axions are 4d strings. To these 4d strings one can associate cosmic string solutions in the spirit of \cite{Greene:1989ya} that describe the backreaction of the strings in the extended four-dimensional space-time. Such string theory solutions have been investigated in detail in \cite{Lanza:2020qmt, Lanza:2021udy, Marchesano:2022avb, Lanza:2022zyg, Grimm:2022sbl, Cota:2022yjw, Martucci:2022krl}. As such string solutions should preserve 2d Poincar\'e invariance along the directions parallel to the string an ansatz for the metric is given by
\begin{equation}
ds^2=-dt^2 +dx^2 +e^{2D}dzd\bar z\,,
\end{equation}
where $z\in \mathbb{C}$ is the coordinate transverse to the string. Supersymmetric solutions to Einsteins equations have to satisfy \cite{Lanza:2020qmt, Lanza:2021udy} 
\begin{equation}\label{SUSYeq}
  \partial_{\bar z} t^a=0 \,,\qquad e^{2D}=|f(z)|e^{-K}\,,
\end{equation}
for a holomorphic function $f(z)$ and $K$ the K\"ahler potential. The profile for $t^a(z)$ should reflect the shift \eqref{eq:axionshift}when encircling  the core of the cosmic string. In the vicinity of the string core located at $z=0$, the local holomorphic profile for a cosmic string with magnetic charge vector  $\underline{\mathbf{e}}=(e^1, \dots, e^n)$ then needs to have to form
\begin{equation}\label{backreact}
 t^a(z) = t^a_0 -\frac{e^a}{2\pi} \log\frac{z}{z_0}\,,
\end{equation}
for $|z|\ll |z_0|$. Here $t^a_0$ are background values for the complex scalar fields and $z_0$ parametrizes the coarseness of the solution. In general this solution receives polynomial corrections in $z$ corresponding to non-perturbative effects. Of particular relevance are therefore the solutions that in the limit $z\rightarrow 0$ flow to regions in the field space where the continuous shift symmetry is approximately unbroken and all instantons charged under the shift symmetry \eqref{eq:axionshift} are suppressed. The strings leading to such solutions have been dubbed EFT strings in \cite{Lanza:2021udy}. The tension of such an EFT string can be estimated by the energy of the backreaction of the string stored in a disk with radius $R$ around the string \cite{Lanza:2021udy}
\begin{equation} \label{Eback}
 \mathcal{E}_{\rm back}(R) = M_{\rm pl}^2  \int_{D(R)}  d^2 z \,\mathbf{t}^*(\mathcal{J}_\mathcal{M})\,. 
\end{equation}
Here, $\mathcal{J}_\mathcal{M}=g_{a\bar b} dt^a d\bar t^{\bar b}$ is the K\"ahler form on the moduli space $\mathcal{M}$ and $\mathbf{t}^*$ denotes the pull-back from $\mathcal{M}$ to the space-time transverse to the string using \eqref{backreact}. In the vicinity of the string core, one can exploit the shift symmetry \eqref{eq:axionshift} to calculate the tension explicitly to be 
\begin{equation}\label{Ebackpert}
 \mathcal{E}_{\rm back}(R) = M_{\rm pl}^2 \left[L_a(R) - L_a(0)\right]\,,
\end{equation}
where $L_a$ are the dual linear multiplets that are related to the $t^a$ via 
\begin{equation}
 L_a =-\frac12 \frac{\partial K}{\partial \text{Re}\,t^a}\,,
\end{equation}
cf. \eqref{eq:defTi}. However, once the shift-symmetry \eqref{eq:axionshift} is broken by non-perturbative effects, it is not possible to dualize the chiral superfields to linear multiplets anymore. In these regimes, the naive profile \eqref{backreact} does in general get corrections due to the non-perturbative effects. Moreover, the tension of the string solution cannot be calculated via \eqref{Ebackpert} and it is not even clear how to calculate the tension of a probe string in a background determined by $t^a_0$. As argued in \cite{Marchesano:2022avb} it is still possible to extend the local solution \eqref{backreact} to a global one precisely by taking into account these non-perturbative effects. In fact \cite{Marchesano:2022avb} conjectured that for elementary strings, i.e. strings with charge vector $\underline{\mathbf{e}}=(\delta_{ab})$ for some $b$, it should always be possible to extend the local solution in a unique way such that the tension of the full solution remains sub-Planckian, $\mathcal{E}_{\rm back}(\infty)<2\pi M_{\rm pl}^2$. 

The global solution then gives rise to a profile $t^a(z)$ yielding a map defined on the entire space $\mathbb{C}$ transverse to the string (or its one-point compactification $\mathbb{P}^1$) to $\cM$. The image of this map is a two-cycle $\Sigma\subset \cM$. This two-cycle in general intersects singular divisors $\{\Delta=0\}\subset \cM$ which, following \cite{Marchesano:2022avb},  can be interpreted as additional strings present in the global solution that regulate the backreaction of the EFT string.  Thus, to each component of the singular locus $\{\Delta=0\}$ one can associate a string. Encircling this string gives rise to a monodromy $M$ similar to the monodromy \eqref{eq:axionshift} induced by the EFT string. Borrowing the results from \cite{Bergshoeff:2006jj}, it is argued in \cite{Marchesano:2022avb} that the order of this monodromy can be associate to the tension localized at the location of the respective string, i.e. to the minimal tension of the string. To get a supersymmetric solution, one has to require that all strings present in the global solution have unipotent monodromy matrices associated to them \cite{Marchesano:2022avb}. On the other hand, the tension associated to the full solution is measured by the order of the monodromy at spatial infinity. The condition that the solution has finite tension then requires the combined monodromies to give rise to a finite order monodromy obtained when encircling the point at infinity, i.e. 
\begin{equation}
 M_\text{\tiny EFT} \prod_{\alpha} M_{\alpha} = M^{-1}_{z=\infty}\,, \qquad \text{with} \qquad M_{z=\infty}^n=\text{Id}\,,
\end{equation} 
for some $n\in \mathbb{N}_{>0}$. Here $\alpha$ scans over all regulator strings present in the solution and $M_\text{\tiny EFT}$ denotes the monodromy around the axionic string at $z=0$. The tension of the full solution is then given by $\mathcal{E}_{\rm back}=\frac{2\pi }{n}M_{\rm pl}^2$. 

Similarly to the calculation of the tension of a global string solution, the picture put forward in \cite{Marchesano:2022avb} can also be used to infer the tension of a probe string in a background determined by a point $t^a_0\in \cM$: Instead of considering an infinitely extended axionic string, one considers the same string wrapped on a loop of radius $R$. From far away, this string now looks like a point-particle and its backreaction dies off quickly converging towards a constant value which determines a point in the scalar field space $t_0^a\in \cM$. We can thus view this configuration as describing a probe string in a background characterized by $t^a_0$. The tension of this probe string is then given by the energy $\mathcal{E}_{\rm back}$ stored in its backreaction. Since the circular string is self-regulating we have $\mathcal{E}_{\rm back} \leq \frac{2\pi}{n} M_{\rm pl}^2$. As long as we choose $t^a_0$ close to the point in moduli space where we have an approximate shift symmetry as in \eqref{eq:axionshift}, $\mathcal{E}_{\rm back}$ indeed calculates the tension of the probe axionic string. However, as we tune $t^a_0$ towards the interior of the moduli space, at some point additional strings nucleate corresponding to the regulator strings associated to the singular divisors of $\cM$. Once this happens, $\mathcal{E}_{\rm back}$ in fact calculates the tension of the bound state of the axionic string and the regulator strings. The tension of the bound state of strings remains sub-Planckian while the tension of the axionic string alone can become super-Planckian. Thus, in case we choose $t^a_0$ to be too far in the interior of $\cM$ away from the point with the approximate axionic shift symmetry, the axionic string, once nucleated, forms a black holes since it is not BPS protected anymore. On the other hand, the bound state of axionic and regulator strings does not form a black hole. This bound state can never be tensionless itself since its associated monodromy corresponds to $M_{z=\infty}$ which is of finite order. Thus, the singularity signals that the probe axionic string ceases to be BPS but instead the relevant BPS object is the bound string. \\

With this preparation, we can now come back to our F-theory setup. For simplicity, let us consider the example with $B_2=\mathbb{P}^2$ and discuss the fate of the D3-brane wrapped on $C^0$ in the vicinity of the singular locus $\{\Delta_P^F=0\}$. Let us denote the D3-brane string by $\mathtt{H}$ and consider the EFT string solution associated to the string $\mathtt{H}$. Consider the string $\mathtt{H}$ wrapped on a loop in a background determined by a point in the $(S_0,T_1)$-plane (cf. figure \ref{fig:modulispace}). If we choose a point in the vicinity of the locus $\text{Re}\,S_0=\infty$, the tension of the string $\mathtt{H}$ is simply given by 
\begin{equation}
 \frac{T_{\mathtt{H}}}{M_{\rm pl}^2}=\frac{1}{\text{Re}\,S_0}\,. 
\end{equation}
Instead, we can also consider a point away from the weak coupling locus and ask about the tension of the D3-brane on $C^0$. As long as we stay below the green line in figure \ref{fig:modulispace} the string tension is given by the energy stored in the backreaction of the  single string. However, if we choose a background determined by a point within the shaded region in figure \ref{fig:modulispace}, the backreaction of the D3-brane on $C^0$ requires the nucleation of a regulator string $\mathtt{P}$ associated to $\{\Delta_{P}^F=0\}$ which itself is tensionless on the singular locus since the monodromy around the singular locus is unipotent. Notice that since the strong coupling singularity is at finite distance, the string $\mathtt{P}$ is a non-critical string with a finite number of particle-like excitations. Once nucleated, the tension stored in the backreaction should be identified with the tension of the bound state of $\mathtt{H}$ and $\mathtt{P}$. On the other hand, the tension of the string $\mathtt{H}$ can be super-Planckian in the strong coupling phase and this string is therefore not part of the BPS spectrum as it becomes unstable against collapsing into a black hole as described in \cite{Banks:2010zn}. Instead it has to be replaced by the finite-tension string corresponding to the bound state
\begin{equation}
\mathtt{H}\rightarrow\mathtt{H}+\mathtt{P}\,. 
\end{equation}
Notice that in order for this to happen, it is important that the discriminant $\Delta_P^F$ depends on $S_0$ because otherwise the backreaction of the string $\mathtt{H}$ would not intersect $\{\Delta_P^F=0\}$. To see this notice that the image of the backreaction of $\mathtt{H}$ in $\cM^F$ are two-cycles $\Sigma$ which, when projected to the real $(\text{Re}\,S_0,\text{Re}\,T_1)$ plane, yield vertical lines in figure \ref{fig:modulispace}. If $\Delta_P^F$ was independent of $S_0$ it would just correspond to another vertical line (just as $\{\Delta_R^F=0\}$ in that figure) which would in general not intersect any other vertical line associated to the backreaction of the string $\mathtt{H}$.

So far, we ignored any contribution to $S_{4d}$ coming from a non-trivial scalar potential. In \cite{Marchesano:2022avb} this was justified by considering either $\cN=2$ theories or heterotic compactifications with standard embedding in the K\"ahler sector described in terms of a GLSM FI-parameter space for which the scalar potential vanishes identically. In our case of interest, the full F-theory field space is, however, not an actual moduli space due to a non-trivial superpotential induced by the $E_8$ gauge instantons which does not play a role in the GLSM analysis of \cite{Marchesano:2022avb}. Strictly speaking in the heterotic case, the analysis of \cite{Marchesano:2022avb} is valid along the locus $S_0=\infty$ where, indeed, the scalar potential vanishes identically. Away from this locus, the heterotic K\"ahler field space is subject to corrections that are not captured by the GLSM analysis as is clear from our analysis in this section. Here we are interested in describing the backreaction of a D3-brane on the curve $C^0$. In this case we cannot rely on a pure GLSM description since the backreaction of this brane necessarily leads away from the $S_0=\infty$ locus. Most strikingly, along the singular locus $\{\Delta_P^F=0\}$ the gauge instanton contribution is unsuppressed. Therefore, unlike in the discussion of \cite{Marchesano:2022avb}, we would have to take into account the effect of the non-trivial scalar potential induced by the superpotential on $\mathcal{E}_{\rm back}$ and thus the tension of the string $\mathtt{H}$. 

Close to the core of the string $\mathtt{H}$, the contribution to the scalar potential is still negligible. Hence the 4d action can still be considered to be of the form as in \eqref{eq:S4d} leading to the supersymmetry condition \eqref{SUSYeq}. Therefore the profile for $S_0$ induced by the backreaction of the string $\mathtt{H}$ is still locally given by \eqref{backreact} for $|z|\ll |z_0|$. However for $|z|\sim |z_0$, the efffect of the non-vanishing scalar potential becomes non-negligible. This in principle affects the supersymmetry condition \eqref{SUSYeq} and hence we might get a deviation from a holomorphic profile. In addition, the energy stored in the backreaction \eqref{Eback} is not simply given by the pull-back of the K\"ahler form since this only captures the effect of instantons correcting the K\"ahler potential, but not those correcting the scalar potential. Therefore, we expect that the regulation of the local string solution is sensitive to the non-perturbative contributions to the superpotential. Calculating the full backreaction for the string would be beyond the scope of this paper. We notice, however, that in the spirit of the analysis in \cite{Mayr:1996sh} one expects that the unsuppressed contributions to the superpotential on singular loci are precisely due to strings that become tensionless along the singular locus. Along these lines we then expect that the string $\mathtt{P}$, that becomes light at $\{\Delta_P^F=0\}$, regulates the backreaction precisely by taking into account also the superpotential. In other words, the regulation of the backreaction in this case should be achieved partially through the scalar potential. It would be interesting to confirm this explicitly and to check whether the full backreaction still has finite tension. For our discussion here the relevant point is that the singular locus $\{\Delta_P^F=0\}$ yields an additional string present in the backreaction of the EFT string obtained from the D3-brane on $C^0$. 

One might wonder whether the string $\mathtt{P}$ associated to the singular locus  $\{\Delta_P^F=0\}$ has a simple geometric origin. A candidate would be for instance the non-critical D3-brane wrapped on a curve in the hyperplane class of the $\mathbb{P}^2$ base. This string carries charge $(-18,1)$ under the axions $(\text{Im}\,S_0, \text{Im}\,T_1)$. To see this we notice that the two generators of the movable cone of $B_3$ can be written as 
\begin{equation}
 C^0=J_1\cdot J_1 \,\quad C^1=\mathcal{S}_+\cdot J_1\,,
\end{equation} 
where $J_1$ is the pull-back of the hyperplane class to the rational fibration. D3-branes on $C^0$ and $C^1$ are thus the primitive EFT strings of $B_3$ \cite{Lanza:2021udy,Cota:2022yjw}. On the other hand, the hyperplane class of $\mathbb{P}^2$ can be written as 
\begin{equation}
\mathcal{C} = C^1 - 18 C^0\,,
\end{equation} 
from which we can read off the charges under the axions. The associated monodromy when encircling the D3-brane on $\mathcal{C}$ is 
\begin{equation}
 M_{(1,-18)} = M_{T_1= \infty} \left(M_{S_0= \infty}\right)^{-18}\,,
\end{equation}
where $M_{S_0= \infty}$ and $M_{T_1= \infty}$ are the monodromies around the large volume divisors, i.e. the horizontal and vertical axis in figure~\ref{fig:modulispace}. This is not the monodromy we would expect from the string asscoiated to $\{\Delta_P^F=0\}$. Indeed, the D3-brane on $\mathcal{C}$ leads to a non-EFT string in the language of \cite{Lanza:2021udy} for which the profile induced by the backreaction is naively given by 
\begin{equation}
 S_0 = S_0^0 +\frac{12}{2 \pi} \log \frac{z}{z_0} \,,\qquad T_1 = T_1^0-\frac{1}{2\pi} \log\frac{z}{z_0}\,,
\end{equation} 
for some background values $S_0^0$ and $T_1^0$ and some parameter $z_0$. As we approach the core of the string we thus classically reach the locus $(\text{Re}\,S_0,\text{Re}\,T_1)=(-\infty, \infty)$ which, as expected for a non-EFT string, is well outside the controlled regime. On the other hand this also does not correspond to the locus $\{\Delta_P^F=0\}$ such that we can rule out the possibility that the string $\mathtt{P}$ is simply describable as a D3-brane on $\mathcal{C}$. After all, it is not surprising that the string $\mathtt{P}$ does not have a geometric interpretation since we know that at $\{\Delta_P^F=0\}$ the geometric description of the $\cN=1$ EFT breaks down. At this point, we do not have a clear microscopic description of this string though it would still be interesting to find a worldsheet description for this string. We leave this task, however, for future work.

To summarize, the picture of \cite{Marchesano:2022avb} suggests that, indeed, the singularity $\Delta_P^F=0$ signals a change in the BPS-string spectrum as $\mathtt{H}$ gets replaced by $\mathtt{H}+\mathtt{P}$. This is reminiscent of the situation in the $\cN=2$ field theory where at the strong coupling singularity also the $W^\pm$ bosons leave the BPS particle spectrum which now consists of the magnetic monopole and the dyon, that is the bound state of $W^\pm$ boson and magnetic monopole. As mentioned above the monodromy around $\{\Delta_P^F=0\}$ is unipotent such that the string $\mathtt{P}$ becomes tensionless on the singular locus. It can therefore indeed be viewed a the analogue of the SW monopole which becomes massless at the strong coupling singularity. The worldsheet perspective and the 4d EFT string analysis of the strings and their backreaction therefore consistently point towards the following interpretation of the strong coupling singularity $\{\Delta_P^F=0\}$: The singular locus $\{\Delta_P^F=0\}$ gives an obstruction to a classical light string limit since the string itself ceases to be BPS in the new strong coupling phase. 
\subsubsection{Comparison to 6d}\label{sec:comparison}
It is instructive to compare the 4d case discussed here to an analogue situation in six dimensions, cf. e.g. \cite{Lee:2018urn, Lee:2018spm} for discussions of string theory limits of F-theory compactifications to 6d. Therefore consider F-theory compactified on a Calabi--Yau three-fold that is a smooth Weierstrass model over the Hirzebruch surface $\mathbb{F}_1$. We already discussed this geometry in the context of heterotic compactifications to four dimension in section \ref{sec:dual}. In 6d we do not have to fear any corrections to the moduli space geometry. Thus, when it comes to tensions of solitonic strings, the tension is completely determined by the geometry of the Calabi--Yau. In the case at hand, the base $B_2$ is a rational fibration over $\mathbb{P}^1$. The K\"ahler form of the base is given by 
\begin{equation}
 J_{\mathbb{F}_1} = t^1 h + t^2 (s_-+h) \,, 
\end{equation} 
where $h$ is the class associated to the fibral $\mathbb{P}^1$ and $s_-$ the class of the zero section associated to the base $\mathbb{P}^1$. The intersection numbers are given by 
\begin{equation}
 s_-\cdot h =1 \,,\qquad s_-\cdot s_-=-1\,,\qquad h\cdot h=0\,,
\end{equation}
The tension of strings $(\mathtt{h},\mathtt{s})$ obtained from D3-branes wrapped on the respective curves are  
\begin{equation}\label{6dtensions}
 \frac{T_\mathtt{h}}{M_{\rm IIB}^2} = 2\pi \, \mathcal{V}_h = 2\pi t^2\,,\qquad \frac{T_{\mathtt{s}}}{M_{\rm IIB}^2} =2\pi \,\mathcal{V}_{s_-}=2\pi t^1\,. 
\end{equation}
The string $\mathtt{h}$ can be identified with the perturbative heterotic string (compactified on K3). The analogue of the strong coupling singularity in 4d now arises at the point at which the zero section $s_-$ vanishes, i.e. at $t^1 = 0$. At this point the string $\mathtt{s}$ becomes tensionless as is clear from \eqref{6dtensions}. For $B_2=\mathbb{F}_1$ this string is a non-critical E-string. Also in 6d the point $t^1=0$ is a strong coupling singularity since any D7-brane gauge theory that we could engineer on $s_-$ would become strongly coupled.\footnote{Notice that unlike in the 4d case, we do not assume standard embedding for the dual heterotic string in the 6d case. Hence, here the perturbative heterotic gauge theory can, in principle, be broken completely.} At this point in moduli space, the tension of the heterotic string is given by
\begin{equation}
\left( \frac{T_\mathtt{h}}{M_{\rm pl}^2} \right)^2=  \frac{\pi  (t^2)^2}{ (t^1 t^2 + \frac{1}{2} (t^2)^2)}\stackrel{t^1 =0}{ \longrightarrow } 2\pi\,.
\end{equation}
Thus also in 6d the perturbative heterotic string has a tension of order of the Planck scale in the vicinity of the strong coupling singularity. In fact the analogy to 4d goes even further: Since the strong coupling singularity corresponds to the blow-down $\mathbb{F}_1\rightarrow \mathbb{P}^2$ the string $\mathtt{h}$ is not part of the BPS spectrum beyond that point since the single divisor class of $\mathbb{P}^2$ corresponds to the class $s_-+h$. Thus, again the perturbative heterotic string gets replaced by a bound state of itself with a non-critical $E$-string that is the relevant BPS state beyond strong coupling. Notice that the string $\mathtt{h}+\mathtt{s}$ can never become tensionless but always has a tension of order of the Planck scale. In \cite{Katz:2020ewz} such strings are called supergravity strings that exist throughout the entire moduli space. On the other hand, strings like the E-string that become tensionless at finite distance can be considered field theory strings since they only give rise to finite number of light degrees of freedom. The middle-ground between these two classes of strings are strings like the heterotic strings obtained from D3-branes wrapping curves with vanishing self-intersection. These can become tensionless in Planck units but only at infinite distance. In 4d the analogue of these strings are given by the axionic strings \cite{Long:2021jlv}.\footnote{Replacing the Planck scale by the 4d species scale allows for a refined classification of the axionic strings in analogy with the 5d supergravity strings \cite{Cota:2022yjw}.} On the other hand, the non-critical string $\mathtt{P}$ associated to the singularities such as $\{\Delta_P^F=0\}$ can be thought of as the analogues of the field theory strings in 6d. Finally, the supergravity strings can be identified in 4d with the bound states of axionic strings and regulator strings. Similar to their 6d cousins, these strings can never become tensionless in Planck units. Following the analysis of \cite{Marchesano:2022avb} this is due to the fact that these latter 4d strings correspond to monodromies in field space with finite order. 

One might have expected that in 4d the relevant object is the D3-brane instanton on $\mathcal{S}_-$ that takes over the role the E-string in 6d. The relevant instanton in this case is the gauge instanton obtained by wrapping a D3-brane on $\mathcal{S}_-$ which gives a non-perturbative contribution to the superpotential 
\begin{equation}
W= M_\text{het}^3 \exp(-\frac{2\pi}{b_{E_8}} S_0) \,,
\end{equation}
where we subtracted a 1-loop term relating the heterotic and the type IIB string scale. Since $S_0=0$ along the singular locus, the scale of the gaugino condensate is of the order of the (naive) heterotic string scale. This is yet another indication that on the singular locus, we are leaving the perturbative heterotic regime. In fact the presence of this unsuppressed D3-brane instanton effect merely indicates that we reach the border of the geometric phase of the theory, but the instanton itself does not give any degrees of freedom associated to this singularity. As discussed already in \cite{Mayr:1996sh} these additional states instead come from a string-like object associated to the singular locus which in our case is the string $\mathtt{P}$ that becomes tensionless along $\{\Delta_P^F=0\}$. The fact that this singularity is associated to a locus where the action of an instanton vanishes tells us that this string is of non-geometric origin. Notice that at the classical level the contribution to the superpotential coming from D3-brane instantons on $\mathcal{S}_-$ might have been thought to be negligibly small since $\text{Re}\,S_0\rightarrow \infty$. However, our analysis revealed that due to the interplay between the F-theory K\"ahler and complex structure sector this conclusion does not persist at the quantum level since in fact the action of the D3-instanton vanishes along $\{\Delta_P^F=0\}$ due to quantum effects. 
\subsubsection{Relation to emergent string conjecture}

We argued that at the strong coupling singularity in the F-theory scalar field space, the D3-brane on $C^0$  leaves the BPS spectrum whereas another, non-geometric, non-critical string becomes tensionless forming a BPS bound state with the critical string. This behavior is interesting from the perspective of the emergent string conjecture\cite{Lee:2019oct}. This conjecture states that any infinite distance limit in the scalar field space of a consistent theory of gravity either corresponds to a limit in which a critical string becomes weakly coupled or to a decompactification limit. Phrased differently, at infinite distance the theory either reduces to a weakly coupled, perturbative string theory, or lifts to a higher dimensional theory.

The string theory limits of F-theory considered in this work correspond to the former case with the critical string corresponding to the D3-brane on $C^0$.  In order for the emergent string conjecture to be realized it is important that the critical string is actually part of the light string spectrum since its perturbative excitations need to furnish the tower of light states required by the Swampland Distance Conjecture. This is indeed the case as long as we are far away from the locus $\{\Delta_P^F=0\}$, i.e. as long as we are below the green, solid line and to the left of the blue, dashed line in figure~\ref{fig:modulispace}. This is the phase of the $\cN=1$ EFT that can effectively be described as a heterotic NLSM with the D3-brane identified with the heterotic string. However, once we move across the singularity $\{\Delta_P^F=0\}$ this string ceases to be BPS. Classically, one would have expected to still find emergent string limits in the green region \cite{Lee:2019jan, Klaewer:2020lfg}. However, the presence of $\{\Delta_P^F=0\}$ obstructs such would-be emergent string limits. The observation that the obstruction of an infinite distance limit and the absence of a critical string go hand-in-hand can in fact be viewed as further evidence for the relation between perturbative strings and infinite distances also in the context of the Distant Axionic String Conjecture \cite{Lanza:2021udy}: If there is no light, perturbative string, there should be no infinite distance and, on the other hand, if there is no infinite distance there should be also no light perturbative string. The analysis of \cite{Marchesano:2022avb} suggests that we can associate boundary divisors of $\cM^F$ with strings. Therefore, also the shaded region in figure~\ref{fig:modulispace} should be associated with a string. This string corresponds to the $\mathtt{H}+\mathtt{P}$ string which we expect to have a mass of order of the Planck scale throughout moduli space. Therefore, there is no infinite distance limit in the green region, i.e. no limit where the string $\mathtt{H}+\mathtt{P}$ becomes light and weakly-coupled.  Let us mention that this is consistent with the \textit{anti-emergence} discussed in \cite{Hamada:2021yxy}. The infinite distance behavior of the field space metric is related to the light states arising in these limits. Once we go away from such limits, the description used in the vicinity of the infinite distance locus in field space eventually breaks down and quantum corrections remove any infinite distance. This is precisely what happens in our case where the infinite distance is linked to the tower of light excitations of the perturbative heterotic string. In regimes of the field space where the description in terms of the perturbative heterotic string breaks down, due to quantum corrections we hence should not encounter an infinite distance. This is essentially what we observe in the F-theory scalar field space. 

Notice that there is an infinite distance limit also to the right of the blue, vertical line in figure \ref{fig:modulispace}. This infinite distance corresponds to the $\text{Re}\,S_0\rightarrow \infty$ divisor for $\text{Re}\,T_1<\text{Re}\,T_1^*$. In this phase, we thus also expect a perturbative string becoming light and weakly-coupled. Again, the string in question is a weakly-coupled heterotic string. This time, however, this heterotic string does not have a worldsheet description in terms of an NLSM with CY target space, but corresponds to a string with a Landau-Ginzburg worldsheet theory. Accordingly, this string does not have a dual description in terms of a D3-brane in F-theory wrapping some curve. Due to the tangency between $\text{Re}\,S_0=\infty$ and $\Delta_P^F=0$ the two phases of the EFT moduli space describable by a critical, perturbative string are separated by a strong coupling region. In particular it is not possible to traverse from the NLSM phase to the LG phase of the $\cN=1$ EFT while keeping the D3-brane on $C^0$ part of the light string spectrum.
%
%
%
%
%
%
%
%
%
%
%
%
%
%
%
%
%
%
%
%
%
%
%
\section{Conclusions} \label{sec:discussion}
In this paper we have investigated the structure of the F-theory scalar field space away from the strict large volume/large complex structure limit.  More precisely, we aimed to identify the border of asymptotic regimes where the asymptotic expansions break down and investigated the physical origin of obstructions \cite{Klaewer:2020lfg} to classically allowed emergent string limits. To facilitate our analysis, we did not attempt to uncover the full interior of the scalar field space, but focused on regimes where the full F-theory is dominated by the physics of a light string. Most importantly, by exploiting the relation to the theory of the light string, we were able to identify a new kind of strong coupling singularity in F-theory that signals the transition to a strong coupling phase of the $\cN=1$ EFT. 

In our analysis we focused on regimes in the F-theory scalar field space, $\cM^F$, in which a critical string, obtained from wrapping a D3-brane on a certain curve in $B_3$, is classically lighter than any other quantum gravity scale in the theory. More precisely, we imposed the decoupling limit $M_\text{string}/M_{\rm IIB}\rightarrow 0$. In this limit the full F-theory reduces to the theory of the string and we can identify the residual field space orthogonal to the $M_\text{string}/M_{\rm IIB}\rightarrow 0$ direction with the deformation space of the theory realized on the light string's worldsheet. The analysis presented in this work focused on the case that the light string is a heterotic string and that its worldsheet theory allows a description in terms of the IR limit of a simple GLSM. Thus, our analysis certainly lacks generality but this choice allowed us to make concrete statements about the structure of $\cM^F$ in the string theory limit. In particular, we were able to use the language of GLSMs to identify singular loci in the residual moduli space that, by duality, also have to be present in the full F-theory field space. We argued that, for the heterotic string with standard embedding, the principal component of the discriminant locus corresponds to a strong coupling singularity for the unbroken $E_8$ gauge group. This statement remains true even when considering gauge bundles that are deformations of the CY tangent bundle. This could be seen most clearly by analyzing the correlators which all become singular along the principal component of the singularity reflecting the strong coupling singularity for the heterotic string. In contrast to that, the heterotic analysis further showed that along the other components of the singular locus (not corresponding to the principal component) only a subset of correlators are singular indicating that only a subsector of the theory is singular. 

To translate the strong coupling singularities and phases into the F-theory language, we used the perturbative corrections to the F-theory scalar field space as guideline. More precisely, we showed that from the analysis of these perturbative corrections the presence of a strong coupling singularity for the unbroken $E_8$ gauge group inside the classical K\"ahler cone of $\cM^F$ can be inferred. Using the perturbative F-theory analysis we were able to deduce the dependence of the singular locus on the tree-level string coupling of the light string. This allowed us to infer the structure of the scalar field space away from the strict weak-coupling limit. Finally, in the spirit of \cite{Marchesano:2022avb} we gave an interpretation of the strong coupling phase in terms of the light BPS-string spectrum. Namely, the strong coupling singularity signals that the D3-brane wrapped on the fiber of $B_3$ leaves the BPS-spectrum and gets replaced by a bound state including a non-critical string associated to the singularity. This non-critical string itself becomes light at the singularity. In this paper, we did not attempt to explicitly describe the full backreaction of the critical D3-brane string. This would require to supplement the analysis of \cite{Marchesano:2022avb} to include a non-trivial scalar potential. We leave this task for future work. However, in the light of the analysis of \cite{Mayr:1996sh}, we argued that the string associated to the strong coupling singularity is closely linked to the presence of unsuppressed contributions to the superpotential and therefore it can indeed serve as a regulator string for the backreaction of the critical string once non-perturbative corrections to the superpotential are taken into account. As we described, this string can be viewed as the 4d avatar of the E-string in 6d F-theory.  Let us stress, however, that the kind of strong coupling discussed here do not simply correspond to a geometric strong coupling point where the classical volume of some divisor hosting a 7-brane gauge theory vanishes. Instead, this strong coupling behavior is indeed "quantum" since it is induced by quantum effects and not visible from a purely geometric analysis. In fact, from the geometric point of view one would have not expected to see such a strong coupling behavior in our limit of field space as the classical volume of the divisor hosting the gauge theory becomes large in this limit. 

The analysis presented in this paper certainly constitutes just a first step towards uncovering the general structure of the F-theory scalar field space. In particular, since we wanted to have computational control over the residual field space in the string theory limit, we restricted to cases with $E_6\times E_8$ gauge group. It would be interesting to investigate other models with different gauge groups along the lines of the analysis presented here. Another simplification arose due to the fact that the residual field space in our case is an actual moduli space and hence we did not need to care about non-perturbative superpotentials other than the $E_8$ gaugino condensate. As a natural generalization one could investigate how our results extend to cases where the residual field space also has a non-vanishing superpotential due to heterotic worldsheet instantons. 

Still, we believe that our analysis provides an interesting insight in the structure of $\cN=1$ field spaces beyond the weak coupling/large volume regime. It illustrates that for scalar field spaces of $\cN=1$, one has to be careful when applying the intuition of $\cN=2$ moduli spaces to them. Not only does one have to worry about the presence of a non-trivial scalar potential that can render some directions massive, but one also has to take into account K\"ahler potential corrections not present in $\cN=2$ theories. In our particular example these corrections spoil the factorization of the scalar field space into K\"ahler and complex structure sector. Therefore regions in the K\"ahler quasi-moduli space that seem to be well under control from our $\cN=2$ intuition can be outside the regime of perturbative control due to the mixing of the complex structure and K\"ahler sector. In our analysis this feature is most striking for the heterotic string coupling $(\text{Re}\,S_0)^{-1}$ which classically is tuned to very small values thus naively allowing for a perturbative description. However, due to the mixing between complex structure and K\"ahler sector this is not the case at the quantum level and we encounter a strong coupling singularity. If we followed our $\cN=2$ moduli space intuition we would have missed such a strong coupling singularity and the associated unsuppressed D3-brane instanton contribution to the scalar potential and would have declared the would-be emergent string limit a perturbatively controlled regime (cf. also the discussion in appendix E of \cite{Klaewer:2020lfg}). This observation can be viewed as an $\cN=1$ analogue of the obstruction observed in the $\cN=2$ hypermultiplet moduli space in \cite{Marchesano:2019ifh, Baume:2019sry}. In the case of type IIB CY threefold compactifications, it was shown that large volume limits in the hypermultiplet moduli space are obstructed at finite 4d string coupling due to quantum corrections. From the perspective of the K\"ahler moduli space of type IIA on the same CY threefold such limits seem to be perfectly valid infinite distance limits. However, in type IIB the K\"ahler moduli space and the moduli space of the 4d dilaton do not factorize thus leading to an obstruction against reaching the classical infinite distance limits. In the 4d $\cN=1$ setting analyzed in this work, we see a similar pattern caused by the additional complication that also the complex structure sector does not factorize from the K\"ahler sector. 

Finally, our results show a very interesting interplay between the spectrum of light, weakly-coupled strings and strong coupling singularities in field space. To be precise, we found that the borders of asymptotic regions associated to an emergent strings are precisely those loci at which the emergent string ceases to be part of the light BPS string spectrum. This observation can be viewed as supporting the Distance Conjecture/Emergent String Conjecture since it clearly shows that if there is no emergent string there is no infinite distance limit and \emph{vice versa}. It would be very interesting to extend this analysis to asymptotic limits that do not qualify as emergent string limits but are genuine decompactification limits to perhaps uncover more of the interior structure of $\cM^F$.

\subsubsection*{Acknowledgements}
It is a pleasure to thank Alberto Castellano, Alvaro Herraez, Severin L\"ust, Fernando Marchesano, Ilarion Melnikov, John Stout, and Timo Weigand for helpful discussions and correspondence. I am particularly grateful to Timo Weigand for useful comments on the draft. This work is supported in part by a grant from the Simons Foundation (602883, CV) and also by the NSF grant PHY-2013858. 
\appendix
\section{Essentials of heterotic/F-theory duality}\label{sec:hetFduality}
Consider F-theory on an elliptically-fibered Calabi--Yau four-fold over some base $B_3$ that has a compatible $K3$-fibration over some base $B_2$, i.e. the base $B_3$ has itself to be a rational fibration over $B_2$, $\rho: B_3\rightarrow B_2$. By heterotic/F-theory duality \cite{Morrison:1996na,Morrison:1996pp} this can be dual to the heterotic string on a Calabi--Yau manifold $Z_3$ that itself is elliptically-fibered over the same base $B_2$:
\begin{equation}
p: T^2\; \rightarrow \;B_2\,.
\end{equation} 
Under heterotic/F-theory duality, the complexified volumes, $t^a$, of the curves inside $B_2$ (in heterotic string units) are identified with the volumes (in type IIB units) of divisors on the F-theory base $B_3$ that are vertical w.r.t  the projection $\rho$. We thus have the classical dictionary between chiral scalar fields in the effective $\cN=1$ action:
\begin{equation}\label{TalphaFhet}
 T_a= -i\eta_{a b} t^b\,.
\end{equation}
Here $\eta_{a b}$ is the intersection form on $B_2$ and the chiral fields $T_a$ are defined in \eqref{eq:defTi}. On the other hand, the volume of the base $B_2$ on the F-theoretic side is classically identified with the heterotic string coupling. The duality requires that if on the F-theory side the gauge group is a subgroup of $E_8\times E_8$ the first factor is realized on the exceptional section $D_0\equiv \mathcal{S}_-$ of $B_3$ and the second factor on $\mathcal{S}_+ = \mathcal{S}_- + \rho^* c_1(\mathcal{T})$. Here $\mathcal{T}$ is the bundle that describes the twist of the $\mathbb{P}^1$ over $B_2$ leading to the rationally fibered $B_3$. We can then identify
\begin{equation}\label{couplings}
\frac{1}{g_{\rm YM,1}^2} = 2 \pi \mathcal{V}_{\mathcal{S}_-} \,,\qquad \frac{1}{g_{\rm YM,2}^2} = 2 \pi \mathcal{V}_{\mathcal{S}_+} \,,\qquad \frac{1}{g_{\rm het}^2} = 2 \pi \left(\mathcal{V}_{\mathcal{S}_-} + \frac{1}{2} \mathcal{V}_{\rho^* c_1(\mathcal{T})}\right)\,, 
\end{equation}
where $g_{{\rm YM},i}$ is the gauge coupling of the gauge theory of the $i$-th factor, and $g_{\rm het}$ is the heterotic string coupling. If we take $\mathcal{V}_{\mathcal{S}_-}$ to be large, these three couplings agree asymptotically up to 1-loop terms as analyzed in \cite{Klaewer:2020lfg}. 

In the main part of the paper we are interested in regions of the F-theory scalar fields space in which $\cM^F$ effectively reduces to the deformation space of the heterotic worldsheet theory. In order to be in such a regime, we need to ensure that the D3-brane wrapped on the generic fiber, $C^0$, of $B_3$ indeed corresponds to a perturbative heterotic string that is weakly coupled and has a stable gauge bundle. The first condition is classically ensured if we take the limit of large base $B_2$ as is clear from \eqref{couplings}. The second condition is more subtle. In section \ref{ssec:strongcoup}, this second condition is crucial to correctly identify the non-perturbative singularity structure in the F-theory moduli space. Therefore let us review in some detail how this is achieved in the context of heterotic/F-theory duality. To that end, we start by recalling that the heterotic bundle data can be mapped to the F-theory compactification data via the so-called spectral cover construction \cite{Friedman:1997yq,Friedman:1997ih} (for reviews cf. \cite{Donagi:2008ca, Weigand:2010wm}). Therefore take the heterotic string compactified on the elliptically-fibered Calabi--Yau three-fold $Z_3$. A gauge bundle $V_1, V_2$ inside the $E_8\times E_8$ gauge group needs to satisfy the Hermitian Yang--Mills equations 
\begin{equation}
F^{0,2}=F^{2,0} =0\,,\qquad g_{i\bar j} F^{i\bar j}=0\,. 
\end{equation}
If restricted to the torus fibers of $Z_3$ these equations imply that a bundle $V$ is flat along the $T^2$ fibers. The moduli space of flat bundles on a $T^2$ is just the torus itself. Hence, for instance a bundle with structure group $U(n)$ can be characterized by a selection of $n$ points on the $T^2$. For an $SU(n)$ bundle one further needs to ensure that the $n$ points sum to zero in the group law of the torus. Fibering these points holomorphically over the base $B_2$ we then get a curve $\mathcal{C}_n$ which is an $n$-fold cover of the base $B_2$. 

For the bundle moduli along the torus to decouple from the geometric moduli one in fact needs to ensure that the torus is large. Thus, we are considering $Z_3$ with an elliptic fibration in the limit of large fiber volume with a selection of $n$ points describing an $n$-fold cover, $\mathcal{C}_n$, of the base $B_2$. Let us assume that $Z_3$ is a Weierstrass model with zero section $E_-$ described by the hypersurface given by 
\begin{equation}
y^2= x^3 + fx z^4 + gz^6\,,
\end{equation}
where $\{x,y,z;f,g\}$ are sections of $\{K_{B_2}^{-2}, K_{B_2}^{-3}, \mathcal{O}; K_{B_2}^{-4}, K_{B_2}^{-6}\}$. The spectral cover $\mathcal{C}_n$ is described by an equation 
\begin{equation}
a_0z^n+a_2 xz^{n-2}+a_3 y z^{n-3} + \dots +a_n x^{n/2} =0\,. 
\end{equation}
Here the $a_i\in \Gamma(Z_3, K_{B_2}^i)$. To define $\mathcal{C}_n$ as we fiber the torus over $B_2$, we need to introduce a line bundle $\mathcal{L}$ such that the $a_i$ are lifted to sections of $\mathcal{L}\otimes K_{B_2}^i$. The class of the spectral cover is then given by 
\begin{equation}
 [\mathcal{C}_n] = n[E_-] + p_*\,c_1({\cal L}) \in H^2(\tilde Z_3, \mathbb{Z}) \,.
\end{equation}
In order to describe how the bundle $V$ varies over the base $B_2$ we need a second ingredient, the spectral line bundle $\cN$ on $\mathcal{C}_n$. This line bundle has the property 
\begin{equation}
p_{n,*} \mathcal{N} = V|_{B_2}\,,
\end{equation}
where $p_n:\,\mathcal{C}_n\rightarrow B_2$ is the natural projection of the spectral cover to the base. In particular, the first Chern class of this line bundle is given by 
\begin{equation}\label{c1N}
c_1(\cN) = -\frac{1}{2}\left(c_1(\mathcal{C}_n) -p^*_nc_1(B_2)\right)+\lambda\left(n\left[E_-\cdot \mathcal{C}_n\right]- p_n^*[c_1(\mathcal{L})-n c_1(B_2)]\right)\,,
\end{equation}
where $\lambda$ is a half-integer. The spectral line bundle is thus specified by a choice for $\lambda$. 

Given the spectral data specified by $(\mathcal{C}_n, \mathcal{N})$ one can recover the heterotic gauge bundle $V$ via a Fourier--Mukai transform. To that end, let us consider the Jacobian fibration $\tilde Z_3$ associated to $Z_3$. The Fourier--Mukai transform can now be defined with respect to the product fibration
\begin{equation}\begin{aligned} 
\begin{tikzcd}
    & Z_3\times_{B_2} \tilde Z_3 \arrow["{\pi_1}",swap, ld] \arrow[d,"{\varrho}"]\arrow["{\pi_2}", rd]&\\
    Z_3 
    & B_2 &\tilde Z_3     
\end{tikzcd}
\end{aligned}\end{equation}
where $\pi_{1(2)}$ are the projections on the first (second) factor and $\varrho$ is the projection on the common base $B_2$. One can further define the space $\widehat{\mathcal{C}}_n = \mathcal{C}_n \times Z_3$. The bundle $V$ is now given by the Fourier--Mukai transform of $\cN$: 
\begin{equation}\label{Fourier-mukai}
V=\pi_{1*} \left(p^*_{\widehat{\mathcal{C}}_n}\mathcal{N} \otimes \mathcal{P}\right)\,, 
\end{equation}
where the kernel of the transform is given by the Poincar\'e sheaf $\cP$ on $\hat{\mathcal{C}}_n$
\begin{equation}
\cP=\cO\left(\Delta-\sigma_1 \times \tilde Z_3 -Z_3 \times \sigma_2\right)\otimes \varrho^*c_1(B_2)|_{\mathcal{C}_n}\,,
\end{equation}
with $\Delta$ the diagonal divisor in $Z_3\times_{B_2} \tilde Z_3$ and $\sigma_{1,2}$ sections of the first and second factor in $Z_3\times_{B_2} \tilde Z_3$, respectively. Calculating the first Chern class of this bundle one confirms \eqref{c1N}. 

Apart from being flat along the torus fiber, there is a second condition on the heterotic K\"ahler moduli that needs to be ensured for the bundle to be stable. The bundles constructed via the spectral cover are holomorphic and flat along the elliptic fibers. To get a stable bundle one needs to ensure that $g_{i\bar j}F^{i\bar j}=0$ on the entire three-fold $Z_3$. This condition can be ensured if we work in the adiabatic limit \cite{Friedman:1997ih, Andreas:2003zb}:
\begin{equation}\label{adiabatic}
 \cV_{B_2}\gg \cV_{T^2}^H \,. 
\end{equation}
 A consequence of this hierarchy is that the volume of $Z_3$, and hence the K\"ahler potential on the heterotic K\"ahler moduli space, factorizes. Denoting the volume of the heterotic fiber by $s^0$ and the volume of the curves $C_a\subset B_2$ by $s^a$ we indeed find 
\begin{equation}
 K_H = - \log\left[\frac{1}{2}s^0 \eta_{ab} s^a s^b + \mathcal{O}\left[( s^0)^2, s^a \right]\right]\,.
\end{equation}

To translate the heterotic spectral data into the geometric data of the F-theory compactification, we have to take the so-called stable degeneration limit in the complex structure moduli space on the F-theory side. In this limit, the K3-fiber of the F-theory four-fold splits into a union of two $dP_9$ surfaces $W_1,W_2$. Since the stable degeneration limit plays an important role in the main text, let us briefly review how this limit is obtained (cf. e.g. \cite{Aspinwall:1997ye, Kachru:1995wm}). Therefore consider a family of elliptic four-folds with Weierstrass equation 
\begin{equation}
 y^2 = x^3 + f_\xi x z^4+g_\xi z^6\,,
\end{equation} 
where $f_\xi, g_\xi$ are polynomials of the coordinates on the base of degree 8 and 12, respectively. Moreover, $\xi$ is parametrizing the family of four-folds and $\xi \rightarrow 0$ corresponds to the stable degeneration limit. Let $v$ denote the projective coordinate on the fibral $\mathbb{P}^1$ of the F-theory base $B_3$ and consider e.g. the family of Weierstrass models 
\begin{equation}
  y^2 = x^3 + f_4v^4 x z^4 + \left( g_5 \xi  v^5 + g_6v^6+ g_7 \eta \xi v^7\right)z^6\,,
\end{equation} 
where $f_i, g_j$ are polynomials in the coordinates of $B_2$. In the limit $\eta \rightarrow 0$ one now obtains a degenerate four-fold with non-minimal singularities at $v=0, \infty$. These singularities can made minimal by blowing up e.g. the point $\xi =v=0$ by substituting 
\begin{equation}
 v\rightarrow v \xi'\,,\quad \xi \rightarrow \xi '\,,
\end{equation}  
and performing the coordinate change $x=\xi'^2 x'\,,\; y=\xi'^3 y$. As a consequence, at $\xi=0$ the fibral $\mathbb{P}^1$ of $B_3$ splits as into a chain of three rational curves. The two curves at the end now host an $II^*$ singularity together with two $I_1$ singularities such that the elliptic fibration over them gives rise to elliptic rational surfaces, i.e. $dP_9$ surfaces. On the other hand there are no singularities on the middle curve such that the elliptic fibration is just a product over this curve. Blowing down the middle curve one eventually arrives at the stable degeneration limit in which the K3-fiber of $Y_4$ degenerates into the union of two $dP_9$ surfaces intersecting along an elliptic curve $E$ which can be identified with the heterotic elliptic fiber. Notice that the complex structure of $E$ is determined by the functions $f_4$ and $g_6$. From here one can extract the data of the spectral cover by considering the defining equation for the $dP_9$ surfaces in Tate form 
\begin{equation}
0=y^2 +x^3+\alpha_1 xyz + \alpha_2 x^2 z^2 + \alpha_3y z^3 +\alpha_4 xz^4 + \alpha_6 z^6  + \sum_i p_i(x,y,z)\xi^i\,.
\end{equation}  
Here the $p_i$ are polynomials of degree $6-i$. For $SU(n)$-bundles we require $p_i=0$ for all $i$ except for 
\begin{equation}
p_1 = z^{5-n}(a_0 z^n + a_2 x z^{n-2} + a_3 y z^{n-3} + a_n x^{n/2})\,,
\end{equation}
which can be recognized as the equation determining the points on the heterotic torus that gave rise to the spectral cover $\mathcal{C}_n$. This illustrates how the spectral cover data can be obtained from the geometric F-theory data in the stable degeneration limit $\xi \rightarrow 0$. On the other hand, the data of the spectral line bundle $\cN$ is mapped to flux on the F-theory compactification which we are not discussing here.

Importantly, the large fiber limit on the heterotic side maps to the stable degeneration limit in F-theory where we can trust the duality between the heterotic string and F-theory. The stable degeneration limit in F-theory is a limit in complex structure moduli space and hence the factorization of the classical heterotic K\"ahler potential is reflected on the F-theory side by the classical factorization between complex structure and K\"ahler moduli space. 

\section{Review of GLSMs} \label{app:GLSMs}
In this appendix, we want to review some basic aspects of $(2,2)$ and $(0,2)$ Gauged Linear Sigma Models (GLSMs) that might be useful to understand the analysis of the heterotic moduli space in section \ref{sec:dual}, see \cite{Melnikov:2019tpl} for a more extensive review. Let us start with the $(2,2)$ Abelian GLSM \cite{Witten:1993yc} which is a two-dimensional $U(1)^d$ gauge theory with $(2,2)$ supersymmetry. Its field content consist of $m$ matter fields $(\Phi^i, \Gamma^i)$, $i=1,\dots, m$, that are coupled to vector multiplets $V_{a,\pm}$ with charges $Q^a_i$, $a=1,\dots, d$. The field strength for these gauge field is part of twisted chiral multiplets $(\Sigma_a, \Upsilon_a)$. In $(0,2)$ language, the charged matter $\Phi^i$ and the neutral matter $\Sigma_a$ are chiral bosonic multiplets, i.e. satisfy 
\begin{equation}
\bar{\mathfrak{D}}_+ \Phi^i =0\,,\qquad \bar{\mathfrak{D}}_+ \Sigma_a=0
\end{equation} 
where ${\mathfrak{D}}_+,\;\bar{\mathfrak{D}}_+$ are the $(0,2)$ superspace derivatives. On the other hand, $\Gamma^i$ are Fermi multiplets satisfying 
\begin{equation}
\bar{\mathfrak{D}}_+ \Gamma^i = \sqrt{2} E^i(\Phi^i,\Sigma_a) \,, 
\end{equation}
for some coupling functions $E^i$, whereas $\Upsilon_a$ are chiral fermionic mulitplets. To preserve $(2,2)$ supersymmetry, the coupling functions $E^i$ have to satisfy 
\begin{equation}\label{app:E2,2}
E^i = i \sqrt{2} \Phi^i \sum_a Q_i^a \Sigma_a \,. 
\end{equation} 
One can now write down a supersymmetric action for these fields which can be split in three components. 
\begin{equation}\begin{aligned}
S_{\rm kin} &= \int d^2z d^2 \theta \left[-\frac{1}{8e_0^2}\bar \Upsilon_a \Upsilon_a  - \frac{i}{2e_0^2}\bar \Sigma_a \partial_- \Sigma_a-\frac{i}{2}\bar \Phi^i\left(\partial_- +i Q_i^a V_{a,-} \right)\Phi^i -\frac{1}{2} \bar \Gamma^i \Gamma^i \right]\,,\\
S_\text{ F-I} &= \frac{1}{8\pi i}\int d^2 z d \theta^+ \Upsilon_a \tau^a|_{\bar \theta^+ =0} + {\rm h.c.}\,,\\
S_{\rm J} &=\int d^2 z d\theta^+\Gamma^i \mathcal{J}_i(\Phi)|_{\bar \theta^+ =0} + {\rm h.c.}\,.
\end{aligned}\end{equation}
Here, we introduced the complex Fayet-Iliopulos terms $\tau^a = \theta^a + i r^a$ and the couplings $\mathcal{J}_i(\Phi)$ are polynomials in $\Phi^i$ of charge $-Q^a_i$. To preserve $(2,2)$ supersymmetry these couplings need to derive from a superpotential $W(\Phi^i)$ as 
\begin{equation}
 \mathcal{J}_i = \frac{\partial W}{\partial \Phi^i}\,.
\end{equation} 
Furthermore, $(0,2)$ supersymmetry requires $E\cdot \mathcal{J}=0$. One now obtains the scalar potential by integrating out all auxiliary fields 
\begin{equation}
 U_{\rm bos} = 2 \sum_{i=1}^n |\phi^i|^2 \left|\sum_{a=1}^d Q_i^a \sigma_a \right|^2 + \sum_{i=1}^n |\mathcal{J}_i|^2 + \frac{e_0^2}{2}\sum_{a=1}^d D_a^2\,,
\end{equation}
where the D-term is given by
\begin{equation}
D_a = \sum_{i=1}^n Q^a_i |\phi^i|^2 -r^a \,. 
\end{equation}
In the above expressions $\phi^i$ and $\sigma_a$ are the leading bosonic components of $\Phi^i$ and $\Sigma_a$, respectively. For large values of the FI-terms, i.e. $r^a \gg 1$, the solution to the low-energy theory is well-approximated by the non-linear sigma model (NLSM) with target space 
\begin{equation}
\cM (r) = \{D_a^{-1}(0)\}/U(1)^d\,.
\end{equation}  
This is obtained by integrating out the $\sigma_a$ fields that become massive as a consequence of solving the D-term constraint giving a vev for the $\phi^i$ fields. For general choices of the charges $Q_i^a$, we have an anomalous $U(1)$ symmetry which prevents us from introducing a superpotential $W(\phi^i)$, such that the $\mathcal{J}_i=0$ in this case. In this case, the target space of the NLSM is a toric variety $V$ and the GLSM is also referred to as the \textit{V-model}. 

Given such a V-model, we can consider introducing an additional matter field $\Phi^0$ with charge $Q^a_0 = -\sum_{i=1}^n Q^a_i$ that cancels the anomaly and allows for the introduction of a superpotential 
\begin{equation}
 W = \Phi^0 P(\Phi^i) \,,
\end{equation}
where $P$ is a polynomial of multi-degree $\sum_i Q_i^a$. The vacuum conditions are now supplemented by
\begin{equation}
 P=0 \,,\qquad \phi^0 \frac{\partial}{\partial \phi^i} P =0\,. 
\end{equation}
For generic choices of the parameters of $P$, the hypersurface $\{P=0\}\subset V$ is smooth (up to orbifold singularities) such that we need to set $\phi^0=0$. In this case the low-energy theory is a NLSM with target space the Calabi--Yau hypersurface $X\subset V$ defined by $P=0$, the so-called \textit{X-model}. 

Apart from the vacua corresponding to the NLSM with target space $V$ there exist additional vacua characterized by large vevs for the $\sigma_a$ fields. These vevs in turn give mass to the charged matter fields. To find these vacua, one considers the effective action 
\begin{equation}
S_{\rm eff} =\int d^z d \theta^+ \Upsilon_a \frac{\partial \tilde{W}}{\partial \Sigma_a} |_{\bar \theta^+ =0} + \rm{h.c.}\,,
\end{equation}
for some twisted effective superpotential $\tilde W$. This superpotential is one-loop exact and can be calculated from the D-term tadpole:
\begin{equation}
\tilde J_a = \frac{\partial \tilde{W}}{\partial \Sigma_a} = -\frac{1}{8\pi i} \log \left[q_a^{-1}\prod_i \left(\frac{Q_i^b \sigma_b}{\mu}\right)^{Q_i^a}  \right]\,,
\end{equation}
where $\mu$ is the one-loop scale. The $\sigma_a$-vacua are obtained whenever $\tilde J_a =0$ for all $a=1,\dots d$. 

If we consider the X-model, the scale $\mu$ drops out due to the condition $Q_0^a = -\sum_i Q_i^a$. In this case there are also no isolated $\sigma$ vacua -- there are either Higgs-$\sigma$ vacua or flat $\sigma$-directions. In this case, we have 
\begin{equation}
 \tilde J_a = -\frac{1}{8\pi i} \log \left[q_a^{-1} (Q_0^a \sigma_a)^{Q_0^a}\prod_i(Q_i^b \sigma_b)^{Q_i^a}\right]\,,
\end{equation}
and the condition to find a $\sigma$ vacuum reads 
\begin{equation}\label{app:singularcondition}
 \prod_i(Q_i^b \sigma_b)^{Q_i^a} = q_a (Q_0^a \sigma_a)^{-Q_0^a}
\end{equation}
For generic values of $q$ this equation has no solution. However, for special values of $q_a$ the above equation is satisfied and corresponds to a flat $\sigma$ direction and hence to a singularity of the theory. The singular locus in the $q$-plane obtained from $\tilde J_a =0$ corresponds to the principal component of the singular divisor that we discuss in detail in the main text. 

One may now consider the standard A-twist of the GLSM which gives a topological field theory for which we want to calculate the correlators of the $\sigma_a$-fields. In the A-twisted model, these correlators localize on the $\sigma$-vacua and are given by \cite{Morrison:1994fr} 
\begin{equation}\label{Vmodelcorrelators}
\langle \sigma_{a_1} \dots \sigma_{a_k} \rangle_V = \sum_{\sigma|d\tilde W =0}\sigma_{a_1} \dots \sigma_{a_k}\left[\det\left(\text{Hess}\, \tilde W(\sigma)\right)\prod_i (Q^b_i \sigma_b)\right]^{-1}\,. 
\end{equation}
The quantum restriction formula \cite{Morrison:1994fr} then relates these correlators to the correlators in the related X-model via 
\begin{equation}\label{quantumrestriction}
\langle \sigma_{a_1} \dots \sigma_{a_k} \rangle_X = \langle \sigma_{a_1} \dots \sigma_{a_k} \frac{-K}{1-K^{-\sum_{a} Q_0^a} }\rangle_V \,,
\end{equation}
where $K= - \sum_{i>0} Q_i^a \sigma_a$ is the operator associated to the anti-canonical class on $V$. This correlator becomes singular whenever $1+K$ is not invertible. One can show that this is the case when $q$ is on the singular locus defined by \eqref{app:singularcondition}. 

One can now identify $\sigma_a$ with classes $\eta_a \in H^{1,1}(X)$ such that the GLSM correlators defined above in fact calculate the correlators of the NLSM on $X$. In particular, the GLSM gives a simple way to compute the loci where the NLSM becomes singular. Notice however, that the GLSM expansion in $q$ corresponding to gauge instanton is not the same as the NLSM instanton expansion corresponding to worldsheet instantons. To match these both one needs to know the precise mirror map. In the particular examples that we consider at least some information about the mirror map is known allowing to match the singular loci in the GLSM FI-parameter space to singularities in heterotic K\"ahler moduli space. 
\newline

So far we only considered the GLSM on the $(2,2)$ locus. In the main text, we also consider deformations away from this locus. The general picture remains the same since for this $(0,2)$ case analogous expressions exist \cite{McOrist:2007kp,McOrist:2008ji}. First, to obtain deformations of the tangent bundle, we can deform the couplings $E$ away from their $(2,2)$ expression \eqref{app:E2,2}. To do that, let us split the matter fields $\Phi^i$ into sets of equal charge 
\begin{equation}
\{\Phi^1, \dots, \Phi^n\} \rightarrow \cup_\alpha \{\Phi^{I_\alpha}_{(\alpha)}, I_{\alpha} = 1, \dots, n_\alpha\}\,,
\end{equation}
such that $\sum_{\alpha} n_\alpha =n$ and $Q_{I_\alpha}^a = Q_{J_\alpha}^a =Q_{(\alpha)}^a$ for all $a$. One can then consider the deformed $E$ couplings 
\begin{equation}
E_{(\alpha)}^{I_\alpha} = i\sqrt{2}\sum_{a=1}^d \Sigma_a \left[A^a_{(\alpha)}\right]^{I_\alpha}_{J_\alpha}\Phi^{I_\alpha}_{(\alpha)}\,.
\end{equation}
One can think of the deformations as being parameterized by the matrices $A^a_{(\alpha)}$. Notice that in order to preserve $(0,2)$ supersymmetry the condition $E\cdot \mathcal{J}=0$ still has to be satisfied. One can summarize the deformations into the matrices 
\begin{equation}\label{app:malpha}
M_{(\alpha)} = \sum_a \Sigma_a A^a_{(\alpha)}\,,
\end{equation}
which are the matrices appearing in \eqref{matricesM}. In terms of these the deformed $\tilde J_\alpha$ (for the $X$-model) reads \cite{McOrist:2007kp}
\begin{equation}
 \tilde J_a = -\frac{1}{8\pi i} \log \left[q_a^{-1} (Q_0^a \sigma_a)^{Q_0^a}\prod_\alpha(\det\,M_{(\alpha)})^{Q_{(\alpha)}^a}\right]\,, 
\end{equation}
and the correlators \eqref{Vmodelcorrelators} become
\begin{equation}
 \langle \sigma_{a_1} \dots \sigma_{a_k} \rangle_V = \sum_{\sigma|d\tilde W =0}\sigma_{a_1} \dots \sigma_{a_k}\left[\det\left(\text{Hess}\, \tilde W(\sigma)\right)\prod_\alpha \det\,M_{(\alpha)}\right]^{-1}\,,
\end{equation} 
from which one can derive the $X$-model correlators via the quantum restriction formula \eqref{quantumrestriction}. 

\bibliography{papers_Max}
\bibliographystyle{JHEP}

\end{document}